\documentclass[aps, prc, twocolumn, tightenlines, letterpaper, amsmath, amssymb, preprintnumbers, superscriptaddress, floatfix, longbibliography, nofootinbib]{revtex4-2}

\usepackage[T1]{fontenc}
\usepackage{bm}
\usepackage{xspace}
\usepackage[dvipsnames]{xcolor}
\usepackage{graphicx}
\usepackage{tabularx}
\usepackage{enumitem}
\usepackage{braket}
\usepackage{dsfont}
\usepackage[normalem]{ulem}

\usepackage{amsmath, nccmath}
\usepackage{lipsum}
\usepackage{cancel}
\usepackage{bm}
\usepackage[linesnumbered,ruled,vlined]{algorithm2e}
\usepackage{algpseudocode}
\usepackage{physics}
\usepackage{amsfonts}

\newcolumntype{Y}{>{\centering\arraybackslash}X}
\usepackage{diagbox}

\DeclareMathAlphabet{\mathpzc}{OT1}{pzc}{m}{it}

\usepackage[hypertexnames=false]{hyperref}
\hypersetup{
    colorlinks=true,       % false: boxed links; true: colored links
    linkcolor=blue,          % color of internal links
    citecolor=blue,        % color of links to bibliography
    filecolor=blue,      % color of file links
    urlcolor=blue           % color of external links
}

\usepackage{orcidlink}

\usepackage{mathtools} %for coloneqq

\begin{document}

\title{Quantum Magic and Multi-Partite Entanglement in the Structure of Nuclei}

\author{Florian Br\"okemeier\,\orcidlink{0009-0005-9290-1686}}
%\email{}
\affiliation{Fakult\"at f\"ur Physik, Universit\"at Bielefeld, D-33615, Bielefeld, Germany}

\author{S. Momme Hengstenberg\,\orcidlink{0009-0009-8764-9980}}
%\email{}
\affiliation{Fakult\"at f\"ur Physik, Universit\"at Bielefeld, D-33615, Bielefeld, Germany}

\author{James W. T. Keeble\,\orcidlink{0000-0002-6248-929X}}
%\email{}
\affiliation{Fakult\"at f\"ur Physik, Universit\"at Bielefeld, D-33615, Bielefeld, Germany}

\author{Caroline E. P. Robin\,\orcidlink{0000-0001-5487-270X}}
\email{crobin@physik.uni-bielefeld.de}
\affiliation{Fakult\"at f\"ur Physik, Universit\"at Bielefeld, D-33615, Bielefeld, Germany}
\affiliation{GSI Helmholtzzentrum f\"ur Schwerionenforschung, Planckstra{\ss}e 1, 64291 Darmstadt, Germany}

\author{Federico Rocco\, \orcidlink{0009-0005-9826-012X}}
%\email{}
\affiliation{Fakult\"at f\"ur Physik, Universit\"at Bielefeld, D-33615, Bielefeld, Germany}

\author{Martin J.~Savage\,\orcidlink{0000-0001-6502-7106}}
\email{mjs5@uw.edu}
\thanks{On leave from the Institute for Nuclear Theory.}
\affiliation{InQubator for Quantum Simulation (IQuS), Department of Physics, University of Washington, Seattle, WA 98195}

\begin{abstract}
Motivated by the Gottesman-Knill theorem, 
we present a detailed study of the quantum complexity of $p$-shell and $sd$-shell nuclei.
Valence-space nuclear shell-model wavefunctions generated by the {\tt BIGSTICK} code are mapped to qubit registers using the Jordan-Wigner mapping
(12 qubits for the $p$-shell and 24 qubits for the $sd$-shell),
from which measures of the many-body entanglement ($n$-tangles)
and magic (non-stabilizerness) are determined. 
While exact evaluations of these measures are possible for nuclei with a modest number of active nucleons, 
Monte Carlo simulations are required for the more complex nuclei.
The broadly-applicable
Pauli-String $\hat I \hat Z$ exact (PSIZe-) MCMC technique is introduced to accelerate the evaluation of 
measures of magic in deformed nuclei (with hierarchical wavefunctions), by factors of $\approx 8$ for some nuclei.
Significant multi-nucleon entanglement is found in the $sd$-shell, dominated by proton-neutron configurations, along with significant measures of magic.  
This is evident not only for the deformed states, but also for nuclei on the path to instability via regions of shape coexistence and level inversion.
These results indicate that quantum-computing resources will accelerate 
precision simulations of such nuclei and beyond.
\end{abstract}

\maketitle

%%%%%%%%%%%%%%%%%%%%%%%%%%%%%%%%
\section{Introduction}
\label{sec:intro}
\noindent
Advances in quantum information science (QIS) have provided new perspectives on quantum many-body systems.
Investigations of entanglement have shed further light onto the structure and dynamics of matter (in nuclear physics, see e.g.~\cite{Beane:2018oxh,Beane:2020wjl,Beane:2021zvo,PhysRevC.107.025204,Bai:2022hfv,Bai:2023rkc,Bai:2023tey,Kirchner:2023dvg,Miller:2023snw,Miller:2023ujx,Johnson:2022mzk,PhysRevC.92.051303,Kruppa:2020rfa,Robin:2020aeh,Kruppa:2021yqs,PhysRevC.105.064308,Pazy:2022mmg,Tichai:2022bxr,Perez-Obiol:2023wdz,Gu:2023aoc,liu2023hints,Bulgac:2022cjg,Bulgac:2022ygo, PhysRevA.103.032426,Faba:2021kop,Faba:2022qop,Hengstenberg:2023ryt,Lacroix:2024drc,Bai:2023hrz,Bai:2024omg,Cervia:2019res,Patwardhan:2021rej,Lacroix:2022krq,Illa:2022zgu,Siwach:2022xhx,Roggero:2022hpy,Martin:2023ljq,Balantekin:2023qvm}), 
and, in turn, have been guiding the development of improved methods for describing quantum many-body systems based on entanglement optimization and/or truncation.
One example is tensor network techniques which can describe classes of weakly-entangled states efficiently with classical computers.
These includes, for example, the density matrix renormalization group (DMRG)~\cite{PhysRevLett.69.2863}, matrix product states (MPS)~\cite{https://doi.org/10.1007/BF02099178,https://doi.org/10.1007/BF01309281,PhysRevLett.91.147902,Perez-Garcia:2006nqo}, projected entangled pair states (PEPS)~\cite{Verstraete:2004cf} and the multi-scale entanglement renormalization ansatz (MERA)~\cite{PhysRevLett.99.220405,PhysRevLett.101.110501}. 
DMRG and variants thereof have been adapted to various kinds of nuclear systems for which a natural weakly-entangled bipartition can be established~\cite{PhysRevC.65.054319,Papenbrock_2005,PhysRevC.67.051303,PhysRevC.69.024312,PhysRevLett.97.110603,PhysRevC.79.014304,Dukelsky:2004vv,PhysRevC.78.041303,PhysRevC.88.044318,PhysRevC.106.034312,PhysRevC.92.051303,Tichai:2022bxr,Tichai:2024cyd,Gorton:2024hbb}.
Tensor-networks states, and other methods related to entanglement re-organization have also been shown to be beneficial in the context of quantum simulations with noisy intermediate-scale quantum (NISQ) computers~\cite{PhysRevResearch.1.023025,Noh_2020,PhysRevX.10.041038,Barratt:2020fjg,PhysRevResearch.3.033002,Chertkov:2021jwg,PhysRevResearch.4.L022020,Miao:2023her,PhysRevResearch.5.033141,PRXQuantum.4.020304,PhysRevA.109.062437,PhysRevA.109.L050402,Robin:2023pgi,Besserve:2024aom,Perez-Obiol:2024vjo}.
\\

Entanglement, however, is not the only necessary ingredient to establish whether a quantum state can be  efficiently simulated classically, or necessitates the use of a quantum computer. 
The degree of non-stabilizerness, or "magic", of a quantum state is also required to establish such distinction. 
Magic quantifies the deviation of a state from stabilizer states, which are a class of quantum states that can be prepared using Clifford operations alone. Stabilizer states can be 
highly entangled, and yet, as encapsulated by the Gottesman-Knill theorem, can be efficiently prepared and simulated with a classical computer~\cite{gottesman1998heisenberg}. 
As such, both entanglement and magic are required to assess the computational complexity and quantum resource requirements for simulating physical systems, and to design an optimal classical/quantum workflow of the simulations.
The exponentially-scaling classical resources required to prepare a given quantum state are 
directly related to the number of T-gates required to prepare the state on a digital quantum computer, 
and hence related to the non-zero magic of the state.
As nuclei, described by non-relativistic nucleons, involve a finite number of particles, the concept 
of asymptotic scaling with systems size, as is used to define complexity classes of problems, 
is replaced by the scaling of resources required to achieve a given level of precision in target observables, e.g., Ref.~\cite{Kashyap:2024wgf}.
Indirectly, this includes the scaling of the active space and the fidelity of the Hamiltonian, 
when compared with that emerging from quantum chromodynamics.
\footnote{A number of properties of nuclei can be comfortably computed with a certain level of precision using classical computing resources alone.  However, systematically reducing 
uncertainties in these properties will become increasingly demanding, and eventually beyond the capabilities of classical computing.}

As with entanglement, measures to quantify the magic in a quantum state have been developed, such as, for example, the minimum relative entropy of magic~\cite{Emerson:2013zse} quantifying the minimum distance between a quantum state and the nearest stabilizer state. Other measures such as, for example, the mana~\cite{Emerson:2013zse} and thauma~\cite{PhysRevLett.124.090505}, the robustness of magic~\cite{PhysRevLett.118.090501}, stabilizer extent~\cite{Bravyi2019simulationofquantum}, stabilizer norm~\cite{PhysRevA.83.032317} and stabilizer nullity~\cite{Beverland:2019jej}, are related to the minimum number of stabilizer states required to expand the quantum state of interest (stabilizer rank), and give an estimate of the computational complexity of Clifford simulations of a quantum circuit.
More easily accessible measures of magic have been introduced, in particular, the stabilizer R\'enyi entropies (SREs)~\cite{Leone:2021rzd} and the Bell magic~\cite{PRXQuantum.4.010301}, which have been shown to be measurable in quantum computing experiments~\cite{Oliviero_2022,PRXQuantum.4.010301,Bluvstein:2023zmt}, and efficiently calculable for MPS~\cite{Haug:2022vpg,Haug:2023hcs,Tarabunga:2024ugl,lami2024quantum}.

While the entanglement structures of various types of physical systems have been extensively examined, the magic properties of quantum many-body systems are much less known.
Investigations of magic in the Ising and Heisenberg models~\cite{Oliviero_2022,Haug:2023hcs,Rattacaso:2023kzm,frau2024nonstabilizerness,Catalano:2024bdh},
in one- and two-dimensional lattice gauge theories~\cite{Tarabunga:2023ggd,Falcao:2024msg},
and in potential simulations of quantum gravity~\cite{Cepollaro:2024qln},
have been performed as examples.

Towards building an understanding of the quantum resources required to simulate
nuclear systems, we have recently investigated the magic-power of the nucleon-nucleon and nucleon-hyperon S-matrices. 
We found that magic and entanglement do not always have the same behaviour in these scattering processes. In particular, certain scattering states were found to exhibit large entanglement and low or zero magic in specific energy regions.
Such differences in the behaviours of entanglement and magic have also been investigated in different contexts.
For example, Ref.~\cite{frau2024nonstabilizerness} showed that, in the one-dimensional Heisenberg model described by MPS, magic typically saturates faster than entanglement with respect to increase of the bound dimension.
In random quantum circuits including measurements, it has been demonstrated that scaling of entanglement and magic with sub-system size undergo phase transitions at different sampling densities, $p$~\cite{Fux:2023brx,Bejan:2023zqm}. Further, it has been demonstrated that the quantum phase transition in the XYZ Heisenberg chain can be signaled by measures of magic, while being undetected by entanglement measures~\cite{Catalano:2024bdh}.

In this work,
as another step towards a more general characterization of the quantum complexity of nuclear systems, 
we investigate and compare magic and entanglement patterns in the ground states of nuclei.
Specifically, we calculate magic and entanglement measures in light ($p$-shell) and mid-mass ($sd$-shell) nuclei, described within the framework of the well-established interacting nuclear shell model~\cite{RevModPhys.77.427}.
We compute SREs to measure the magic content of the wave functions, and the $n$-tangles to estimate detailed multipartite entanglement. Such $n$-tangles are well-suited measures to characterize shell-model wave functions which include refined multi-nucleon correlations in restricted model spaces.
While the $n$-tangles can be calculated exactly for the present model-space sizes, magic, which involves the evaluation of an exponentially-large number of Pauli strings, can only be calculated in an approximate way in typical $sd$-shell nuclei. Motivated by the work in Ref.~\cite{Tarabunga:2023ggd}, which employed Markov Chain Monte Carlo (MCMC) techniques to sample the Pauli strings and calculate the SREs in 2D lattice gauge theories, 
we have
performed an extensive and detailed suite of MCMC evaluations of Pauli strings for each nucleus, 
and have developed
a new MCMC algorithm, which we call PSIZe-MCMC, 
to accelerate sampling Pauli string expectation values 
within wave function exhibiting collectivity, such as many $sd$-shell nuclei.

Nuclei exhibit an impressive range of shape deformations.  Even light nuclei can be significantly deformed in their ground and excited states, induced by the two-nucleon tensor force and higher.  This is particularly pronounced in $sd$-shell nuclei, peaking in the vicinity of Mg. 
While the deformation parameters become smaller with increasing neutron number for a given proton number $Z$,
the systems may enter a region of shape-co-existence on their way to instability.  
An example of this behavior can be found in the Mg isotopes, where $^{24}$Mg is 
maximally deformed in the isotopic chain, and with $\beta$ decreasing with increasing neutron number. 
$^{28}$Mg is suspected to be in the region of shape co-existence on the road from normal (shell model) state ordering to inverted orderings around $^{32}$Mg (see, for example, Ref.~\cite{McNeel:2020rfd}),
brought about by a closing gap to the $fp$-shell.  This evolution is expected to arise from significant multi-particle correlations, both classical and quantum, in the nuclear wavefunctions.  
We take steps toward systematically quantifying the quantum correlations in these nuclei.

The manuscript is organized as follows: In section~\ref{sec:shell_model} we provide a brief reminder of the shell-model description of nuclear wave functions, and in section~\ref{sec:SMmap} we describe the mapping of these wave functions to qubits, which will serve for the subsequent calculations of entanglement and magic measures. 
In section~\ref{sec:multientanglement} we describe and present calculations of multipartite entanglement in $p$- and $sd$-shell nuclei via computations of $n$-tangles. 
In section~\ref{sec:magic} we briefly review the stabilizer formalism leading to the definition of SREs as a measure of magic, and present results for $p$- and $sd$-shell nuclei obtained via exact or MCMC computations using the newly-introduced PZIe-MCMC algorithm. 
In section~\ref{sec:comps} we examine, in more detail, the behaviour of entanglement and magic in comparison with deformation of nuclei of the Ne and Mg chains. 
Finally, conclusions and outlook are presented in section~\ref{sec:conclusions}.

\section{Relevant Aspects of the Spherical Shell-Model Description of Nuclei} 
\label{sec:shell_model}
\noindent
The interacting shell model~\cite{RevModPhys.77.427} is an active-space configuration-interaction method that has been successfully used for decades in the description of 
the structure of nuclei. 
Based on the large energy gaps between major single-particle (harmonic oscillator) shells, this framework assumes that only nucleons around the Fermi level are active and interact within a valence shell, while nucleons below a major shell gap can be considered frozen. 
The single-particle space is thus divided into three parts: 
the inert fully-occupied core of orbitals, 
the active partially-filled valence space, 
and the space of remaining empty orbitals.

A nuclear state is described as linear combination of Slater determinants, each representing a configuration of protons and neutrons within the valence space:
\begin{equation}
    \ket{\Psi} = \sum_{\alpha_\pi \alpha_\nu} A_{{\alpha_\pi}{\alpha_\nu}} \ket{\Phi}_{\alpha_\pi} \otimes \ket{\Phi}_{\alpha_\nu} \; ,
\label{eq:wf}
\end{equation}
where $\ket{\Phi}_{\alpha_\pi}$ and $\ket{\Phi}_{\alpha_\nu}$ respectively denote proton and neutron configurations which are obtained by creating protons and neutrons on top of  
a fully-filled core $\ket{0}_\pi \otimes \ket{0}_\nu$. 
Labeling the single-particle states $i\equiv \{ \mathpzc{n}_i, l_i, j_i, j_{z_i}, t_{z_i}\}$, where $\mathpzc{n}_i$ is the main quantum number, $l_i$ and $j_i$ are orbital and total angular momenta, $j_{z_i}$ is the projection of the total angular momentum and $t_{z_i}$ is the isospin projection, these configurations can be written as
\begin{align}
\ket{\Phi}_{\alpha_\pi} = \prod_{i \in \alpha_\pi} a^\dagger_i \ket{0}_\pi \; , 
\hspace{0.2cm}
\ket{\Phi}_{\alpha_\nu} = \prod_{i \in \alpha_\nu} a^\dagger_i \ket{0}_\nu \; ,
\end{align}
where $a^\dagger_i$ is the operator creating a nucleon in state $i$, which is restricted to the  valence space.

The valence nucleons interact via a two-body Hamiltonian of the form
\begin{align}
    \hat{H} 
    =\sum_i \varepsilon_i a^\dagger_i a_i \ +\  \frac{1}{4} \sum_{ijkl} \widetilde{v}_{ijkl} 
    \ a^\dagger_i a^\dagger_j a_l a_k \; .
    \label{eq:SM_Hami}
\end{align}
In order to compensate for the truncated Hilbert space, the (anti-symmetrized) interaction matrix elements $\widetilde{v}_{ijkl} = v_{ijkl}- v_{ijlk}$ and single-particle energies $\varepsilon_i$ are phenomenologically adjusted, {\it i.e.} they are fit to a set of experimental data, in order to reproduce properties (such as binding and excitation energies) of known nuclei. This provides a way to implicitly encompass the effect of neglected configurations, as well as the effect of many-nucleon (in particular three-nucleon) interactions. In this work we focus on $p$-shell and $sd$-shell nuclei, and use the corresponding Cohen-Kurath~\cite{COHEN19651} and usdb~\cite{PhysRevC.74.034315} interactions.

The expansion coefficients $\{ A_{\alpha_\pi \alpha_\nu}\}$ of the nuclear ground state in 
Eq.~(\ref{eq:wf}) are determined by diagonalizing the Hamiltonian in the space of many-body configurations $\{ \ket{\Phi}_{\alpha_\pi} \otimes \ket{\Phi}_{\alpha_\nu} \}$, which is equivalent to applying a variational principle to the energy $E(\Psi) = \langle \Psi | \hat H|\Psi \rangle $ of the system.

For the present study of quantum information and quantum correlations in light nuclei, 
we have used the {\tt BIGSTICK} shell-model code~\cite{JOHNSON20132761,Johnson:2018hrx} to perform this task, and generate the ground-state wave functions of $p$- and $sd$-shell nuclei, {\it i.e.} nuclei with a number of protons and neutrons $ 2 \leq Z,N \leq 20$. 
This code uses the so-called "M-scheme" where the configurations $\ket{\Phi}_{\alpha_\pi}$ and $\ket{\Phi}_{\alpha_\nu}$ have a good projection of total angular momentum $J_z$, but not necessarily a good total $J$. The nuclear wave function $\ket{\Psi}$ in Eq.~(\ref{eq:wf}) is however ensured to have a good $J$ because both the model space and the nuclear interaction are rotationally invariant. 
In this work we will generally choose the wave function in the maximum $J_z = J$ sector, and will explore other $J_z$ values for selected cases.
We also note that both the Cohen-Kurath~\cite{COHEN19651} and usdb~\cite{PhysRevC.74.034315} interactions are isospin symmetric, 
and that we do not include the Coulomb interaction between protons
nor strong-isospin breaking effects from the quark masses in quantum chromodynamics (QCD) 
and the electroweak sector. 
Therefore our results are fully symmetric under interchange of protons and neutrons.

\color{black}

%%%%%%%%%%%%%%%%%%%%%%%%%%%%%%%%%%%%%
\section{Mapping the Shell-Model Basis States to Qubits}
\label{sec:SMmap}
\noindent
The entanglement and magic measures 
that we present
below require 
a mapping of the shell-model wave functions to quantum registers.
While there are multiple ways to map the Hilbert space of the shell model to qubits, 
we have chosen to follow a common quantum chemistry (Jordan-Wigner) 
mapping where the occupancy 
of each single-nucleon orbital is defined by the orientation of a single qubit\footnote{
% There are a number of potential ways to map the Hilbert space spanned by the shell-model basis states to the registers of quantum computers.
% For example, building upon the work in Refs.~\cite{Illa:2023scc,Illa:2024kmf},
% the occupations of two $2s$ states can be mapped to the states of a $d=4$ qudit, a qu4it, 
% corresponding to the occupations $\{ (0,0),(0,1),(1,0),(1,1) \}$ of the 
% $(2s_{1/2, -1/2}, 2s_{1/2,+1/2})$ substates.  Similarly, the occupations of the four $1p_{3/2}$ 
% states can be mapped to a qudit with $d=16$. 
%
We note that the Jordan-Wigner mapping used here is of special significance, because this is also the way that the classical shell model codes, such as BIGSTICK, encode the many-body configurations. Therefore, the magic computed with such mapping is directly related to the computational complexity within the shell-model framework. Moreover, with such mapping, the $n$-tangles provide direct information about multi-nucleon entanglement, as noted in section~\ref{sec:entang} below, and allow for detailed analysis of which orbitals are involved in physical phenomena and collective behaviors.
Nonetheless, there are a number of other potential ways to map the Hilbert space spanned by the shell-model basis states to the registers of quantum computers, using, for example, alternative mappings to qubits (parity encoding~\cite{Seeley:2012zpt}, Bravyi-Kitaev~\cite{Bravyi:2000vfj}, binary encoding, Gray code~\cite{Gray_code}, and more) which would involve similar scalings. Alternatively one can employ mappings to higher dimensional qudits. For example, 
building upon the work in Refs.~\cite{Illa:2023scc,Illa:2024kmf},
the occupations of two $2s$ states can be mapped to the states of a $d=4$ qudit, a qu4it, 
corresponding to the occupations $\{ (0,0),(0,1),(1,0),(1,1) \}$ of the 
$(2s_{1/2, -1/2}, 2s_{1/2,+1/2})$ substates.  Similarly, the occupations of the four $1p_{3/2}$ 
states can be mapped to a qudit with $d=16$.
While such mappings to qudits are typically advantageous in terms of quantum resources~\cite{Illa:2023scc,Illa:2024kmf}, the calculations of quantum complexity measures on classical computers become more involved~\cite{Chernyshev:2024pqy}.
In any case, using such mappings is outside the scope of the present study and we leave it for future work.}
.
This mapping 
has been used previously in studies using the shell model, see e.g. Refs.~\cite{ovrum2007,PhysRevC.105.064308,PhysRevC.105.064317,PhysRevC.106.034325,PhysRevC.108.064305,Perez-Obiol:2023wdz,Perez-Obiol:2023vod},
as part of the growing effort in using quantum computers to simulate nuclei~\cite{Holland:2019zju,Roggero:2020sgd,Baroni:2021xtl,
Rigo:2022ces,Li:2023eyg,Turro:2023dhg,PhysRevLett.120.210501,PhysRevA.100.012320,PhysRevC.105.064308,DiMatteo:2020dhe,Robin:2023pgi,Illa:2023scc,Siwach_2022,Lacroix:2020nhy,RuizGuzman:2021cdj,RuizGuzman:2021qyj,Lacroix:2022vmg,Ayral:2023ron,Mangin-Brinet:2023sey,Beaujeault-Taudiere:2023mju,Zhang:2024uxp}.
This means that to define a $p$-shell nucleus, 
four qubits are required to define the occupancy of the $1p_{3/2}$ neutron orbitals and 
four are required for the protons.
Similarly, two are required for the $1p_{1/2}$ neutrons and two for the protons.
This gives a total of six 
for the neutrons and six for the protons, and hence twelve qubits  for $p$-shell nuclei.

\begin{figure}[h]
\includegraphics[width=\columnwidth]{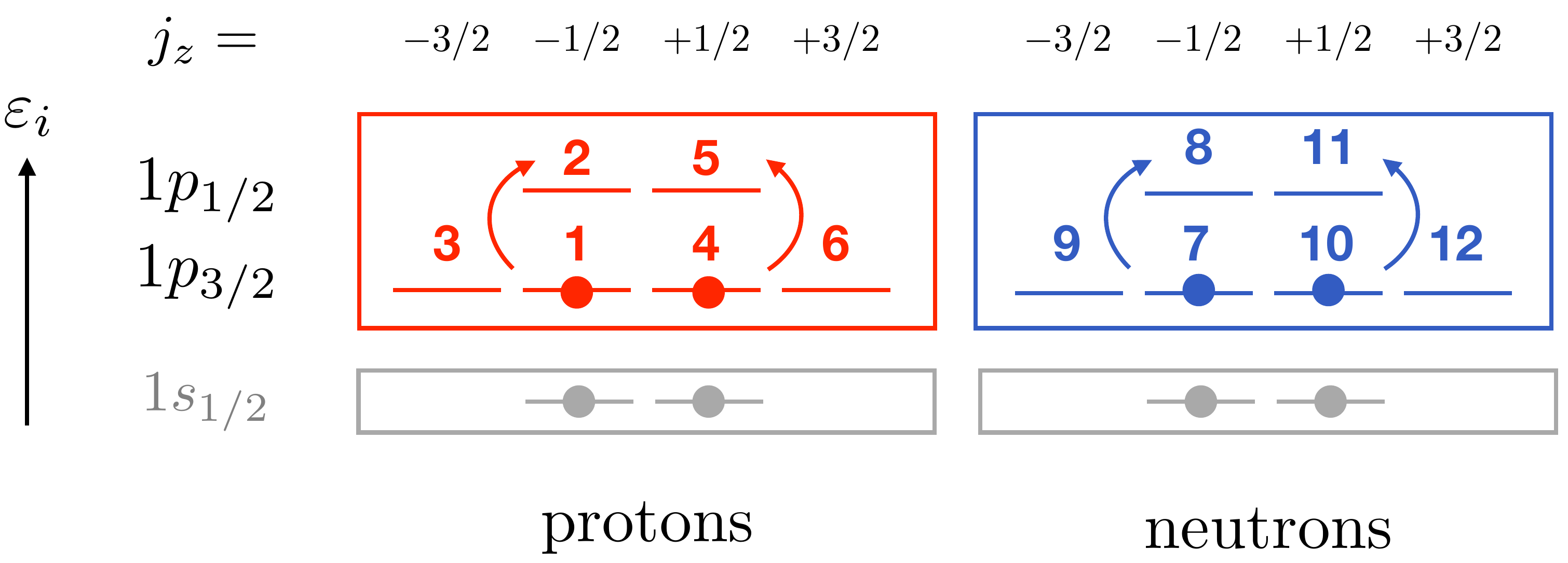}
\caption{ Mapping of the $p$-shell valence space in the {\tt BIGSTICK} code to qubits. The filled $^4$He core (two protons and two neutrons in the $1s_{1/2}$) is also shown in grey. 
}
\label{f:bigstick_mapping_p}
\end{figure}

As an explicit example, the  states in the wavefunctions produced by the {\tt BIGSTICK} code  are ordered in terms of single-particle quantum numbers in the following way\footnote{We note that the ordering of the single-particle states is in principle arbitrary and does not impact the calculation of the $n$-tangles and stabilizer R\'enyi entropies calculated in the next sections.}:
\begin{eqnarray}
    p{\rm -shell\  basis} & = & \{
    1p_{{3\over 2},-{1\over 2}} , 
    1p_{{1\over 2},-{1\over 2}} , 
    1p_{{3\over 2},-{3\over 2}} , 
    \nonumber\\
    & & \ 
    1p_{{3\over 2},+{1\over 2}} , 
    1p_{{1\over 2},+{1\over 2}} , 
    1p_{{3\over 2},+{3\over 2}} , 
    \}
    \ ,
    \label{eq:pshellbasis}
\end{eqnarray}
for  $p$-shell protons, and similarly for the  $p$-shell neutrons,  as shown in Fig.~\ref{f:bigstick_mapping_p}.

The shell-model wavefunction for the ground-state  of $^6{\rm Be}$ 
(two protons in the  $p$-shell on a $^4{\rm He}$ core),
 characterized by $J=J_z=0$, is
\begin{eqnarray}
|^6{\rm Be}\rangle_{\rm gs} & = & 
\alpha | 1p_{{1\over 2},-{1\over 2}} , 1p_{{1\over 2},+{1\over 2}} \rangle_\pi \otimes |0\rangle_\nu
\nonumber\\
& + & 
\beta | 1p_{{3\over 2},-{1\over 2}} , 1p_{{3\over 2},+{1\over 2}} \rangle_\pi \otimes |0\rangle_\nu
\nonumber\\
& + & 
\gamma | 1p_{{3\over 2},-{3\over 2}} , 1p_{{3\over 2},+{3\over 2}} \rangle_\pi \otimes |0\rangle_\nu
\nonumber\\
& = & 
\alpha |010010 000000\rangle
\nonumber\\
& + & 
\beta |100100 000000\rangle
\nonumber\\
& + & 
\gamma |001001 000000\rangle
    \ ,
\end{eqnarray}
where 
the last equality shows the wavefunction in terms of binary orbital occupations of the 
twelve $p$-shell orbitals given in Eq.~(\ref{eq:pshellbasis}). \\

\begin{figure}[h]
\includegraphics[width=\columnwidth]{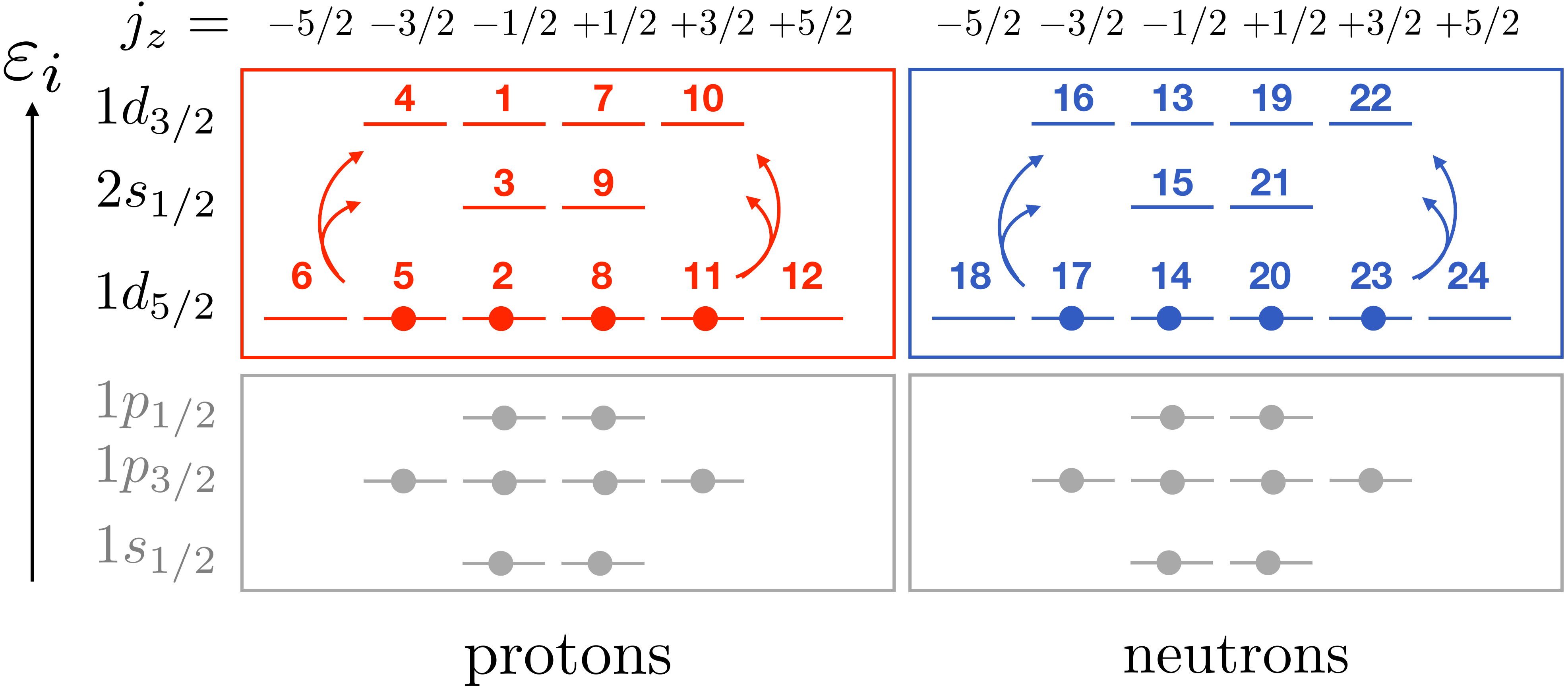}
\caption{Mapping of the $sd$-shell active space in the {\tt BIGSTICK} code. 
The fully-filled $^{16}$O core is shown in grey. }
\label{f:bigstick_mapping_sd}
\end{figure}

Similarly, $1d_{5/2}$-$2s_{1/2}$-$1d_{3/2}$ ($sd$-shell) nuclei are described with a total of 24 qubits. 
The mapping of the $sd$-shell proton basis states to qubits consistent with the ordering of outputs from the {\tt BIGSTICK} code is given by,
\begin{eqnarray}
    sd-{\rm shell\  basis} & = & \{
    1{d}_{{3\over 2},-{1\over 2}} , 
    1{d}_{{5\over 2},-{1\over 2}} , 
    2{s}_{{1\over 2},-{1\over 2}} , \nonumber \\
&&    1{d}_{{3\over 2},-{3\over 2}} , 
    1{d}_{{5\over 2},-{3\over 2}} , 
    1{d}_{{5\over 2},-{5\over 2}} , 
    \nonumber\\
&&  1{d}_{{3\over 2},+{1\over 2}} , 
    1{d}_{{5\over 2},+{1\over 2}} , 
    2{s}_{{1\over 2},+{1\over 2}} , \nonumber \\
&&    1{d}_{{3\over 2},+{3\over 2}} , 
    1{d}_{{5\over 2},+{3\over 2}} , 
    1{d}_{{5\over 2},+{5\over 2}} 
    \}
    \ ,
    \label{eq:sdshellbasis}
\end{eqnarray}
and similarly for the $sd$-shell neutron basis states. 
This is shown in Fig.~\ref{f:bigstick_mapping_sd}.

\section{Multi-Partite Entanglement in Nuclei}
\label{sec:entang}
\noindent
Various measures of entanglement in quantum many-body systems have been developed, for different possible partitionings of the wave function. For example, the von Neumann entanglement entropy~\cite{von1955mathematical} is a way to quantify how two subsystems of a bipartite pure state are entangled with each other. The shell-model nuclear state in Eq.~(\ref{eq:wf}) exhibits a natural bi-partitioning in terms of proton and neutron components, and the corresponding entanglement entropy has been investigated in Ref.~\cite{Johnson:2022mzk} which found low proton-neutron entanglement for nuclei away from $N=Z$. Entanglement entropy in the shell-model framework was also studied in Refs.~\cite{PhysRevC.105.064308,Perez-Obiol:2023wdz} using a different bi-partitioning of the nuclear states. Specifically the authors calculated single-orbital entanglement entropy and two-orbital mutual information in shell-model nuclei. The mutual information, which encompasses both classical and quantum correlations, was found to be dominant in the proton-proton and neutron-neutron sector, while correlations between one neutron and one proton orbitals were found to be weak.

To complement these studies, and because shell-model wave functions are able to capture detailed many-nucleon correlations within a given valence space, we find it informative to compute a multi-partite entanglement measure, specifically the $n$-tangles, which can quantify how $n \leq n_Q$ qubits are entangled within a larger $n_Q$-qubit system~\cite{PhysRevLett.78.5022,PhysRevLett.80.2245,PhysRevA.61.052306,PhysRevA.63.044301}. 
The $n$-tangle $\tau^{(n)}$ is defined as
\begin{align}
    \tau^{(n)}_{(i_1 ... i_n)} &= |\langle \Psi | \hat{\sigma}_y^{(i_1)} \otimes ... \otimes \hat{\sigma}_y^{(i_n)}| \Psi^* \rangle|^2 \; ,
\label{eq:n-tangle}
\end{align}
where $\sigma_y^{(i_k)}$ is the Pauli matrix acting on qubit $i_k$. 
The expression of the $n$-tangles above exhibits the anti-symmetric generator of $SO(2)$ ($\sigma_y$ Pauli matrix), ensuring that the $2$-tangle coincides with the squared concurrence for $2$-qubit systems, see {\it e.g.} Ref.~\cite{PhysRevLett.109.200503}. The generalizations to larger values of $n$ have been introduced in Refs.~\cite{PhysRevA.61.052306,PhysRevA.63.044301} and have been shown to constitute a measure of multipartite entanglement for even $n$.
\footnote{
There are several helpful examples that can be considered to verify that $n$-tangles are useful measures of multipartite entanglement. For example, the $|GHZ\rangle = (|000\rangle + |111\rangle)/\sqrt{2}$ and $|W\rangle = (|001\rangle + |010\rangle + |100\rangle)/\sqrt{3}$ states are well-known entangled states of three qubits. In addition to 3-qubit entanglement, the W state possesses 2-qubit entanglement, while the GHZ state does not. This is verified as $\tau_2(W)=4/9$ and $\tau_2(GHZ)=0$. 
One can find further applications to $4$-qubit entangled states in Ref.~\cite{Verstraete_2002}, as well as $n$-qubit $GHZ$ and $W$ states in Ref.~\cite{PhysRevA.63.044301}.}
The definition of $n$-tangles for odd values of $n$ are in general not straightforward, see {\it e.g.} Refs.~\cite{PhysRevA.63.044301,Li_2011}, and will not be considered in the present study.
It should also be reminded that the $n$-tangles for $n \geq 4$, may contain contribution of $m$-way entanglement (with $m<n$), i.e. they are not, {\it a priori}, measures of  “irreducible” $n$-way entanglement~\cite{PhysRevA.63.044301}.

Since the coefficients of the shell-model wave function in Eq.~(\ref{eq:wf}) are taken to be real\footnote{This is possible since the Hamiltonian is invariant under $S_y = \hat T \hat \Pi_y^{-1}$, where $\hat T$ is the time reversal operator and $\hat \Pi_y = \hat P e^{-i\pi \hat J_y}$ is the reflection with respect to the $xz$ plane.}, 
$\ket{\Psi^*} = \ket{\Psi}$.
Using the orbital-to-qubit Jordan-Wigner (JW) mapping defined in section~\ref{sec:SMmap}, the $n$-tangles quantify the entanglement between $n$ of the valence orbitals. With such JW mapping, the nucleon creation and annihilation operators $a^\dagger_i$ and $a_i$ are mapped to, for example,
\begin{align}
    a^\dagger_i \rightarrow \left( \prod_{j < i} \sigma_z^{(j)} \right)  \sigma_-^{(i)} 
    \ \  , \ \ 
    a_i \rightarrow \left( \prod_{j < i} \sigma_z^{(j)} \right) \sigma_+^{(i)} 
    \; ,
\end{align}
where $\sigma_\pm = (\sigma_x \pm i \sigma_y)/2 $. 
Thus, the 2-tangle operator $\sigma_y^{(i_1)} \sigma_y^{(i_2)}$ has contributions from $ a^\dagger_{i_1} a_{i_2}$ and $ a_{i_1} a^\dagger_{i_2}$ (as particle number is conserved) and represents a one-body (one-nucleon) operator. 
Since proton and neutron numbers are 
individually conserved,
there are no proton-neutron 2-tangles (i.e., $i_1$ and $i_2$ have the same isospin projections).
As mentioned above, we do not consider $n$-tangles with odd values of $n$ (they would also vanish due to particle-number conservation).
Thus, the 4-tangle, which represents a two-body operator, is the lowest-order tangle capturing many-body entanglement.

%%%%%%%%%%%%%%%%
\subsection{Multi-orbital entanglement in $p$-shell nuclei} 
\label{sec:multientanglement}
\begin{figure}[h]
\includegraphics[width=\columnwidth]{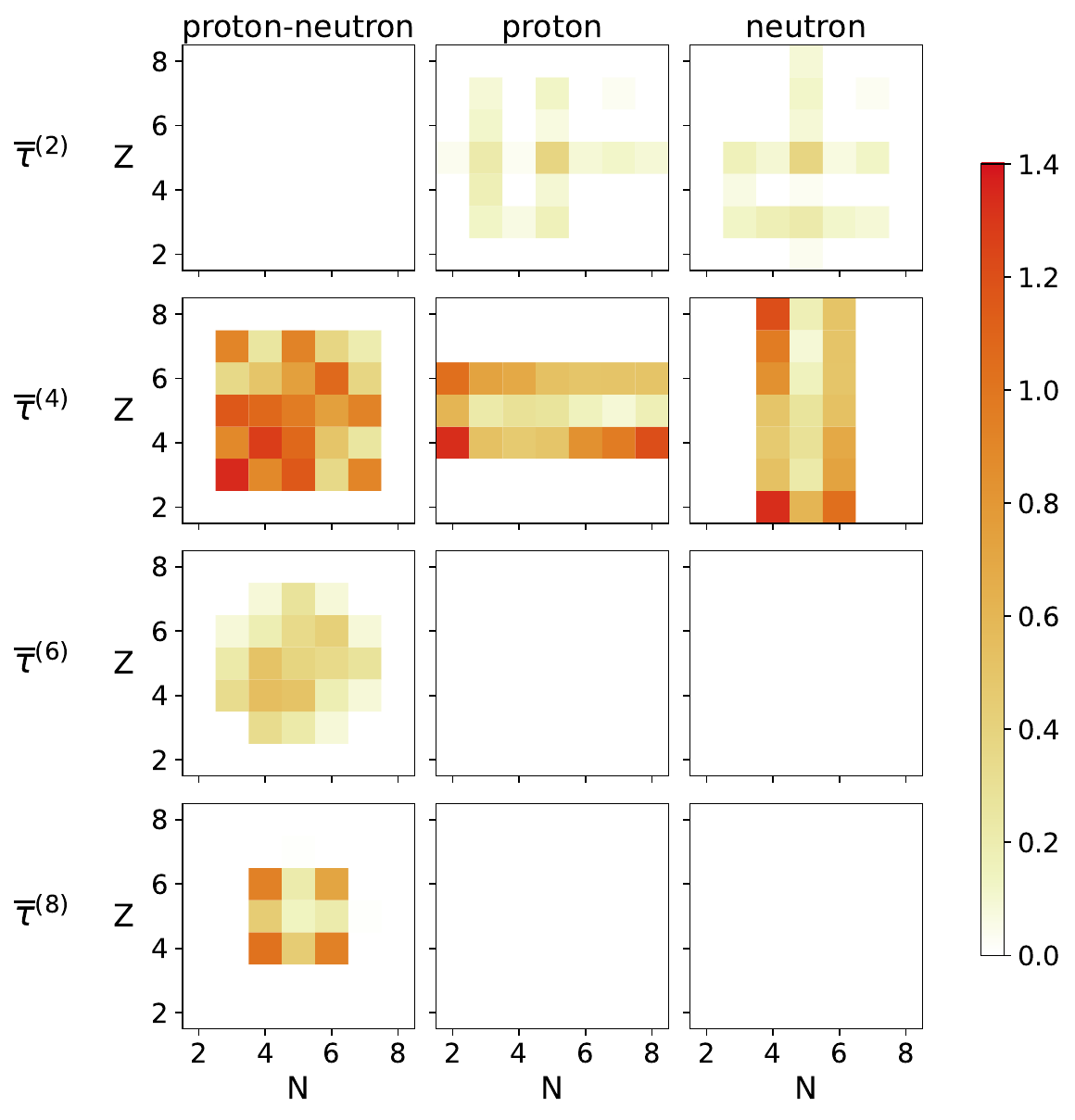}
\caption{Values of the proton-neutron (left), pure proton (middle), and pure neutron (right)
summed $n$-tangles $\overline{\tau}^{(n)}_{\pi\nu}$, $\overline{\tau}^{(n)}_{\pi}$, $\overline{\tau}^{(n)}_{\nu}$, for $n=2,4,6,8$, as defined in Eqs.~\eqref{eq:ntangles_sum} for $p$-shell nuclei.}
\label{f:p_4-6-8-tangles}
\end{figure}

\noindent
Due to the large number of $n$-tangles corresponding to all possible combinations of single-particle states (allowed by the symmetry selection rules), in order to appreciate and compare the importance of many-body entanglement in various nuclei, we first consider the summation of $n$-tangles $\overline{\tau}^{(n)}$ for a given value of $n$, in the  proton, neutron and mixed proton-neutron sectors:
\begin{align}
\overline{\tau}^{(n)}_{\pi}  &\equiv \sum_{\substack{ i_1, i_2, .. i_n \\ \text{all protons} } } \tau^{(n)}_{(i_1 i_2 .. i_n)} \; , \nonumber \\
\overline{\tau}^{(n)}_{\nu}  &\equiv \sum_{\substack{ i_1, i_2, .. i_n \\ \text{all neutrons} } } \tau^{(n)}_{(i_1 i_2 .. i_n)} \; , \nonumber \\
\overline{\tau}^{(n)}_{\pi\nu}  &\equiv \sum_{\substack{ i_1, i_2, .. i_n \\ \text{mixed } } } \tau^{(n)}_{(i_1 i_2 .. i_n)} 
\; .
\label{eq:ntangles_sum}
\end{align}
The results for $p$-shell nuclei are shown in Fig.~\ref{f:p_4-6-8-tangles}.
\begin{figure*}[ht]
\includegraphics[width=\textwidth]{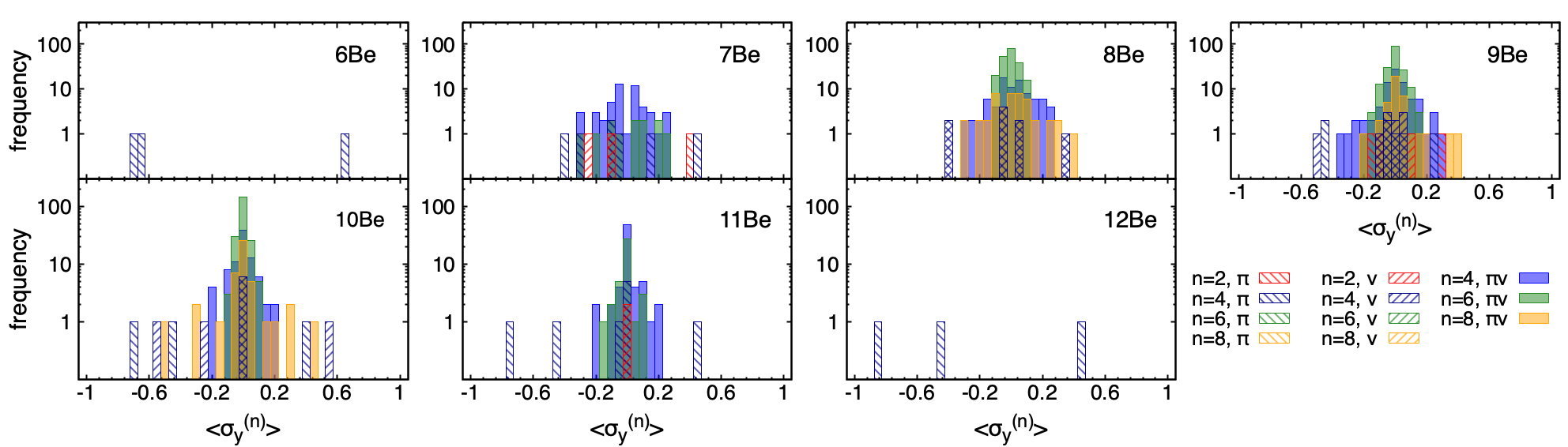}
\caption{
Distributions of 
Pauli-string expectation values $\langle \hat{\sigma}_y^{(n)} \rangle \equiv \langle \Psi | \hat{\sigma}_y^{(i_1)} \otimes ... \otimes \hat{\sigma}_y^{(i_n)}| \Psi \rangle $ with $|\langle \hat{\sigma}_y^{(n)} \rangle| \geq 10^{-4}$ for $n=2,4,6,8$ in Be isotopes. The pure proton ($\pi$), pure neutron ($\nu$) and mixed proton-neutron ($\pi\nu)$ Pauli strings are shown separately. 
Bin widths of $0.05$ were used.}
\label{f:tangles_distrib_Be}
\end{figure*}
As stated above, the proton-neutron $2$-tangles cancel due to conservation of proton and neutron numbers.
Further, due to the reduced size of the active space and rotational symmetry selection rules, 
there are also no pure proton or pure neutron $n$-tangles with $n \geq 6$ in the $p$-shell. 
It appears that the two-body entanglement, captured by $\overline{\tau}^{(4)}$, 
and the proton-neutron four-body entanglement (when permitted by the model space), captured by $\overline{\tau}^{(8)}_{\pi\nu}$, are the largest.
Overall, proton-neutron multi-body entanglement appears to be significant in the middle of the shell, while like-particle (pure proton or pure neutron) entanglement is typically larger near the shell closures. \\

It is interesting to examine these results further for a few select nuclei.
The Beryllium nuclei are particularly interesting as they are known to display unique cluster and molecular-like structures, leading to large deformations~\cite{VONOERTZEN200643,RevModPhys.90.035004}. For example, $^8$Be ($Z=N=4$), which is characterized by a very short half-life of $\sim 10^{-17}$ s~\cite{TILLEY2004155}, is understood to be a clustered nucleus composed of two $\alpha$ particles. Adding neutrons to these $\alpha$ clusters can result in molecular-like arrangements, where the neutrons 
establish the "binding" between the $\alpha$'s, similarly to electrons binding atoms into molecules. In particular, $^9$Be and $^{10}$Be, which have infinite and very long lifetime ($> 10^{6}$ y), respectively~\cite{TILLEY2004155}, have been observed to display $\alpha-n-\alpha$ and $\alpha-2n-\alpha$ molecular-like structures in their ground states~\cite{PhysRevC.91.024610,PhysRevLett.131.212501}. 
While the presently-used shell-model framework may not be able to fully capture such clustered and molecular structures, the correlations present in shell-model wave functions may exhibit signs of these underlying structures.
From Fig.~\ref{f:p_4-6-8-tangles}, it appears that proton entanglement, captured by $\overline{\tau}^{(4)}_{\pi}$, increases as the neutron number $N$ grows for $N \geq 8$, via proton-neutron interactions. This is in accordance with Ref.~\cite{Perez-Obiol:2023wdz} which calculated the one-orbital von Neumann entropy and mutual information (MI) within the same framework as the present one. That reference however found very weak MI between proton and neutron orbitals. 
The von Neumann entropy and MI, however, 
do not furnish
information about multipartite correlations. 
Indeed, 
as seen
in Fig.~\ref{f:p_4-6-8-tangles}, 
large multi-body proton-neutron entanglement is in fact present. 
In contrast
to $\overline{\tau}^{(4)}_{\pi}$, 
$\overline{\tau}^{(4)}_{\pi\nu}$ tends to decrease with larger $N$ in $^{8-12}$Be.

Figure~\ref{f:tangles_distrib_Be} displays the distributions of $n$-tangles for the Be chain.
More precisely, Fig.~\ref{f:tangles_distrib_Be} shows the distributions of expectations value 
of the Pauli strings $\langle \Psi | \hat{\sigma}_y^{(i_1)} \otimes ... \otimes \hat{\sigma}_y^{(i_n)}| \Psi \rangle$ (which includes the information about the sign) for these nuclei. 
The frequency corresponds to the number of strings within a given interval. 
We differentiate again between the pure proton, pure neutron and mixed proton-neutron Pauli strings,
and use a logarithmic scale to enhance the small like-particle contributions.
%%%%%%%%%%%% 
$^6$Be, which has an empty neutron valence space, and two active protons features a wave function with only three many-body configurations, leading to three non-zero (proton) tangles of large values.
When adding neutrons, the nuclear wave functions becomes fragmented and the distribution of the $n$-tangles changes dramatically, displaying a large number of small elements peaked around zero.
In $^8$Be, we note a large contribution of proton-neutron $8$-tangles, which is related to $4$-body proton-neutron entanglement, and thus could signal the two-$\alpha$ structure displayed by this nucleus.
We note, however, that $n$-tangles with $n \geq 4$ in general do not provide a fully irreducible measure of multipartite entanglement, and thus may encompass contributions from lower-body tangles~\cite{li2010relationship}. In $^8$Be we observe that $2$-tangles vanish and thus the $4$-tangles capture genuine $4$-orbital entanglement. The $8$-tangles, however, may contain contributions from products of $4$-tangles. 
It is clear that the increase in proton entanglement with growing $N$ that was seen in Fig.~\ref{f:p_4-6-8-tangles} is due to a few contributions of large magnitudes. 
On the other hand, the distribution of proton-neutron entanglement is drastically different across the chain, as it displays a large number of small contributions, 
thus presenting a more collective behaviour. \\

\begin{figure}[ht]
\includegraphics[width=\columnwidth]{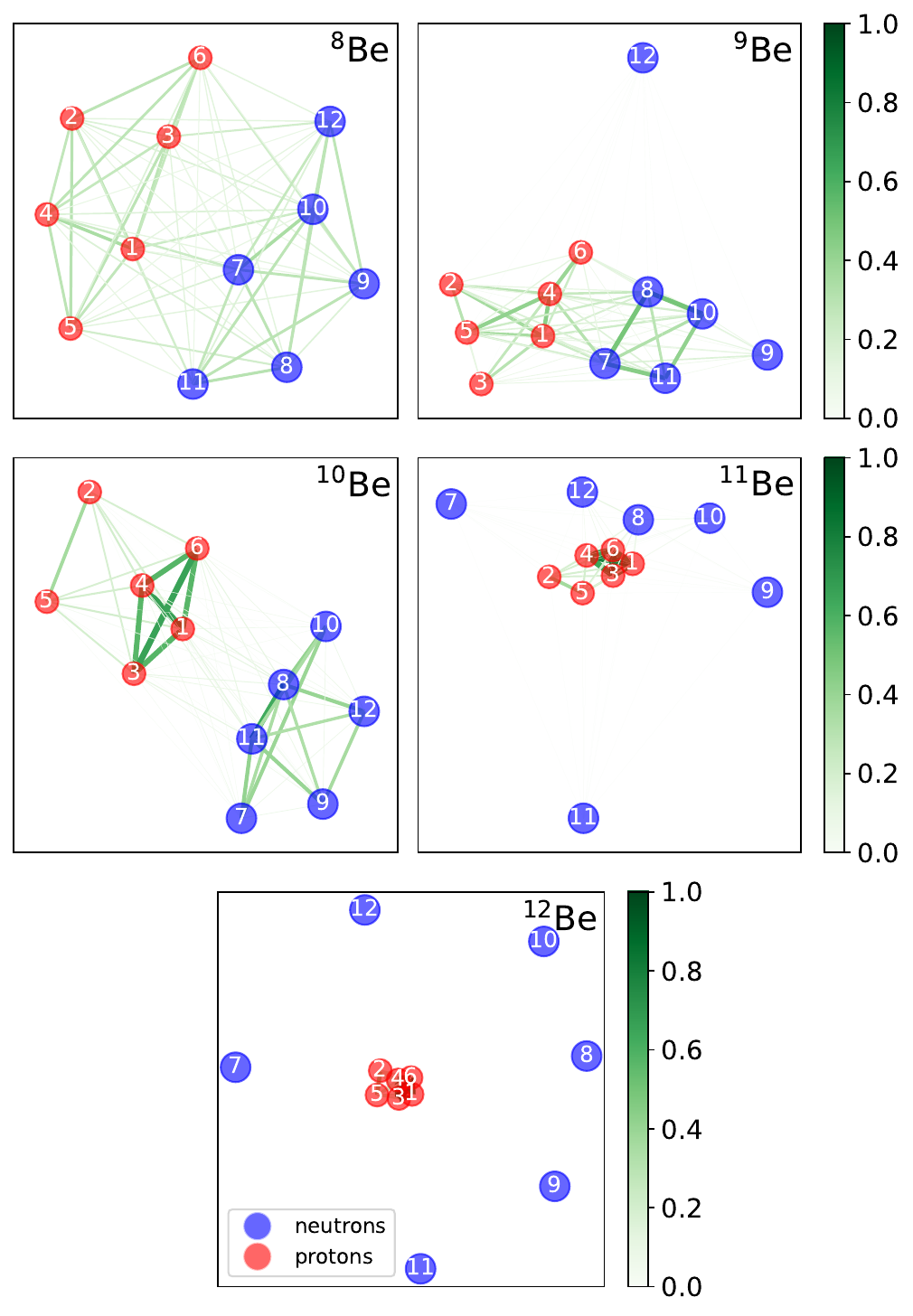}
\caption{Network representation of the 4-tangles $\tau_{(i_1,...,i_4)}^{(4)}$in $^{8-12}$Be. The nodes represent the single-particle orbitals and are labeled as in Fig.~\ref{f:bigstick_mapping_p}. The values of the edges, representing the entanglement, are determined as in Eq.~\eqref{eq:edge_4tangle} and are indicated by both the darkness and the thickness of the lines. Orbitals that are shown closer together are also more entangled. The plots have been generated with NetworkX~\cite{SciPyProceedings_11}.}
\label{f:4tangles_network_Be}
\end{figure}

In Fig.~\ref{f:4tangles_network_Be},
in order to dissect these results further, 
we examine
the details of the 4-tangles between single-particle orbitals in $^{8-12}$Be
using network plots.
These network plots have been generated using NetworkX~\cite{SciPyProceedings_11}. The nodes represent each single-particle orbital and the value of each edge $e^{(4)}_{i_1 i_2}$ between two nodes $i_1$ and $i_2$ has been obtained by summing the value of the 4-tangles linking these two nodes, {\it i.e.}
\begin{align}
    e^{(4)}_{i_1 i_2} = \sum_{i_3 < i_4}   \tau^{(4)}_{(i_1, i_2, i_3, i_4)} \; ,
    \label{eq:edge_4tangle}
\end{align}
since $\tau^{(4)}_{(i_1, i_2, i_3, i_4)} = \tau^{(4)}_{\mathcal{P}(i_1, i_2, i_3, i_4)}$ with $\mathcal{P}(i_1, i_2, i_3, i_4)$ representing any permutation of the indices $(i_1, i_2, i_3, i_4)$.
This summation makes it easier to visualize the entanglement between orbitals. 
The edge value is represented by both the color and the thickness of the edge, 
and, to some extent, 
the distance between the nodes (orbitals). That is, more entangled orbitals are connected by darker and thicker edges, and are closer to each other on each figure.
The numerical values of the edges can be found in Ref.~\cite{supplemental}.

Overall it is seen that the neutrons progressively disentangle from the system as $N$ increases, while the protons appear closer together and thus become more entangled.
The large size of the entangled network in $^8$Be is indicative of the collectivity in this nucleus, with entanglement being distributed between all proton and neutron orbitals, as seen from the large number of edges between them.
Adding a neutron to $^8$Be results in one proton orbital to be almost fully occupied, and disentangling from the rest of the system. Both remaining sets of proton and neutron orbitals are closer together 
(in the figure),
signaling that they are more individually entangled.
In $^{10}$Be, the two sets of proton and neutron states disentangle from each other, 
but remain entangled among themselves. In particular, the proton $1p_{3/2}$ subshell is the most entangled.
As  neutrons are added, the neutron orbitals become filled and disentangle from the system, while further entangling the protons. In $^{12}$Be, the protons are entangled via the presence of the neutron mean field.

%%%%%%%%%%%%%%%%
\subsection{Multi-orbital entanglement in $sd$-shell nuclei} 

\begin{figure}[h]
\includegraphics[width=\columnwidth]{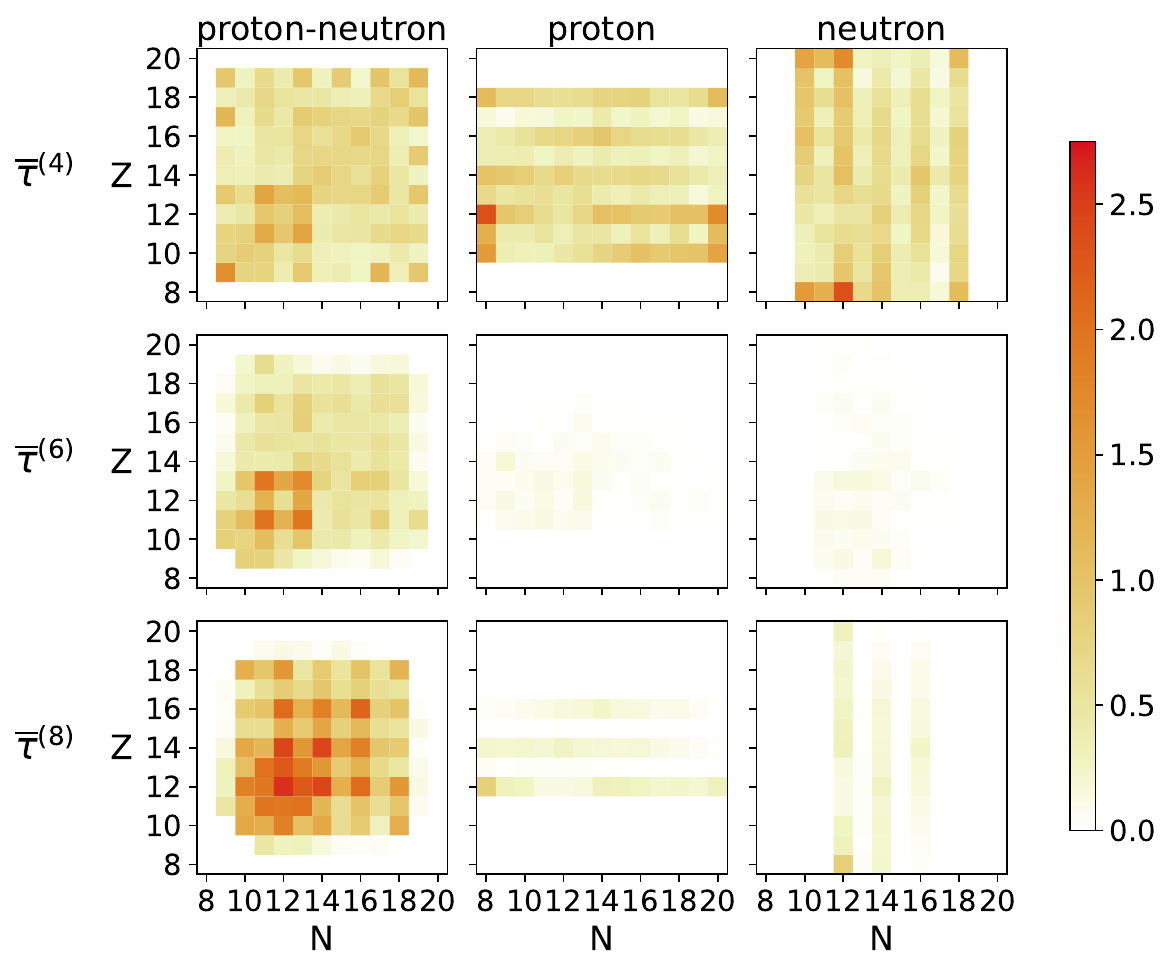}
\caption{Values of the proton-neutron, pure proton, and pure neutron summed $n$-tangles, $\overline{\tau}^{(n)}_{\pi\nu}$, $\overline{\tau}^{(n)}_{\pi}$, $\overline{\tau}^{(n)}_{\nu}$, as defined in Eq.~\eqref{eq:ntangles_sum}, for $sd$-shell nuclei.}
\label{f:sd_4-6-8-tangles}
\end{figure}

Nuclei in the $sd$-shell also exhibit a large variety of structure properties, in particular, various deformation features.
The Ne chain ($Z=10$), for example, is known to exhibit a transition in shape as $N$ increases, evolving from a large prolate deformation in $^{20-23}$Ne to nearly-degenerate oblate-prolate shape in $^{24-25}$Ne, to a spherical shape or possible shape coexistence in heavier isotopes near the island of inversion~\cite{PhysRevC.87.014333,IWASAKI2005118,WANG2023138038,Xue:2024sup}. 
Similarly to $^{20}$Ne, other nuclei with $N=Z$, such as $^{24}$Mg and $^{28}$Si display large prolate and oblate deformation, respectively~\cite{VONOERTZEN200643}.
Figure~\ref{f:sd_4-6-8-tangles} shows
the summed $n$-tangles with $n=4,6,8$ in the proton, neutron and proton-neutron sectors for $sd$-shell nuclei. 
In the pure proton and pure neutron sectors, entanglement is largely limited to $4$-tangles, corresponding to 4-orbital, or 2-body, entanglement, and displays a similar trend as in the $p$-shell.
The behaviour of the mixed proton-neutron entanglement, however, differs. 
Interestingly, the summation of proton-neutron $4$-tangles,
$\overline{\tau}^{(4)}_{\pi\nu}$, 
appears to be rather homogeneous and weaker
compared to the
higher $n$-tangles, which present more distinct structures. The $6$-tangles, $\overline{\tau}^{(6)}_{\pi\nu}$, related to 3-body entanglement, 
present large contributions in a few odd-mass nuclei, while the $8$-tangles, $\overline{\tau}^{(8)}_{\pi\nu}$, 
largely dominate in a region around $^{24}$Mg, 
as well as a region of even-even nuclei around the center of the shell.

\begin{figure}[ht]
\includegraphics[width=\columnwidth]{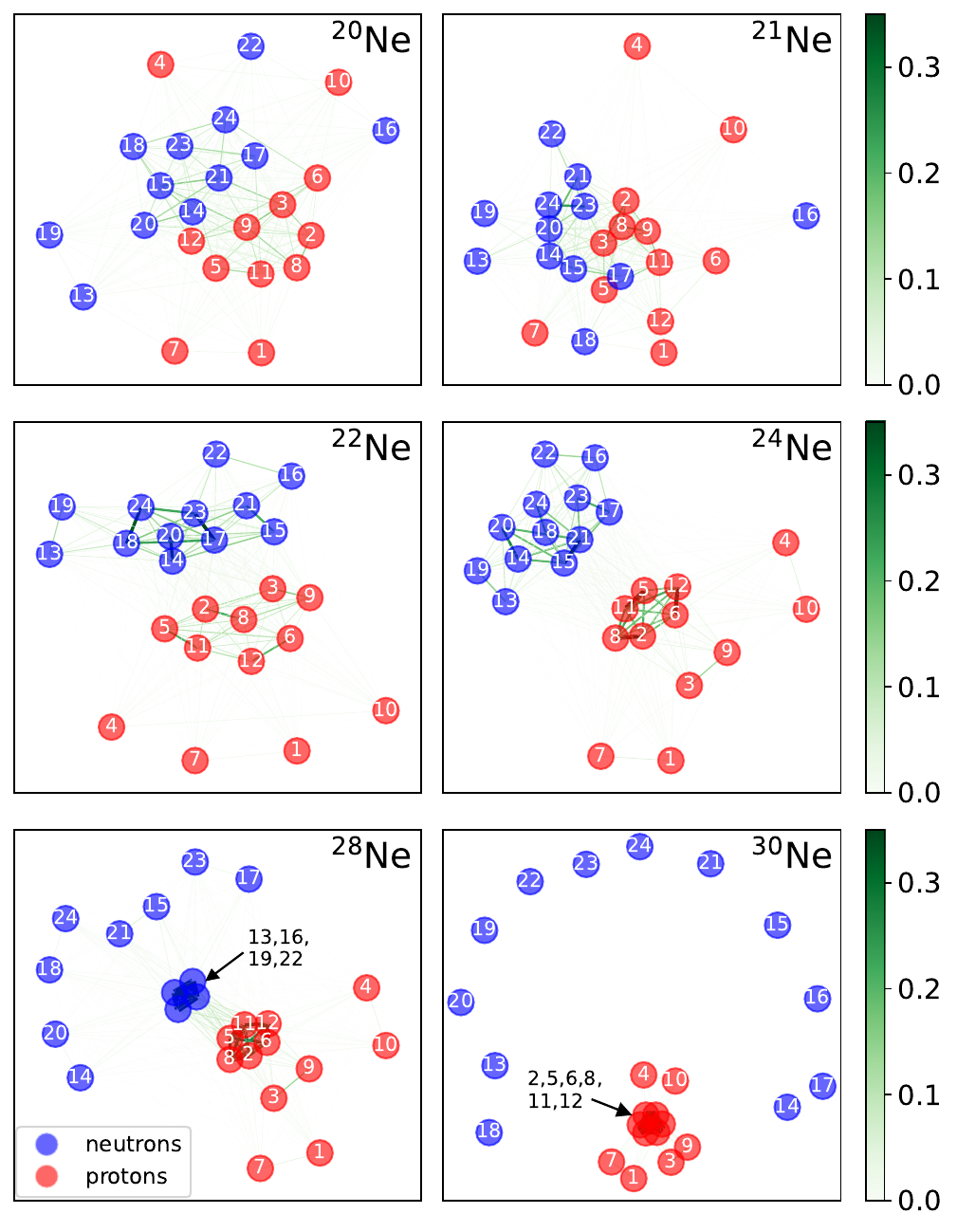}
\caption{Network representation of the 4-tangles $\tau_{(i_1,...i_4)}^{(4)}$ in Ne isotopes. The nodes represent the single-particle orbitals and are labeled as in Fig.~\ref{f:bigstick_mapping_p}. The values of the edges, representing the entanglement, are determined as in Eq.~\eqref{eq:edge_4tangle} and are indicated by both the darkness and the thickness of the lines. For clarity, the values of some edges in $^{28}$Ne and $^{30}$Ne are larger than the maximal value of the colorbar, they have values up to $0.86$. The networks have been generated with NetworkX~\cite{SciPyProceedings_11}.}
\label{f:4tangles_network_Ne}
\end{figure}

As an example, we examine the Ne chain in more detail.
The 4-tangles for these isotopes present a similar behaviour 
to those in the Be chain: the two-body proton-neutron entanglement is strongest around $N=Z$, while the pure proton component dominates in isotopes with neutron excess. This is again understood as adding neutrons redistributes the protons over the orbitals, via proton-neutron interaction.
The details of these 4-tangles can be seen in Fig.~\ref{f:4tangles_network_Ne}, in the same network form as described above. 
It is seen that around $N=Z$, the entanglement is largely shared between proton and neutron orbitals and collectively distributed among many components, while same-isospin entanglement dominates in nuclei away from $N=Z$, and is characterised by fewer but larger components

\begin{figure}[ht]
\includegraphics[width=\columnwidth]{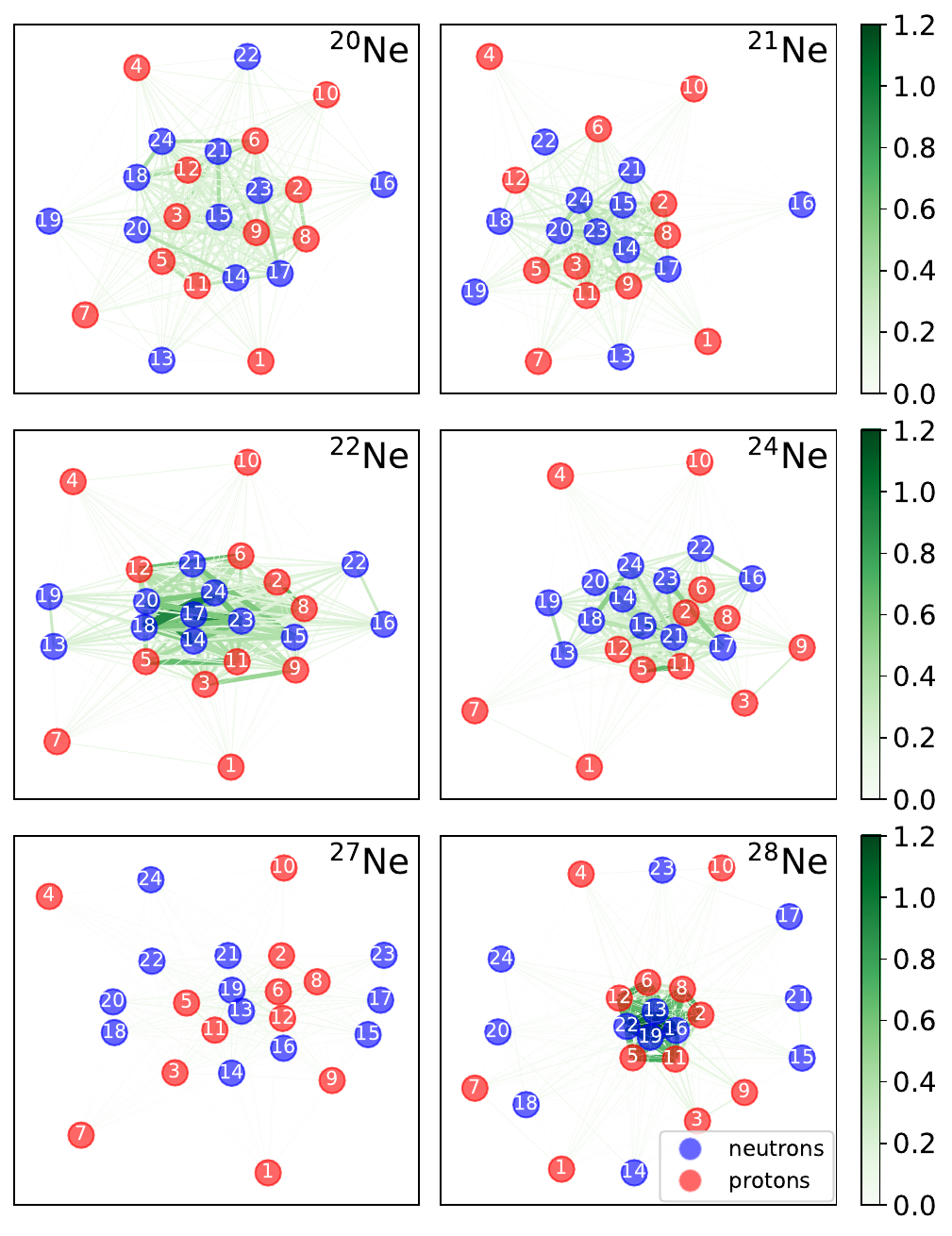}
\caption{Network representation of the 8-tangles $\tau_{i_1,...,i_8}^{(8)}$ in Ne isotopes. The nodes represent the single-particle orbitals and are labeled as in Fig.~\ref{f:bigstick_mapping_p}. The values of the edges, representing the entanglement, are determined as in Eq.~\eqref{eq:edge_8tangle} and are indicated by both the darkness and the thickness of the lines. 
The networks have been generated with NetworkX~\cite{SciPyProceedings_11}.}
\label{f:8tangles_network_Ne}
\end{figure}

The 8-tangles present a rather different behaviour, as we observe a strong proton-neutron component around $N=Z$, which may again signal $\alpha$-particle
correlations. Interestingly, such large proton-neutron 8-tangle component persists in even-even isotopes all the way to $^{28}$Ne, a nucleus at the boundary of the island of inversion, predicted to exhibit possible shape coexistence~\cite{IWASAKI2005118,WANG2023138038,Xue:2024sup}. 
In comparison, pure proton and pure neutron components are almost negligible along the chain.
Fig.~\ref{f:8tangles_network_Ne} shows the network representation of these 8-tangles. 
Similarly to the 4-tangles, the value of each edge $e^{(8)}_{i_1 i_2}$ between two nodes $i_1$ and $i_2$ has been obtained by summing the value of the 8-tangles linking these two nodes, {\it i.e.},
\begin{align}
    e^{(8)}_{i_1 i_2} = \sum_{i_3 < i_4 < i_5 < i_6 < i_7 < i_8}   \tau^{(8)}_{(i_1, i_2, i_3, i_4, i_5, i_6, i_7, i_8)} \; .
    \label{eq:edge_8tangle}
\end{align}
In $^{20}$Ne, the entanglement is mainly distributed among the $d_{5/2}$ and $s_{1/2}$ proton and neutron orbitals. As neutrons are added to the system, the neutron $d_{3/2}$ becomes more occupied and entangled with the rest. In $^{28}$Ne, the proton-neutron 8-tangles are largely contained within the proton $d_{5/2}$ and neutron $d_{3/2}$. Neon isotopes with odd neutron numbers, appear to have low proton-neutron 8-orbital entanglement. 
The numerical values of the edges in Figs.~\ref{f:4tangles_network_Ne} and \ref{f:8tangles_network_Ne} can be found in Ref.~\cite{supplemental}.

%%%%%%%%%%%%%%%%%%%%%%%%%%%%%%%%%%%%%%%
\section{Magic Measures in Shell-Model Nuclei}
\label{sec:magic}
\noindent
Magic, or non-stabilizerness, is a notion rooted in the stabilizer formalism, 
of which we briefly review the relevant  aspects below.
The stabilizer formalism is centered around the Pauli group, which, for a system of  $n_Q$ qubits, is the group $\mathcal{G}_{n_Q}$ of Pauli-string operators with multiplicative phases:
\begin{equation}
    \mathcal{G}_{n_Q} = \lbrace \varphi \, \hat{\sigma}^{(1)} \otimes \hat{\sigma}^{(2)} \otimes ... \otimes \hat{\sigma}^{(n_Q)} \rbrace \; ,
    \label{eq:Gn}
\end{equation}
where $\hat{\sigma}^{(j)} \in \lbrace \mathds{1}^{(j)}, \hat{\sigma}_x^{(j)}, \hat{\sigma}_y^{(j)}, \hat{\sigma}_z^{(j)} \rbrace$ act on qubit 
(shell-model orbital) 
$j$ and $\varphi \in \lbrace \pm 1 , \pm i \rbrace$.
For an arbitrary $n_Q$-qubit pure state $\ket{\Psi}$, the
elements $\hat P \in \mathcal{G}_{n_Q}$ which stabilize $\ket{\Psi}$, {\it i.e.}, such that $\hat{P} \ket{\Psi} = \ket{\Psi}$, form an Abelian group $\mathcal{S}(\ket{\Psi})$ called the Pauli stabilizer group of $\ket{\Psi}$~\cite{Gottesman:1996rt}.
The state $\ket{\Psi}$ is called a stabilizer state if $\mathcal{S}(\ket{\Psi})$ 
contains exactly $d = 2^{n_Q}$ elements, and
is fully specified by its Pauli stabilizer group.
It is known that stabilizer states can be prepared with circuits comprised 
of Hadamard (H), phase (S), and CNOT gates only, 
which generate the group of Clifford operations. 
While these gates are able to generate entanglement, they can be efficiently simulated with classical computers~\cite{gottesman1998heisenberg}.
Non-stabilizer states (magic states) require supplementing the above set of gates 
with a non-Clifford operation, such as the T gate, which is then sufficient for universal quantum computation.
Consequently, the number of T gates, which are the key to realizing quantum advantages, quantifies the actual resource requirement in a quantum computation\footnote{The number of CNOT gates in a circuit is also relevant in the context of computation on NISQ devices 
(as CNOT gates provide the dominant source of errors in 
current simulations, 
and dominate the time required to implement a quantum circuit).}.

As mentioned in the introduction, various measures have been developed to quantify the amount of magic in a quantum state, which is related to the number of T gates required to prepare that state~\cite{Beverland:2019jej,Leone:2021rzd,Garcia:2024scq}.
In this work we have chosen to compute the stabilizer R\'enyi entropies (SREs)~\cite{Leone:2021rzd}
of nuclear shell-model (active-space) wavefunctions. 
Starting from a general expansion of the density matrix of an arbitrary state $\ket{\Psi}$:
\begin{equation}
    \hat{\rho} = \ket{\Psi} \bra{\Psi} = 
    \frac{1}{d} \sum_{\hat P \in \widetilde{\mathcal{G}}_{n_Q}} \langle \Psi |\hat{P} | \Psi \rangle \, \hat{P} 
    \ =\ 
    \frac{1}{d} \sum_{\hat P \in \widetilde{\mathcal{G}}_{n_Q}} c_P \, \hat{P} 
    \; ,
\end{equation}
where $c_P \equiv \langle \Psi |\hat{P} | \Psi \rangle$ and $\widetilde{\mathcal{G}}_{n_Q}$ is the subgroup of $\mathcal{G}_{n_Q}$ 
in Eq.~(\ref{eq:Gn})
with phases $\varphi = +1$,
the authors of Ref.~\cite{Leone:2021rzd} showed that the quantity 
\begin{equation}
    \Xi_P \equiv   \frac{c_P^2 }{d} \; ,
\end{equation}
is a probability distribution for pure states,
corresponding to the probability for $\hat{\rho}$ to be in $\hat{P}$.
It was shown that $\ket{\Psi}$ is a stabilizer state if and only if the expansion coefficients 
$c_P = \pm 1$ for $d$ commuting Pauli strings $\hat P \in \widetilde{\mathcal{G}}_{n_Q}$, 
and $c_P = 0$ for the remaining 
$d^2-d$ strings~\cite{zhu2016clifford}. 
Thus, $\Xi_P = 1/d$ or $0$ for a 
(qubit) stabilizer state, and the stabilizer $\alpha$-R\'enyi entropies~\cite{Leone:2021rzd},
\begin{equation} 
\mathcal{M}_{\alpha}(\ket{\Psi})= -\log_2 d + \frac{1}{1-\alpha} \log_2 
\left( \sum_{\hat{P} \in \widetilde{\mathcal{G}}_{n_Q}} \Xi_P^{\alpha} \right) \; ,
\label{eq:Renyi_entropy_def1}
\end{equation}
provide a measure of magic in $\ket{\Psi}$. 
The constant offset, $- \log_2 d$, ensures that the SREs 
vanish for stabilizer states. 
It was demonstrated that SREs with $\alpha \geq 2$ constitute magic monotones for pure states, in contrast to $\alpha < 2$~\cite{Leone:2024lfr,Haug:2023hcs}.\\

It has also recently been shown~\cite{Haug:2024ptu} that SREs probe different aspects of magic, depending on the value of $\alpha$. Specifically, that SREs with $\alpha > 1$ 
provide a
measure of
the distance to the nearest stabilizer state, while those with $\alpha < 1$ are related to the stabilizer rank and complexity of Clifford (classical) simulations of the quantum state. 
This is accordance with the earlier demonstration that ${\cal M}_{1/2}$ is twice the logarithm of the stabilizer norm~\cite{Leone:2021rzd}. 
In the present study, we calculate the SREs ${\cal M}_{\rm lin}$,  ${\cal M}_1$ and 
${\cal M}_2$.
In the $\alpha=1$ limit, ${\cal M}_1$ corresponds to the 
Shannon entropy,
\begin{eqnarray}
    {\cal M}_1 = -\sum_{\hat{P} \in \widetilde{\mathcal{G}}_{n_Q}}  
    \Xi_P\log_2 d \ \Xi_P
    \ ,
    \label{eq:Renyi_entropy_defM1}
\end{eqnarray}
which can be 
derived
from Eq.~(\ref{eq:Renyi_entropy_def1}) by taking 
$\alpha=1-\epsilon$, 
followed by the $\epsilon\rightarrow 0$ limit.
Expanding the logarithm in Eq.~(\ref{eq:Renyi_entropy_defM1}) around
$d \ \Xi_P = 1$, defines the linear magic
${\cal M}_{\rm lin}$, 
\begin{eqnarray}
    {\cal M}_{lin} = 1 - d \sum_{\hat{P} \in \widetilde{\mathcal{G}}_{n_Q}} \Xi_P^2 \; .
    \label{eq:Renyi_entropy_def3}
\end{eqnarray}
The $\alpha=2$ measure of magic, ${\cal M}_2$, corresponds to 
\begin{eqnarray}
    {\cal M}_2 & = &  -
    \log_2 \ d \sum_{\hat{P} \in \widetilde{\mathcal{G}}_{n_Q}}  \Xi_P^2 \; .
    \label{eq:M2_def}
\end{eqnarray}

The exponential scaling of the number of Pauli strings with the number of qubits, $d^2$, 
implies a practical upper limit to the size of the active shell-model space for which all of
the $\Xi_P$ can be computed exactly.
We find that for all of the $p$-shell nuclei, along with the $sd$-shell nuclei with either vanishing protons or vanishing neutrons, 
the SREs can be evaluated exactly, involving $n_Q=12$ qubits 
($d=4096$ and $d^2\approx 16.8 \times 10^6$),
while exact evaluations of SREs for $sd$-shell with both active protons and neutrons, 
mapped to  $n_Q=24$ qubits
($d\approx 16.8 \times 10^6$ and $d^2\approx 2.8 \times 10^{14}$),
are beyond what is practical with a single-processor.
The situation is somewhat better than this because the nuclear structure and nature of the nuclear Hamiltonian gives rise to many vanishing matrix elements. 
However, the impracticality persists even with this consideration.
For this reason, and following Ref.~\cite{Tarabunga:2023ggd},
Markov-Chain Monte-Carlo (MCMC)
was employed to provide statistical estimates of the SREs in the larger active model spaces.
Details of the MCMC sampling technique that we employed are given in App.~\ref{appendix:magic}.
In the non-spherical nuclei, the distribution of amplitudes in the wavefunction was such that thermalization of the chains was slow, 
and the subsequent
multi-chain sampling was slow to converge.
Matrix elements of the  
$d$ Pauli strings composed of $\hat I$ and $\hat Z$ operators are typically $\Xi_P\approx 1$,
while those of the other 
$d^2-d$ strings are $\Xi_P \ll 1$, creating the situation in which an exponentially small number of samples are much larger than the rest.
To improve sampling efficiency and reduce classical resource requirements for such nuclei, we have
introduced the PSIZe-MCMC algorithm, in which the 
matrix elements of the  
$\hat I$ and $\hat Z$ strings are computed exactly, and the contributions from the other strings are evaluated using MCMC.  This is detailed in App.~\ref{appendix:psize}.

\begin{figure}[!ht]
    \centering
    \includegraphics[width=\columnwidth, keepaspectratio]{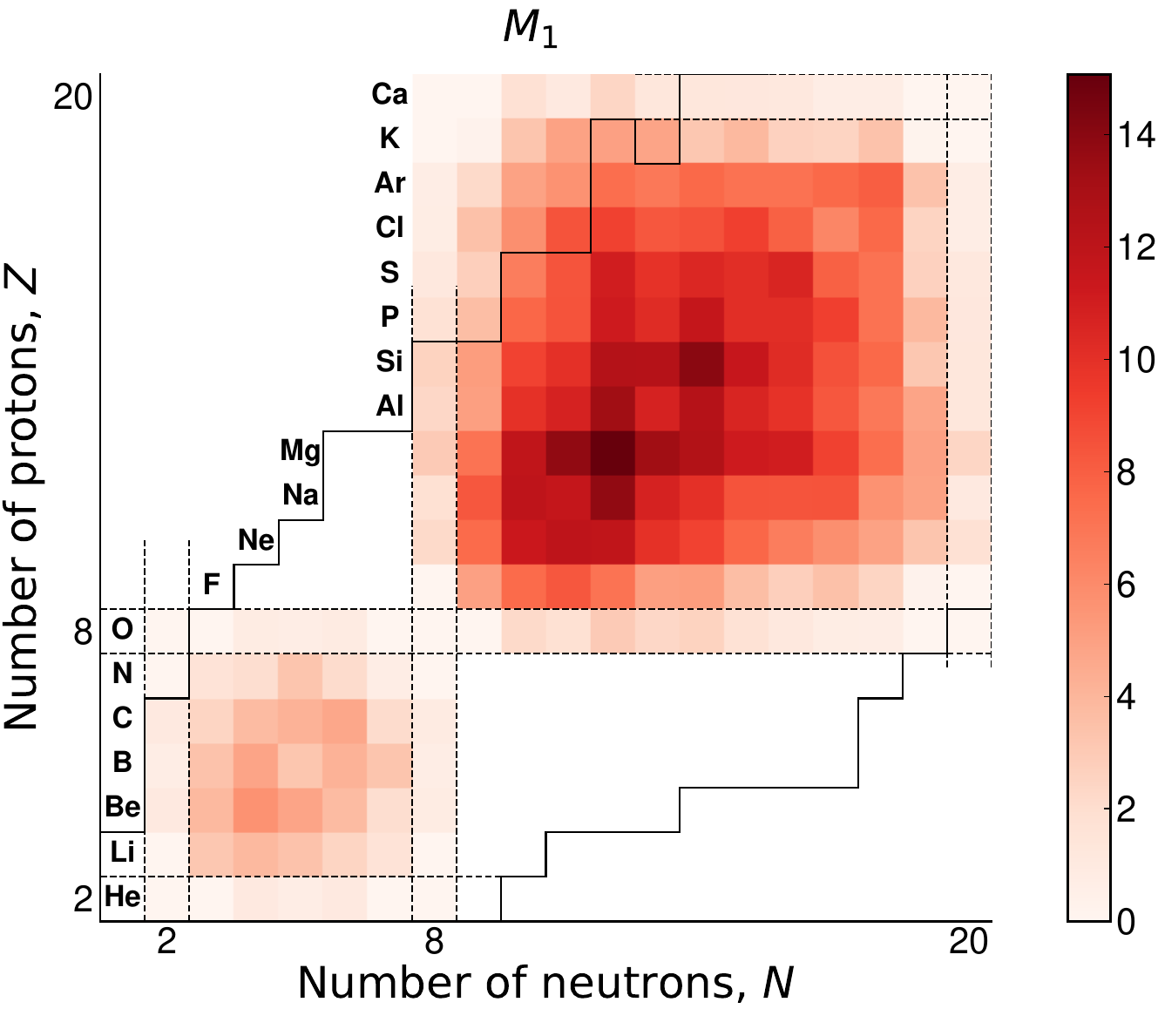} 
    \includegraphics[width=\columnwidth, keepaspectratio]{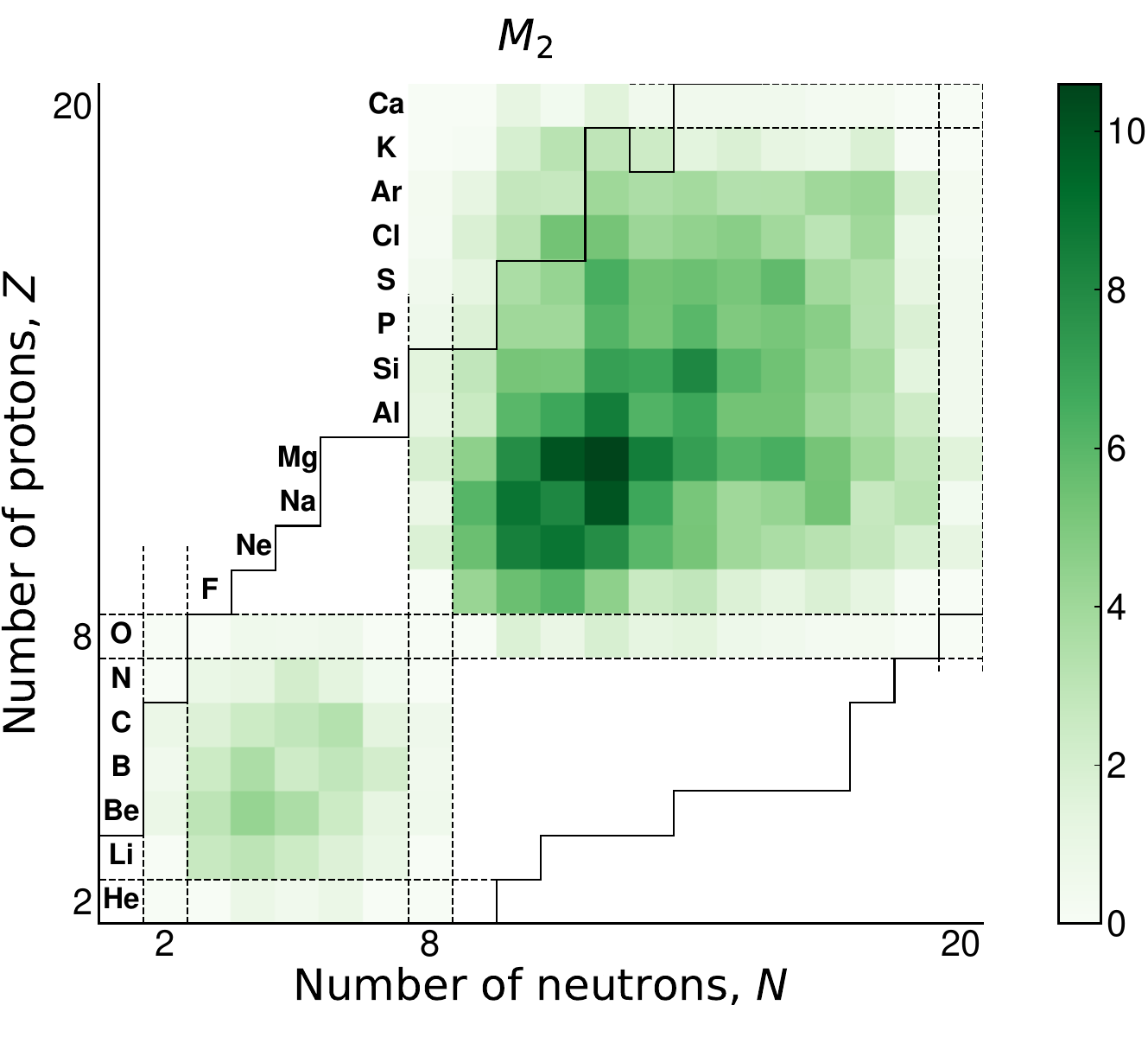}
    \caption{
    The chart of the ${\cal M}_1$ (upper) and ${\cal M}_2$  (lower)
    measures of magic of the $p$-shell and $sd$-shell nuclei computed from their active-space nuclear shell-model wavefunctions calculated using the {\tt BIGSTICK} code.
    The solid-gray lines denote the limits of stability, while the dashed-gray lines denote the closed shells in the spherical nuclear shell model.
    The numerical values used to generate these figures can be found in Table~\ref{tab:magicresultsHeLiBeB}- Table~\ref{tab:magicresultsCa} in App.~\ref{appendix:resulttables}.
    While it is the experimentally determined dripline that is displayed~\cite{nndc}, our results are obtained from an isospin-symmetric nuclear interaction without Coulomb,
    and hence a meaningful comparison would involve modifications that include
    a shift toward neutron excess due to the Coulomb interaction. } 
    \label{fig:M12chart}
\end{figure}
The results of our computations of ${\cal M}_1$ (upper) and ${\cal M}_2$ (lower) in the $p$-shell and 
$sd$-shell are shown in Fig.~\ref{fig:M12chart}.
While the maximum magic is found to coincide with the maximum deformation, $\beta$,
in each isotopic chain,  the magic is found to persist at a large value beyond where the deformation becomes small, indicating quantum complexity extends at a significant level through the region of shape co-existence, as transition into the region of level inversion.
%%%
%\sout{As this is also the pattern exhibited by the multi-partite $n$-tangles, this suggests that quantumcomputers will likely be able to provide a helpful acceleration of no-core nuclear structure and reaction calculations.}
This is also the pattern exhibited by the multi-partite proton-neutron $n$-tangles (see {\it e.g.} Fig.~\ref{fig:comparison} below and appendix~\ref{appendix:tangles}). These correlations between large quantum complexity (magic and entanglement) and collectivity is in accordance with the fact that capturing collectivity is known to be a challenge for {\it ab-initio} methods such as the no-core shell model, and suggest that quantum computers will likely be able to provide a helpful acceleration of precision no-core nuclear structure and reaction calculations.

It is helpful to consider representative 
examples of the distributions of Pauli strings for select nuclei.
\begin{figure}[!ht]
    \centering
    \includegraphics[width=\columnwidth]{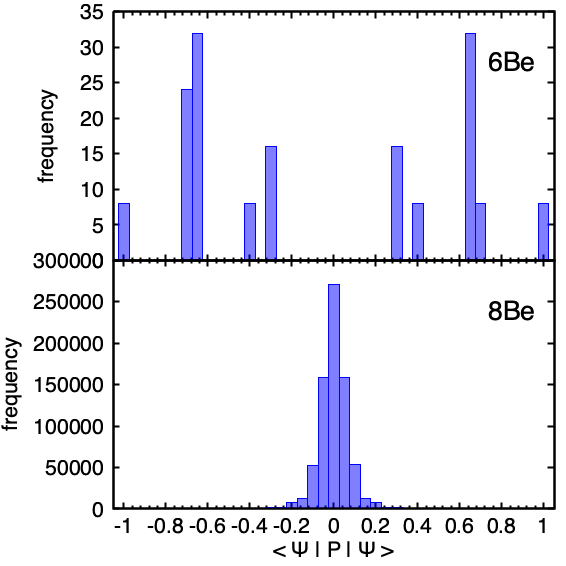}
    \caption{
    Histograms of the (non-zero) 
    Pauli-string expectation values $\langle \Psi |\hat P | \Psi \rangle$
    obtained from the {\tt BIGSTICK} shell-model wavefunctions for $^6$Be and $^8$Be.
    A bin width of $0.05$ has been used.
    }
    \label{fig:cPBe}
\end{figure}
Figure~\ref{fig:cPBe} shows the distribution of values of $c_P$ obtained for $^6$Be and $^8$Be
(recall that for a stabilizer state,
such a  histogram would have support only at $c_P=\pm 1$ for $d$ of the possible $d^2$ strings).
These distributions give measures of magic for $^6$Be of ${\cal M}_1=1.0422$ and 
${\cal M}_2=0.8465$, which are smaller than those of
$^8$Be of ${\cal M}_1=5.6194$ and ${\cal M}_2=4.2940$.
It is clear from these distributions and the corresponding measures of magic that $^6$Be is closer to a stabilizer state than $^8$Be.
That is to say that, in the spherical shell-model basis, $^8$Be has substantially 
more quantum complexity than $^6$Be.
From a physics perspective, this is consistent with $^6$Be closely resembling two protons 
on a $^4$He core, while 
$^8$Be has significant collective structure, consistent with two $^4$He nuclei near threshold.

The results that we have presented so far are for the ground states of nuclei with $J_z=J$.
It is interesting to examine the behavior of the measures of magic for different spatial orientations 
of the nuclei, corresponding to the different $J_z$ states.
As an example, consider $^7$Li with one proton and two neutrons in the $p$-shell.
Exact calculation of the ${\cal M}_1$s associated with the 
$|J_z|= {3\over 2}, {1\over 2} $ states give
$3.8834$ and $4.0700$, respectively.
Similarly, in the $sd$-shell, 
$^{19}$O that has a ground state with spin and parity $J^\pi={5\over 2}^+$, and 
exact calculations of the ${\cal M}_2$ associated with the 
$|J_z|={5\over 2}, {3\over 2}, {1\over 2} $ states give $0.9677, 0.9663, 0.9690$, respectively.
The magic does depend upon the $J_z$ value of the state, but it is a small to modest-sized effect.

%%%%%%%%%%%%%%%%%%%%%%%%%%%%%%%%%%%%%%%
\section{Comparisons}
\label{sec:comps}
\begin{figure}[!ht]
    \centering
    \includegraphics[width=\columnwidth]{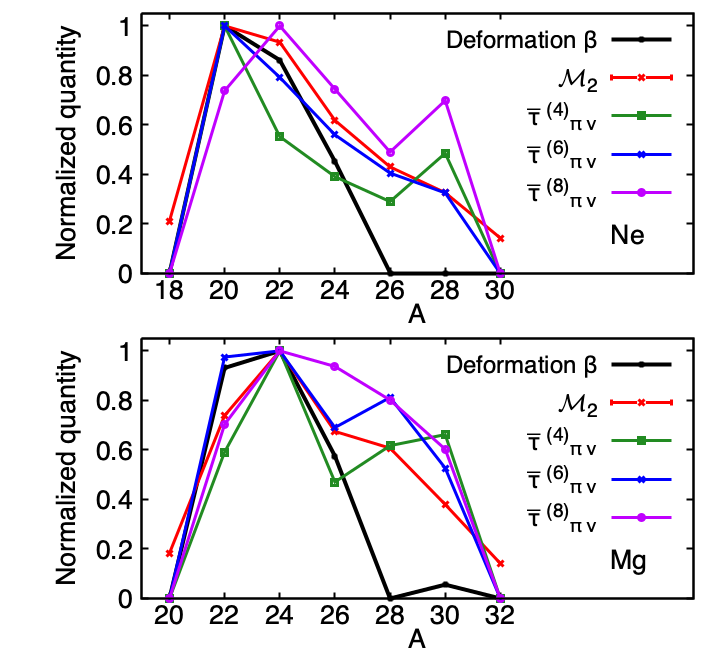}
    \caption{
   The magic ${\cal M}_2$, the $n=4,6,8$-tangles $\bar{\tau}_{\pi\nu}^{(n)}$ in the proton-neutron sector and the deformation parameter $\beta$ in the
   neon isotopic chain $^{18}$Ne -  $^{30}$Ne (upper) and the 
   magnesium isotope chain $^{20}$Mg -  $^{32}$Mg (lower). 
   The values of $\beta$ were taken from 
   \href{https://www-phynu.cea.fr/science_en_ligne/carte_potentiels_microscopiques/tables/HFB-5DCH-table_eng.htm}{Summary Tables}~\cite{PhysRevC.81.014303} reproduced at
   the website~\cite{phynu-cea}.
   Each quantity has been normalized to its maximum value in the chain.
    }
    \label{fig:comparison}
\end{figure}

\noindent
It is interesting to compare the behavior of the quantum information with the 
shape parameters in an isotopic chain.  
As specific examples, 
we examine the behavior of ${\cal M}_2$, the summed proton-neutron $n=2,4,6$-tangles $\bar{\tau}_{\pi\nu}^{(n)}$ and the $\beta$ deformation parameter for 
$^{18}$Ne -  $^{30}$Ne and 
$^{20}$Mg -  $^{32}$Mg. 
Those deformation parameters, which are not accessible within the spherical shell model framework, have been taken from deformed Hartree-Fock-Bogoliubov generator coordinate 
calculations 
of Ref.~\cite{PhysRevC.81.014303}, provided in Ref.~\cite{phynu-cea}. 
We selected the proton-neutron part of the $n$-tangles as proton-neutron correlations are usually considered to be related to deformation.
For convenience, we have normalized each quantity to its maximum value,
as shown in Fig.~\ref{fig:comparison}.
In the Mg isotopic chain, while $\beta$ drops to zero 
for $^{28}$Mg, a nucleus in the shape coexistence region at the boundary of the island of inversion, the magic and proton-neutron $n$-tangles remain significant, before becoming small (magic) or vanishing ($n$-tangles) at the neutron shell closure.
The same behaviour is exhibited by the Ne chain.
The trend of the $n$-tangles in the proton-neutron sector confirms the long-thought crucial role of the proton-neutron correlations in driving deformation.
As shown in appendix~\ref{appendix:tangles}, the pure proton and pure neutron $n$-tangles do not exhibit such clear correlation with nuclear shape.

The correlation between magic and nuclear shape, possibly including shape co-existence, suggests a connection between the deformation of a nucleus and the classical resources required to compute its ground state wavefunction using a spherical basis.  
As the required classical computing resources scale with the exponential of the magic~\cite{gottesman1998heisenbergrepresentationquantumcomputers,Aaronson_2004}, 
our results suggest that they scale exponentially with the 
(non-trivial)
``shape-complexity'' of the nucleus, 
something that is not captured by a single shape-parameter alone, but appears to imprint a signature in the higher-body multi-partite $n$-tangles.
It would be interesting to compare the classical resources required for full-space calculations of 
these nuclei to better examine the relation between nuclear shapes, entanglement and magic.

%%%%%%%%%%%%%%%%%%%%%%%%%%%%%%
\section{Conclusions}
\label{sec:conclusions}
\noindent
Advances in quantum information science are transforming our understanding of quantum many-body systems,
and are providing new techniques and algorithms for predicting their properties and dynamics that are out of reach of experiment and of classical computing alone.
This new technology provides opportunities to further improve our understanding of nuclei and nuclear reactions.  
Further, the now anticipated fault-tolerant quantum computers and processing units should provide 
computational capabilities that were impractical to consider seriously just a few years ago.

Nuclei are particularly interesting self-bound systems of two species of fermions (protons and neutrons)
with strong short-range central and tensor two-body forces, strong three-body forces, 
and long-range electromagnetic interactions (neglecting the weak interactions).   
Beyond electromagnetism, the two species of fermions are nearly identical, 
but because of fine-tunings in the Standard Model, 
the small differences in quark masses are significantly amplified,
for instance furnishing a two-nucleon system near unitarity.
Combined, these features give rise to remarkable structures and complexities of nuclei, 
including of light nuclei.

In the context of computational complexity, 
the non-stabilizerness of a quantum state, 
encapulated by the measures of magic, 
determines the quantum resources that are 
required to prepare the state, beyond the classical resources.
Entanglement alone is insufficient to define a need for quantum resources, 
as some entangled states are accessible via a classical gate set, 
as encapsulated in the Gottesman-Knill 
theorem~\cite{gottesman1998heisenbergrepresentationquantumcomputers} and codified 
by Aaronson and Gottesman~\cite{Aaronson_2004}
\footnote{Examples of further advances can be found in, e.g., Refs.~\cite{Anders_2006,Bravyi_2016,Gidney_2021}.}.
It is the combination of non-stabilizerness with large-scale multi-partite entanglement that drives the need for quantum computing resources to prepare and manipulate a quantum state.

In this work, we have examined the quantum complexity in light and mid-mass nuclei, 
focusing on the entanglement structure and magic of the active nucleons in the 
$p$-shell and $sd$-shells of the spherical nuclear shell model.
We have found,
unsurprisingly,
that the known complexity of these nuclei,
including collective effects such as shape deformation and shape co-existence,
which present challenges for the spherical shell model,
are reflected in measures of multi-nucleon entanglement and magic.
In deformed nuclei and isotopes on the path to instability,
the higher-body entanglement, including collective proton-neutron entanglement, 
is prominent, as are the measures of magic.  The relatively large values of these quantities persist
for isotopes beyond those that are deformed.

Implicit in our studies is the computation of matrix elements of strings of Pauli operators in
the ground-state wavefunctions of $p$-shell and $sd$-shells nuclei that are mapped to qubits,
with each qubit defining the occupancy of a single-particle shell model state.
Calculations in the $p$-shell and in  $sd$-shell nuclei with only one specie present are performed exactly,
while for the typical $sd$-shell nuclei, 
the measures of magic are evaluated using an extensive 
suite of detailed MCMC evaluations. 
To accelerate the convergence of these evaluations in deformed nuclei, we introduced the 
PSIZe-MCMC algorithm,
where the $d$ matrix elements of Pauli strings of $\hat I, \hat Z$
operators are evaluated exactly, 
while the remaining matrix elements are evaluated using MCMC.

From a theoretical perspective, there is a path to be pursued in which transformations among the basis states and Hamiltonian are identified that reduce the magic and multi-partite entanglement in the ground state wavefunctions.  
Repeating our calculations using a deformed/collective basis is expected to yield results with less 
quantum complexity.
This is  along the lines of work we and others have pursued in entanglement re-arrangement, e.g., Ref.~\cite{Robin:2020aeh}, 
aligning with QIS methods such as MERA.

Quantum information science techniques,
specifically measures of multi-body entanglement and magic,
are providing new insights about the structure of nuclei, and opening new paths forward for theoretical and numerical techniques to improve predictions of their structure.

\color{black}

%%%%%%%%%%%%%%%%%%%%%%%%%%%%%%
\begin{acknowledgements}
We thank Thomas Neff for providing the matrix elements of the shell-model interactions.
We would like to also thank Marc Illa for providing feedback on the manuscript, Niklas M\"uller and Henry Froland for useful discussions, as well as Emanuele Tirrito for his inspiring presentation at the IQuS workshop {\it Pulses, Qudits and Quantum Simulations}\footnote{\url{https://iqus.uw.edu/events/pulsesquditssimulations/}}, 
co-organized by Yujin Cho, Ravi Naik, Alessandro Roggero and Kyle Wendt, and for subsequent discussions,
an also related discussions with Alessandro Roggero and Kyle Wendt.
We would further thank Thomas Papenbrock and Rahul Trivedi for useful discussions
during the 
IQuS workshop {\it Entanglement in Many-Body Systems: From Nuclei to Quantum Computers and Back}\footnote{\url{https://iqus.uw.edu/events/entanglementinmanybody/}},
co-organized by Mari Carmen Banuls, Susan Coppersmith, Calvin Johnson and Caroline Robin.
This work was supported, in part, by Universit\"at Bielefeld (Caroline, Federico), 
by the Deutsche Forschungsgemeinschaft (DFG, German Research Foundation) through the CRC-TR 211 'Strong-interaction matter under extreme conditions'– project number 315477589 – TRR 211 (Momme),
by ERC-885281-KILONOVA Advanced Grant (Caroline), 
and by the MKW NRW under the funding code NW21-024-A (James).
This work was also supported by U.S. Department of Energy, Office of Science, Office of Nuclear Physics, InQubator for Quantum Simulation (IQuS)\footnote{\url{https://iqus.uw.edu}} under Award Number DOE (NP) Award DE-SC0020970 via the program on Quantum Horizons: QIS Research and Innovation for Nuclear Science\footnote{\url{https://science.osti.gov/np/Research/Quantum-Information-Science}}, and by the Department of Physics\footnote{\url{https://phys.washington.edu}}
and the College of Arts and Sciences\footnote{\url{https://www.artsci.washington.edu}} at the University of Washington (Martin). 
This research used resources of the National Energy Research
Scientific Computing Center, a DOE Office of Science User Facility
supported by the Office of Science of the U.S. Department of Energy
under Contract No. DE-AC02-05CH11231 using NERSC awards
NP-ERCAP0027114 and NP-ERCAP0029601.
Some of the computations in this work were performed
on the GPU cluster at Bielefeld University. We
thank the Bielefeld HPC.NRW team for their support.
This research was also partly supported by the cluster computing resource provided by the IT Department at the GSI Helmholtzzentrum für Schwerionenforschung, Darmstadt, Germany.

\end{acknowledgements}

\bibliography{biblio_paper}

\onecolumngrid

\appendix

%%%%%%%%%%%%%%%%%%%%%%%
\color{black}
\section{$n$-tangles}
\label{appendix:tangles}

In Table \ref{tab:tanglesHeLiBeB}-\ref{tab:tanglesCa}, we provide the values of $\overline{\tau}^{(n)}_{\pi\nu}$, $\overline{\tau}^{(n)}_{\pi}$ and $\overline{\tau}^{(n)}_{\nu}$ for $n=4,6,8$ for $p$-shell and $sd$-shell nuclei (corresponding to the numbers displayed in Figs.~\ref{f:p_4-6-8-tangles} and \ref{f:sd_4-6-8-tangles}).

\begin{table}[!t]
\caption{
Summed values of $n$-tangles, $\overline{\tau}^{(n)}_{\pi\nu}$, $\overline{\tau}^{(n)}_{\pi}$ and $\overline{\tau}^{(n)}_{\nu}$, for $n=2,4,6,8$, in the $J_z=J$ ground states of the $p$-shell nuclei
helium, lithium, beryllium and  boron.
The implicit isospin symmetry of the Hamiltonian implemented in {\tt BIGSTICK} 
gives rise to exact relations between the $n$-tangles among different nuclei. 
All entries result from exact calculations.
}
\renewcommand{\arraystretch}{1.4}
\begin{tabularx}{\textwidth}{|Y| Y | Y || Y || Y | Y || Y| Y| Y|| Y| Y| Y|| Y| Y| Y|  }
  \hline
 Z & N & A & J &  \multicolumn{2}{c||}{summed $2$-tangle}  &  \multicolumn{3}{c||}{summed $4$-tangle}   &   \multicolumn{3}{c||}{summed $6$-tangle}   &   \multicolumn{3}{c|}{summed $8$-tangle} \\
   &  &  &  & $\overline{\tau}^{(2)}_{\pi}$    &   $\overline{\tau}^{(2)}_{\nu}$ &  $\overline{\tau}^{(4)}_{\pi\nu}$   &   $\overline{\tau}^{(4)}_{\pi}$    &   $\overline{\tau}^{(4)}_{\nu}$ &  $\overline{\tau}^{(6)}_{\pi\nu}$   &   $\overline{\tau}^{(6)}_{\pi}$   &   $\overline{\tau}^{(6)}_{\nu}$ &  $\overline{\tau}^{(8)}_{\pi\nu}$    &   $\overline{\tau}^{(8)}_{\pi}$   &   $\overline{\tau}^{(8)}_{\nu}$\\
\hline
\hline 
\multicolumn{3}{|c||}{ Helium (He) }  \\ 
\hline 
\hline
2 & 2 & 4 & 0 & 0 & 0 & 0 & 0 & 0 & 0 & 0 & 0 & 0 & 0 & 0  \\
  & 2+1 & 5 & 3/2 & 0 & 0 & 0 & 0 & 0 & 0 & 0 & 0 & 0 & 0 & 0  \\
  & 2+2 & 6 & 0 & 0 & 0 & 0 & 0 & 1.3318 & 0 & 0 & 0 & 0 & 0 & 0  \\
  & 2+3 & 7 & 3/2 & 0 & 0.0368 & 0 & 0 & 0.6054 & 0 & 0 & 0 & 0 & 0 & 0  \\
  & 2+4 & 8 & 0 & 0 & 0 & 0 & 0 & 1.0357 & 0 & 0 & 0 & 0 & 0 & 0  \\
  & 2+5 & 9 & 1/2 & 0 & 0 & 0 & 0 & 0 & 0 & 0 & 0 & 0 & 0 & 0  \\
  & 2+6 & 10 & 0 & 0 & 0 & 0 & 0 & 0 & 0 & 0 & 0 & 0 & 0 & 0  \\
\hline
\hline 
\multicolumn{3}{|c||}{ Lithium (Li) }  \\ 
\hline 
\hline
2+1 & 2 & 5 & 3/2 & 0 & 0 & 0 & 0 & 0 & 0 & 0 & 0 & 0 & 0 & 0  \\
  & 2+1 & 6 & 1 & 0.1204 & 0.1204 & 1.3415 & 0 & 0 & 0 & 0 & 0 & 0 & 0 & 0  \\
  & 2+2 & 7 & 3/2 & 0.0644 & 0.1770 & 0.8911 & 0 & 0.5255 & 0.3287 & 0 & 0 & 0 & 0 & 0  \\
  & 2+3 & 8 & 2 & 0.1610 & 0.2141 & 1.1578 & 0 & 0.2232 & 0.2207 & 0 & 0 & 0 & 0 & 0  \\
  & 2+4 & 9 & 3/2 & 0.0045 & 0.1127 & 0.3543 & 0 & 0.7270 & 0.0850 & 0 & 0 & 0 & 0 & 0  \\
  & 2+5 & 10 & 1 & 0.0025 & 0.0900 & 0.9132 & 0 & 0 & 0 & 0 & 0 & 0 & 0 & 0  \\
  & 2+6 & 11 & 3/2 & 0 & 0 & 0 & 0 & 0 & 0 & 0 & 0 & 0 & 0 & 0  \\
\hline
\hline 
\multicolumn{3}{|c||}{ Beryllium (Be) }  \\ 
\hline 
\hline
2+2 & 2 & 6 & 0 & 0 & 0 & 0 & 1.3318 & 0 & 0 & 0 & 0 & 0 & 0 & 0  \\
  & 2+1 & 7 & 3/2 & 0.1770 & 0.0644 & 0.8911 & 0.5255 & 0 & 0.3287 & 0 & 0 & 0 & 0 & 0  \\
  & 2+2 & 8 & 0 & 0 & 0 & 1.2750 & 0.4610 & 0.4610 & 0.5452 & 0 & 0 & 1.0180 & 0 & 0  \\
  & 2+3 & 9 & 3/2 & 0.0963 & 0.0291 & 1.0756 & 0.4924 & 0.2943 & 0.5114 & 0 & 0 & 0.4447 & 0 & 0  \\
  & 2+4 & 10 & 0 & 0 & 0 & 0.4959 & 0.8419 & 0.6871 & 0.1897 & 0 & 0 & 0.9311 & 0 & 0  \\
  & 2+5 & 11 & 1/2 & 0.0002 & 0 & 0.2524 & 0.9634 & 0 & 0.0847 & 0 & 0 & 0 & 0 & 0  \\
  & 2+6 & 12 & 0 & 0 & 0 & 0 & 1.2037 & 0 & 0 & 0 & 0 & 0 & 0 & 0  \\
\hline
\hline 
\multicolumn{3}{|c||}{ Boron (B) }  \\ 
\hline 
\hline
2+3 & 2 & 7 & 3/2 & 0.0368 & 0 & 0 & 0.6054 & 0 & 0 & 0 & 0 & 0 & 0 & 0  \\
  & 2+1 & 8 & 2 & 0.2141 & 0.1610 & 1.1578 & 0.2232 & 0 & 0.2207 & 0 & 0 & 0 & 0 & 0  \\
  & 2+2 & 9 & 3/2 & 0.0291 & 0.0963 & 1.0756 & 0.2943 & 0.4924 & 0.5114 & 0 & 0 & 0.4447 & 0 & 0  \\
  & 2+3 & 10 & 3 & 0.3826 & 0.3826 & 0.9651 & 0.2719 & 0.2719 & 0.3874 & 0 & 0 & 0.1370 & 0 & 0  \\
  & 2+4 & 11 & 3/2 & 0.0882 & 0.0668 & 0.7462 & 0.1522 & 0.5304 & 0.3468 & 0 & 0 & 0.2105 & 0 & 0  \\
  & 2+5 & 12 & 1 & 0.1171 & 0.1231 & 0.9231 & 0.0905 & 0 & 0.2773 & 0 & 0 & 0.0061 & 0 & 0  \\
  & 2+6 & 13 & 3/2 & 0.0911 & 0 & 0 & 0.1800 & 0 & 0 & 0 & 0 & 0 & 0 & 0  \\
\hline
\hline
\end{tabularx}
 \label{tab:tanglesHeLiBeB}
\end{table}
\begin{table}[!t]
\caption{
Summed values of $n$-tangles, $\overline{\tau}^{(n)}_{\pi\nu}$, $\overline{\tau}^{(n)}_{\pi}$ and $\overline{\tau}^{(n)}_{\nu}$, for $n=2,4,6,8$, in the $J_z=J$ ground states of the $p$-shell nuclei
carbon, nitrogen and oxygen.
The implicit isospin symmetry of the Hamiltonian implemented in {\tt BIGSTICK} 
gives rise to exact relations between the $n$-tangles among different nuclei. All entries result from exact calculations.
}
\renewcommand{\arraystretch}{1.4}
\begin{tabularx}{\textwidth}{|Y| Y | Y || Y || Y | Y || Y| Y| Y|| Y| Y| Y|| Y| Y| Y|  }
  \hline
 Z & N & A & J &  \multicolumn{2}{c||}{summed $2$-tangle}  &  \multicolumn{3}{c||}{summed $4$-tangle}   &   \multicolumn{3}{c||}{summed $6$-tangle}   &   \multicolumn{3}{c|}{summed $8$-tangle} \\
   &  &  &  & $\overline{\tau}^{(2)}_{\pi}$    &   $\overline{\tau}^{(2)}_{\nu}$ &  $\overline{\tau}^{(4)}_{\pi\nu}$   &   $\overline{\tau}^{(4)}_{\pi}$    &   $\overline{\tau}^{(4)}_{\nu}$ &  $\overline{\tau}^{(6)}_{\pi\nu}$   &   $\overline{\tau}^{(6)}_{\pi}$   &   $\overline{\tau}^{(6)}_{\nu}$ &  $\overline{\tau}^{(8)}_{\pi\nu}$    &   $\overline{\tau}^{(8)}_{\pi}$   &   $\overline{\tau}^{(8)}_{\nu}$\\
\hline
\hline 
\multicolumn{3}{|c||}{ Carbon (C) }  \\ 
\hline 
\hline
2+4 & 2 & 8 & 0 & 0 & 0 & 0 & 1.0357 & 0 & 0 & 0 & 0 & 0 & 0 & 0  \\
  & 2+1 & 9 & 3/2 & 0.1127 & 0.0045 & 0.3543 & 0.7270 & 0 & 0.0850 & 0 & 0 & 0 & 0 & 0  \\
  & 2+2 & 10 & 0 & 0 & 0 & 0.4959 & 0.6871 & 0.8419 & 0.1897 & 0 & 0 & 0.9311 & 0 & 0  \\
  & 2+3 & 11 & 3/2 & 0.0668 & 0.0882 & 0.7462 & 0.5304 & 0.1522 & 0.3468 & 0 & 0 & 0.2105 & 0 & 0  \\
  & 2+4 & 12 & 0 & 0 & 0 & 1.0704 & 0.4931 & 0.4931 & 0.4167 & 0 & 0 & 0.7160 & 0 & 0  \\
  & 2+5 & 13 & 1/2 & 0.0010 & 0 & 0.3753 & 0.4991 & 0 & 0.0824 & 0 & 0 & 0 & 0 & 0  \\
  & 2+6 & 14 & 0 & 0 & 0 & 0 & 0.5041 & 0 & 0 & 0 & 0 & 0 & 0 & 0  \\
\hline
\hline 
\multicolumn{3}{|c||}{ Nitrogen (N) }  \\ 
\hline 
\hline
2+5 & 2 & 9 & 1/2 & 0 & 0 & 0 & 0 & 0 & 0 & 0 & 0 & 0 & 0 & 0  \\
  & 2+1 & 10 & 1 & 0.0900 & 0.0025 & 0.9132 & 0 & 0 & 0 & 0 & 0 & 0 & 0 & 0  \\
  & 2+2 & 11 & 1/2 & 0 & 0.0002 & 0.2524 & 0 & 0.9634 & 0.0847 & 0 & 0 & 0 & 0 & 0  \\
  & 2+3 & 12 & 1 & 0.1231 & 0.1171 & 0.9231 & 0 & 0.0905 & 0.2773 & 0 & 0 & 0.0061 & 0 & 0  \\
  & 2+4 & 13 & 1/2 & 0 & 0.0010 & 0.3753 & 0 & 0.4991 & 0.0824 & 0 & 0 & 0 & 0 & 0  \\
  & 2+5 & 14 & 0 & 0.0320 & 0.0320 & 0.1984 & 0 & 0 & 0 & 0 & 0 & 0 & 0 & 0  \\
  & 2+6 & 15 & 1/2 & 0 & 0 & 0 & 0 & 0 & 0 & 0 & 0 & 0 & 0 & 0  \\
\hline
\hline 
\multicolumn{3}{|c||}{ Oxygen (O) }  \\ 
\hline 
\hline
2+6 & 2 & 10 & 0 & 0 & 0 & 0 & 0 & 0 & 0 & 0 & 0 & 0 & 0 & 0  \\
  & 2+1 & 11 & 3/2 & 0 & 0 & 0 & 0 & 0 & 0 & 0 & 0 & 0 & 0 & 0  \\
  & 2+2 & 12 & 0 & 0 & 0 & 0 & 0 & 1.2037 & 0 & 0 & 0 & 0 & 0 & 0  \\
  & 2+3 & 13 & 3/2 & 0 & 0.0911 & 0 & 0 & 0.1800 & 0 & 0 & 0 & 0 & 0 & 0  \\
  & 2+4 & 14 & 0 & 0 & 0 & 0 & 0 & 0.5041 & 0 & 0 & 0 & 0 & 0 & 0  \\
  & 2+5 & 15 & 1/2 & 0 & 0 & 0 & 0 & 0 & 0 & 0 & 0 & 0 & 0 & 0  \\
  & 2+6 & 16 & 0 & 0 & 0 & 0 & 0 & 0 & 0 & 0 & 0 & 0 & 0 & 0  \\
\hline
\hline
\end{tabularx}
 \label{tab:tanglesCNO}
\end{table}
\begin{table}[!t]
\caption{
Summed values of $n$-tangles, $\overline{\tau}^{(n)}_{\pi\nu}$, $\overline{\tau}^{(n)}_{\pi}$ and $\overline{\tau}^{(n)}_{\nu}$, for $n=4,6,8$, in the $J_z=J$ ground states of the $sd$-shell nuclei
oxygen, fluorine and neon.
The implicit isospin symmetry of the Hamiltonian implemented in {\tt BIGSTICK} 
gives rise to exact relations between the $n$-tangles among different nuclei. 
All entries result from exact calculations.
}
\renewcommand{\arraystretch}{1.4}
\begin{tabularx}{\textwidth}{|Y| Y | Y || Y || Y| Y| Y|| Y| Y| Y|| Y| Y| Y|  }
\hline
 Z & N & A & J &  \multicolumn{3}{c||}{summed $4$-tangle}   &   \multicolumn{3}{c||}{summed $6$-tangle}   &   \multicolumn{3}{c|}{summed $8$-tangle} \\
   &  &  &  &  $\overline{\tau}^{(4)}_{\pi\nu}$   &   $\overline{\tau}^{(4)}_{\pi}$    &   $\overline{\tau}^{(4)}_{\nu}$ &  $\overline{\tau}^{(6)}_{\pi\nu}$   &   $\overline{\tau}^{(6)}_{\pi}$   &   $\overline{\tau}^{(6)}_{\nu}$ &  $\overline{\tau}^{(8)}_{\pi\nu}$    &   $\overline{\tau}^{(8)}_{\pi}$   &   $\overline{\tau}^{(8)}_{\nu}$\\
\hline
\hline 
\multicolumn{3}{|c||}{ Oxygen (O) }  \\ 
\hline 
\hline
8 & 8 & 16 & 0 & 0 & 0 & 0 & 0 & 0 & 0 & 0 & 0 & 0  \\
  & 8+1 & 17 & 5/2 & 0 & 0 & 0 & 0 & 0 & 0 & 0 & 0 & 0  \\
  & 8+2 & 18 & 0 & 0 & 0 & 1.5349 & 0 & 0 & 0 & 0 & 0 & 0  \\
  & 8+3 & 19 & 5/2 & 0 & 0 & 1.2560 & 0 & 0 & 0.0248 & 0 & 0 & 0  \\
  & 8+4 & 20 & 0 & 0 & 0 & 2.3574 & 0 & 0 & 0.0530 & 0 & 0 & 0.8180  \\
  & 8+5 & 21 & 5/2 & 0 & 0 & 0.6817 & 0 & 0 & 0.0455 & 0 & 0 & 0.0475  \\
  & 8+6 & 22 & 0 & 0 & 0 & 1.0005 & 0 & 0 & 0.0459 & 0 & 0 & 0.2076  \\
  & 8+7 & 23 & 1/2 & 0 & 0 & 0.3687 & 0 & 0 & 0.0221 & 0 & 0 & 0.0179  \\
  & 8+8 & 24 & 0 & 0 & 0 & 0.3599 & 0 & 0 & 0.0040 & 0 & 0 & 0.0355  \\
  & 8+9 & 25 & 3/2 & 0 & 0 & 0.1640 & 0 & 0 & 0.0018 & 0 & 0 & 0  \\
  & 8+10 & 26 & 0 & 0 & 0 & 1.1027 & 0 & 0 & 0 & 0 & 0 & 0  \\
  & 8+11 & 27 & 3/2 & 0 & 0 & 0 & 0 & 0 & 0 & 0 & 0 & 0  \\
  & 8+12 & 28 & 0 & 0 & 0 & 0 & 0 & 0 & 0 & 0 & 0 & 0  \\
\hline
\hline 
\multicolumn{3}{|c||}{ Fluorine (F) }  \\ 
\hline 
\hline
8+1 & 8 & 17 & 5/2 & 0 & 0 & 0 & 0 & 0 & 0 & 0 & 0 & 0  \\
  & 8+1 & 18 & 1 & 1.6633 & 0 & 0 & 0 & 0 & 0 & 0 & 0 & 0  \\
  & 8+2 & 19 & 1/2 & 0.7535 & 0 & 0.3587 & 0.8130 & 0 & 0 & 0 & 0 & 0  \\
  & 8+3 & 20 & 2 & 0.7628 & 0 & 0.3973 & 0.7790 & 0 & 0.0710 & 0.4599 & 0 & 0  \\
  & 8+4 & 21 & 5/2 & 0.3827 & 0 & 0.9564 & 0.4667 & 0 & 0.1286 & 0.2850 & 0 & 0.2910  \\
  & 8+5 & 22 & 4 & 0.9194 & 0 & 0.4848 & 0.2570 & 0 & 0.0443 & 0.3002 & 0 & 0.0261  \\
  & 8+6 & 23 & 5/2 & 0.3497 & 0 & 0.9299 & 0.1672 & 0 & 0.1682 & 0.1518 & 0 & 0.1953  \\
  & 8+7 & 24 & 3 & 0.3790 & 0 & 0.3705 & 0.0768 & 0 & 0.0454 & 0.0529 & 0 & 0.0200  \\
  & 8+8 & 25 & 5/2 & 0.2481 & 0 & 0.4352 & 0.0446 & 0 & 0.0191 & 0.0358 & 0 & 0.0515  \\
  & 8+9 & 26 & 1 & 1.1628 & 0 & 0.0776 & 0.1656 & 0 & 0.0022 & 0.0537 & 0 & 0  \\
  & 8+10 & 27 & 5/2 & 0.3440 & 0 & 0.7345 & 0.0518 & 0 & 0 & 0 & 0 & 0  \\
  & 8+11 & 28 & 3 & 0.9394 & 0 & 0 & 0 & 0 & 0 & 0 & 0 & 0  \\
  & 8+12 & 29 & 5/2 & 0 & 0 & 0 & 0 & 0 & 0 & 0 & 0 & 0  \\
\hline
\hline 
\multicolumn{3}{|c||}{ Neon (Ne) }  \\ 
\hline 
\hline
8+2 & 8 & 18 & 0 & 0 & 1.5349 & 0 & 0 & 0 & 0 & 0 & 0 & 0  \\
  & 8+1 & 19 & 1/2 & 0.7535 & 0.3587 & 0 & 0.8130 & 0 & 0 & 0 & 0 & 0  \\
  & 8+2 & 20 & 0 & 0.8629 & 0.3151 & 0.3151 & 0.7491 & 0 & 0 & 1.3475 & 0 & 0  \\
  & 8+3 & 21 & 3/2 & 0.8044 & 0.3300 & 0.3929 & 1.0760 & 0 & 0.0895 & 1.2836 & 0 & 0  \\
  & 8+4 & 22 & 0 & 0.4770 & 0.4531 & 0.8436 & 0.5934 & 0 & 0.0573 & 1.8262 & 0 & 0.2556  \\
  & 8+5 & 23 & 5/2 & 0.6365 & 0.5760 & 0.5563 & 0.9642 & 0 & 0.0784 & 1.0614 & 0 & 0.0358  \\
  & 8+6 & 24 & 0 & 0.3374 & 0.7577 & 0.8387 & 0.4195 & 0 & 0.0604 & 1.3597 & 0 & 0.2145  \\
  & 8+7 & 25 & 1/2 & 0.2936 & 0.8789 & 0.3826 & 0.4407 & 0 & 0.0386 & 0.6251 & 0 & 0.0218  \\
  & 8+8 & 26 & 0 & 0.2506 & 1.0400 & 0.5564 & 0.3037 & 0 & 0.0092 & 0.8956 & 0 & 0.0812  \\
  & 8+9 & 27 & 3/2 & 0.3103 & 0.9113 & 0.1661 & 0.4228 & 0 & 0.0047 & 0.2941 & 0 & 0  \\
  & 8+10 & 28 & 0 & 0.4190 & 0.9628 & 0.7319 & 0.2438 & 0 & 0 & 1.2756 & 0 & 0  \\
  & 8+11 & 29 & 3/2 & 0.2866 & 1.0046 & 0 & 0.2110 & 0 & 0 & 0 & 0 & 0  \\
  & 8+12 & 30 & 0 & 0 & 1.4111 & 0 & 0 & 0 & 0 & 0 & 0 & 0  \\
\hline
\hline
\end{tabularx}
\label{tab:tanglesOFNe}
\end{table}

\begin{table}[!t]
\caption{
Summed values of $n$-tangles, $\overline{\tau}^{(n)}_{\pi\nu}$, $\overline{\tau}^{(n)}_{\pi}$ and $\overline{\tau}^{(n)}_{\nu}$, for $n=4,6,8$, in the $J_z=J$ ground states of the $sd$-shell nuclei
sodium, magnesium and aluminum.
The implicit isospin symmetry of the Hamiltonian implemented in {\tt BIGSTICK} 
gives rise to exact relations between the $n$-tangles among different nuclei. 
All entries result from exact calculations.
}
\renewcommand{\arraystretch}{1.4}
\begin{tabularx}{\textwidth}{|Y| Y | Y || Y || Y| Y| Y|| Y| Y| Y|| Y| Y| Y|  }
\hline
 Z & N & A & J &  \multicolumn{3}{c||}{summed $4$-tangle}   &   \multicolumn{3}{c||}{summed $6$-tangle}   &   \multicolumn{3}{c|}{summed $8$-tangle} \\
   &  &  &  &  $\overline{\tau}^{(4)}_{\pi\nu}$   &   $\overline{\tau}^{(4)}_{\pi}$    &   $\overline{\tau}^{(4)}_{\nu}$ &  $\overline{\tau}^{(6)}_{\pi\nu}$   &   $\overline{\tau}^{(6)}_{\pi}$   &   $\overline{\tau}^{(6)}_{\nu}$ &  $\overline{\tau}^{(8)}_{\pi\nu}$    &   $\overline{\tau}^{(8)}_{\pi}$   &   $\overline{\tau}^{(8)}_{\nu}$\\
\hline
\hline 
\multicolumn{3}{|c||}{ Sodium (Na) }  \\ 
\hline 
\hline
8+3 & 8 & 19 & 5/2 & 0 & 1.2560 & 0 & 0 & 0.0248 & 0 & 0 & 0 & 0  \\
  & 8+1 & 20 & 2 & 0.7628 & 0.3973 & 0 & 0.7790 & 0.0710 & 0 & 0.4599 & 0 & 0  \\
  & 8+2 & 21 & 3/2 & 0.8044 & 0.3929 & 0.3300 & 1.0760 & 0.0895 & 0 & 1.2836 & 0 & 0  \\
  & 8+3 & 22 & 3 & 1.3388 & 0.5290 & 0.5290 & 1.9895 & 0.1278 & 0.1278 & 1.9716 & 0 & 0  \\
  & 8+4 & 23 & 3/2 & 0.9262 & 0.3509 & 0.6348 & 1.2223 & 0.0716 & 0.1109 & 1.9351 & 0 & 0.1228  \\
  & 8+5 & 24 & 4 & 1.3970 & 0.4897 & 0.6000 & 1.9530 & 0.0960 & 0.1285 & 2.0015 & 0 & 0.0354  \\
  & 8+6 & 25 & 5/2 & 0.3957 & 0.4887 & 0.6835 & 0.3968 & 0.0166 & 0.0485 & 1.1630 & 0 & 0.1852  \\
  & 8+7 & 26 & 1 & 0.4762 & 0.3220 & 0.2562 & 0.5593 & 0.0143 & 0.0196 & 0.6337 & 0 & 0.0146  \\
  & 8+8 & 27 & 3/2 & 0.4893 & 0.5446 & 0.6901 & 0.4521 & 0.0234 & 0.0221 & 1.0073 & 0 & 0.1144  \\
  & 8+9 & 28 & 2 & 0.6603 & 0.3235 & 0.1679 & 0.8140 & 0.0421 & 0.0183 & 0.5993 & 0 & 0  \\
  & 8+10 & 29 & 5/2 & 0.7149 & 0.6217 & 0.6249 & 0.3209 & 0.0122 & 0 & 0.9665 & 0 & 0  \\
  & 8+11 & 30 & 2 & 0.6562 & 0.2752 & 0 & 0.6458 & 0.0161 & 0 & 0.0737 & 0 & 0  \\
  & 8+12 & 31 & 5/2 & 0 & 1.0971 & 0 & 0 & 0.0114 & 0 & 0 & 0 & 0  \\
\hline
\hline 
\multicolumn{3}{|c||}{ Magnesium (Mg) }  \\ 
\hline 
\hline
8+4 & 8 & 20 & 0 & 0 & 2.3574 & 0 & 0 & 0.0530 & 0 & 0 & 0.8180 & 0  \\
  & 8+1 & 21 & 5/2 & 0.3827 & 0.9564 & 0 & 0.4667 & 0.1286 & 0 & 0.2850 & 0.2910 & 0  \\
  & 8+2 & 22 & 0 & 0.4770 & 0.8436 & 0.4531 & 0.5934 & 0.0573 & 0 & 1.8262 & 0.2556 & 0  \\
  & 8+3 & 23 & 3/2 & 0.9262 & 0.6348 & 0.3509 & 1.2223 & 0.1109 & 0.0716 & 1.9351 & 0.1228 & 0  \\
  & 8+4 & 24 & 0 & 0.8064 & 0.5063 & 0.5063 & 0.6085 & 0.0471 & 0.0471 & 2.5972 & 0.1278 & 0.1278  \\
  & 8+5 & 25 & 5/2 & 1.0691 & 0.7316 & 0.4866 & 1.3734 & 0.1423 & 0.0829 & 2.2773 & 0.1469 & 0.0339  \\
  & 8+6 & 26 & 0 & 0.3794 & 1.0476 & 0.8202 & 0.4201 & 0.0313 & 0.0503 & 2.4338 & 0.3190 & 0.2621  \\
  & 8+7 & 27 & 1/2 & 0.4266 & 1.0090 & 0.3817 & 0.5348 & 0.0314 & 0.0546 & 1.2695 & 0.2960 & 0.0261  \\
  & 8+8 & 28 & 0 & 0.4978 & 0.9034 & 0.7205 & 0.4941 & 0.0349 & 0.0198 & 2.0812 & 0.2536 & 0.1472  \\
  & 8+9 & 29 & 1/2 & 0.4419 & 0.8832 & 0.2554 & 0.5203 & 0.0563 & 0.0117 & 0.8666 & 0.2067 & 0  \\
  & 8+10 & 30 & 0 & 0.5338 & 1.0284 & 0.5817 & 0.3205 & 0.0245 & 0 & 1.5659 & 0.2224 & 0  \\
  & 8+11 & 31 & 3/2 & 0.3932 & 1.0275 & 0 & 0.2854 & 0.0382 & 0 & 0.1251 & 0.1883 & 0  \\
  & 8+12 & 32 & 0 & 0 & 1.7202 & 0 & 0 & 0.0232 & 0 & 0 & 0.3099 & 0  \\
\hline
\hline 
\multicolumn{3}{|c||}{ Aluminum (Al) }  \\ 
\hline 
\hline
8+5 & 8 & 21 & 5/2 & 0 & 0.6817 & 0 & 0 & 0.0455 & 0 & 0 & 0.0475 & 0  \\
  & 8+1 & 22 & 4 & 0.9194 & 0.4848 & 0 & 0.2570 & 0.0443 & 0 & 0.3002 & 0.0261 & 0  \\
  & 8+2 & 23 & 5/2 & 0.6365 & 0.5563 & 0.5760 & 0.9642 & 0.0784 & 0 & 1.0614 & 0.0358 & 0  \\
  & 8+3 & 24 & 4 & 1.3970 & 0.6000 & 0.4897 & 1.9530 & 0.1285 & 0.0960 & 2.0015 & 0.0354 & 0  \\
  & 8+4 & 25 & 5/2 & 1.0691 & 0.4866 & 0.7316 & 1.3734 & 0.0829 & 0.1423 & 2.2773 & 0.0339 & 0.1469  \\
  & 8+5 & 26 & 5 & 1.1478 & 0.6046 & 0.6046 & 1.7216 & 0.1478 & 0.1478 & 1.9045 & 0.0427 & 0.0427  \\
  & 8+6 & 27 & 5/2 & 0.7606 & 0.4066 & 0.6608 & 0.7601 & 0.0626 & 0.1122 & 1.5589 & 0.0239 & 0.1857  \\
  & 8+7 & 28 & 2 & 0.7160 & 0.3417 & 0.2794 & 0.4937 & 0.0305 & 0.0355 & 0.8873 & 0.0216 & 0.0166  \\
  & 8+8 & 29 & 5/2 & 0.7356 & 0.4329 & 0.8259 & 0.8282 & 0.0438 & 0.0728 & 1.2460 & 0.0227 & 0.1640  \\
  & 8+9 & 30 & 3 & 0.8816 & 0.3463 & 0.2184 & 0.8133 & 0.0321 & 0.0352 & 0.6749 & 0.0119 & 0  \\
  & 8+10 & 31 & 5/2 & 0.4705 & 0.3292 & 0.6386 & 0.4902 & 0.0296 & 0 & 0.4796 & 0.0115 & 0  \\
  & 8+11 & 32 & 2 & 0.9363 & 0.1559 & 0 & 0.1656 & 0.0106 & 0 & 0.0877 & 0.0062 & 0  \\
  & 8+12 & 33 & 5/2 & 0 & 0.2706 & 0 & 0 & 0.0132 & 0 & 0 & 0.0081 & 0  \\
\hline
\hline
\end{tabularx}
\label{tab:tanglesNaMgAl}
\end{table}
\begin{table}[!t]
\caption{
Summed values of $n$-tangles, $\overline{\tau}^{(n)}_{\pi\nu}$, $\overline{\tau}^{(n)}_{\pi}$ and $\overline{\tau}^{(n)}_{\nu}$, for $n=4,6,8$, in the $J_z=J$ ground states of the $sd$-shell nuclei
silicon, phosphorus and sulfur.
The implicit isospin symmetry of the Hamiltonian implemented in {\tt BIGSTICK} 
gives rise to exact relations between the $n$-tangles among different nuclei. 
All entries result from exact calculations.
}
\renewcommand{\arraystretch}{1.4}
\begin{tabularx}{\textwidth}{|Y| Y | Y || Y || Y| Y| Y|| Y| Y| Y|| Y| Y| Y|  }
\hline
 Z & N & A & J &  \multicolumn{3}{c||}{summed $4$-tangle}   &   \multicolumn{3}{c||}{summed $6$-tangle}   &   \multicolumn{3}{c|}{summed $8$-tangle} \\
   &  &  &  &  $\overline{\tau}^{(4)}_{\pi\nu}$   &   $\overline{\tau}^{(4)}_{\pi}$    &   $\overline{\tau}^{(4)}_{\nu}$ &  $\overline{\tau}^{(6)}_{\pi\nu}$   &   $\overline{\tau}^{(6)}_{\pi}$   &   $\overline{\tau}^{(6)}_{\nu}$ &  $\overline{\tau}^{(8)}_{\pi\nu}$    &   $\overline{\tau}^{(8)}_{\pi}$   &   $\overline{\tau}^{(8)}_{\nu}$\\
\hline
\hline 
\multicolumn{3}{|c||}{ Silicon (Si) }  \\ 
\hline 
\hline
8+6 & 8 & 22 & 0 & 0 & 1.0005 & 0 & 0 & 0.0459 & 0 & 0 & 0.2076 & 0  \\
  & 8+1 & 23 & 5/2 & 0.3497 & 0.9299 & 0 & 0.1672 & 0.1682 & 0 & 0.1518 & 0.1953 & 0  \\
  & 8+2 & 24 & 0 & 0.3374 & 0.8387 & 0.7577 & 0.4195 & 0.0604 & 0 & 1.3597 & 0.2145 & 0  \\
  & 8+3 & 25 & 5/2 & 0.3957 & 0.6835 & 0.4887 & 0.3968 & 0.0485 & 0.0166 & 1.1630 & 0.1852 & 0  \\
  & 8+4 & 26 & 0 & 0.3794 & 0.8202 & 1.0476 & 0.4201 & 0.0503 & 0.0313 & 2.4338 & 0.2621 & 0.3190  \\
  & 8+5 & 27 & 5/2 & 0.7606 & 0.6608 & 0.4066 & 0.7601 & 0.1122 & 0.0626 & 1.5589 & 0.1857 & 0.0239  \\
  & 8+6 & 28 & 0 & 0.8761 & 0.6367 & 0.6367 & 0.6561 & 0.0761 & 0.0761 & 2.4488 & 0.1736 & 0.1736  \\
  & 8+7 & 29 & 1/2 & 0.7355 & 0.6777 & 0.3989 & 0.5513 & 0.0542 & 0.0690 & 1.3849 & 0.1710 & 0.0289  \\
  & 8+8 & 30 & 0 & 0.5702 & 0.7088 & 0.9618 & 0.4104 & 0.0373 & 0.0219 & 1.8442 & 0.1731 & 0.2281  \\
  & 8+9 & 31 & 3/2 & 0.7868 & 0.6695 & 0.4314 & 0.5525 & 0.0608 & 0.0213 & 0.9523 & 0.1252 & 0  \\
  & 8+10 & 32 & 0 & 0.4776 & 0.5530 & 0.7703 & 0.4299 & 0.0301 & 0 & 0.8962 & 0.0896 & 0  \\
  & 8+11 & 33 & 3/2 & 0.2899 & 0.4321 & 0 & 0.0680 & 0.0220 & 0 & 0.0417 & 0.0532 & 0  \\
  & 8+12 & 34 & 0 & 0 & 0.3501 & 0 & 0 & 0.0077 & 0 & 0 & 0.0314 & 0  \\
\hline
\hline 
\multicolumn{3}{|c||}{ Phosphorus (P) }  \\ 
\hline 
\hline
8+7 & 8 & 23 & 1/2 & 0 & 0.3687 & 0 & 0 & 0.0221 & 0 & 0 & 0.0179 & 0  \\
  & 8+1 & 24 & 3 & 0.3790 & 0.3705 & 0 & 0.0768 & 0.0454 & 0 & 0.0529 & 0.0200 & 0  \\
  & 8+2 & 25 & 1/2 & 0.2936 & 0.3826 & 0.8789 & 0.4407 & 0.0386 & 0 & 0.6251 & 0.0218 & 0  \\
  & 8+3 & 26 & 1 & 0.4762 & 0.2562 & 0.3220 & 0.5593 & 0.0196 & 0.0143 & 0.6337 & 0.0146 & 0  \\
  & 8+4 & 27 & 1/2 & 0.4266 & 0.3817 & 1.0090 & 0.5348 & 0.0546 & 0.0314 & 1.2695 & 0.0261 & 0.2960  \\
  & 8+5 & 28 & 2 & 0.7160 & 0.2794 & 0.3417 & 0.4937 & 0.0355 & 0.0305 & 0.8873 & 0.0166 & 0.0216  \\
  & 8+6 & 29 & 1/2 & 0.7355 & 0.3989 & 0.6777 & 0.5513 & 0.0690 & 0.0542 & 1.3849 & 0.0289 & 0.1710  \\
  & 8+7 & 30 & 1 & 0.7039 & 0.3204 & 0.3204 & 0.4763 & 0.0361 & 0.0361 & 0.8121 & 0.0211 & 0.0211  \\
  & 8+8 & 31 & 1/2 & 0.6915 & 0.3299 & 0.7319 & 0.4703 & 0.0368 & 0.0152 & 1.1307 & 0.0214 & 0.1576  \\
  & 8+9 & 32 & 1 & 0.7300 & 0.2511 & 0.2159 & 0.5847 & 0.0265 & 0.0112 & 0.6019 & 0.0139 & 0  \\
  & 8+10 & 33 & 1/2 & 0.3598 & 0.3237 & 0.7703 & 0.4528 & 0.0293 & 0 & 0.5203 & 0.0149 & 0  \\
  & 8+11 & 34 & 1 & 0.8913 & 0.1872 & 0 & 0.1289 & 0.0071 & 0 & 0.1259 & 0.0079 & 0  \\
  & 8+12 & 35 & 1/2 & 0 & 0.2530 & 0 & 0 & 0.0098 & 0 & 0 & 0.0089 & 0  \\
\hline
\hline 
\multicolumn{3}{|c||}{ Sulfur (S) }  \\ 
\hline 
\hline
8+8 & 8 & 24 & 0 & 0 & 0.3599 & 0 & 0 & 0.0040 & 0 & 0 & 0.0355 & 0  \\
  & 8+1 & 25 & 5/2 & 0.2481 & 0.4352 & 0 & 0.0446 & 0.0191 & 0 & 0.0358 & 0.0515 & 0  \\
  & 8+2 & 26 & 0 & 0.2506 & 0.5564 & 1.0400 & 0.3037 & 0.0092 & 0 & 0.8956 & 0.0812 & 0  \\
  & 8+3 & 27 & 3/2 & 0.4893 & 0.6901 & 0.5446 & 0.4521 & 0.0221 & 0.0234 & 1.0073 & 0.1144 & 0  \\
  & 8+4 & 28 & 0 & 0.4978 & 0.7205 & 0.9034 & 0.4941 & 0.0198 & 0.0349 & 2.0812 & 0.1472 & 0.2536  \\
  & 8+5 & 29 & 5/2 & 0.7356 & 0.8259 & 0.4329 & 0.8282 & 0.0728 & 0.0438 & 1.2460 & 0.1640 & 0.0227  \\
  & 8+6 & 30 & 0 & 0.5702 & 0.9618 & 0.7088 & 0.4104 & 0.0219 & 0.0373 & 1.8442 & 0.2281 & 0.1731  \\
  & 8+7 & 31 & 1/2 & 0.6915 & 0.7319 & 0.3299 & 0.4703 & 0.0152 & 0.0368 & 1.1307 & 0.1576 & 0.0214  \\
  & 8+8 & 32 & 0 & 0.9048 & 0.6363 & 0.6363 & 0.4576 & 0.0119 & 0.0119 & 2.1530 & 0.1350 & 0.1350  \\
  & 8+9 & 33 & 3/2 & 0.6957 & 0.6139 & 0.2839 & 0.4440 & 0.0244 & 0.0096 & 0.8100 & 0.1070 & 0  \\
  & 8+10 & 34 & 0 & 0.3249 & 0.6328 & 0.7766 & 0.3683 & 0.0125 & 0 & 0.9936 & 0.1014 & 0  \\
  & 8+11 & 35 & 3/2 & 0.2088 & 0.4540 & 0 & 0.0606 & 0.0090 & 0 & 0.0325 & 0.0484 & 0  \\
  & 8+12 & 36 & 0 & 0 & 0.3719 & 0 & 0 & 0.0038 & 0 & 0 & 0.0311 & 0  \\
\hline
\hline
\end{tabularx}
\label{tab:tanglesSiPS}
\end{table}
\begin{table}[!t]
\caption{
Summed values of $n$-tangles, $\overline{\tau}^{(n)}_{\pi\nu}$, $\overline{\tau}^{(n)}_{\pi}$ and $\overline{\tau}^{(n)}_{\nu}$, for $n=4,6,8$, in the $J_z=J$ ground states of the $sd$-shell nuclei
chlorine, argon and potassium.
The implicit isospin symmetry of the Hamiltonian implemented in {\tt BIGSTICK} 
gives rise to exact relations between the $n$-tangles among different nuclei. 
All entries result from exact calculations.
}
\renewcommand{\arraystretch}{1.4}
\begin{tabularx}{\textwidth}{|Y| Y | Y || Y || Y| Y| Y|| Y| Y| Y|| Y| Y| Y|  }
\hline
 Z & N & A & J &  \multicolumn{3}{c||}{summed $4$-tangle}   &   \multicolumn{3}{c||}{summed $6$-tangle}   &   \multicolumn{3}{c|}{summed $8$-tangle} \\
   &  &  &  &  $\overline{\tau}^{(4)}_{\pi\nu}$   &   $\overline{\tau}^{(4)}_{\pi}$    &   $\overline{\tau}^{(4)}_{\nu}$ &  $\overline{\tau}^{(6)}_{\pi\nu}$   &   $\overline{\tau}^{(6)}_{\pi}$   &   $\overline{\tau}^{(6)}_{\nu}$ &  $\overline{\tau}^{(8)}_{\pi\nu}$    &   $\overline{\tau}^{(8)}_{\pi}$   &   $\overline{\tau}^{(8)}_{\nu}$\\
\hline
\hline 
\multicolumn{3}{|c||}{ Chlorine (Cl) }  \\ 
\hline 
\hline
8+9 & 8 & 25 & 3/2 & 0 & 0.1640 & 0 & 0 & 0.0018 & 0 & 0 & 0 & 0  \\
  & 8+1 & 26 & 1 & 1.1628 & 0.0776 & 0 & 0.1656 & 0.0022 & 0 & 0.0537 & 0 & 0  \\
  & 8+2 & 27 & 3/2 & 0.3103 & 0.1661 & 0.9113 & 0.4228 & 0.0047 & 0 & 0.2941 & 0 & 0  \\
  & 8+3 & 28 & 2 & 0.6603 & 0.1679 & 0.3235 & 0.8140 & 0.0183 & 0.0421 & 0.5993 & 0 & 0  \\
  & 8+4 & 29 & 1/2 & 0.4419 & 0.2554 & 0.8832 & 0.5203 & 0.0117 & 0.0563 & 0.8666 & 0 & 0.2067  \\
  & 8+5 & 30 & 3 & 0.8816 & 0.2184 & 0.3463 & 0.8133 & 0.0352 & 0.0321 & 0.6749 & 0 & 0.0119  \\
  & 8+6 & 31 & 3/2 & 0.7868 & 0.4314 & 0.6695 & 0.5525 & 0.0213 & 0.0608 & 0.9523 & 0 & 0.1252  \\
  & 8+7 & 32 & 1 & 0.7300 & 0.2159 & 0.2511 & 0.5847 & 0.0112 & 0.0265 & 0.6019 & 0 & 0.0139  \\
  & 8+8 & 33 & 3/2 & 0.6957 & 0.2839 & 0.6139 & 0.4440 & 0.0096 & 0.0244 & 0.8100 & 0 & 0.1070  \\
  & 8+9 & 34 & 1 & 0.7841 & 0.1964 & 0.1964 & 0.5849 & 0.0067 & 0.0067 & 0.5885 & 0 & 0  \\
  & 8+10 & 35 & 3/2 & 0.6815 & 0.2490 & 0.5298 & 0.5609 & 0.0096 & 0 & 0.5418 & 0 & 0  \\
  & 8+11 & 36 & 2 & 0.9671 & 0.1194 & 0 & 0.1425 & 0.0026 & 0 & 0.0128 & 0 & 0  \\
  & 8+12 & 37 & 3/2 & 0 & 0.1564 & 0 & 0 & 0.0013 & 0 & 0 & 0 & 0  \\
\hline
\hline 
\multicolumn{3}{|c||}{ Argon (Ar) }  \\ 
\hline 
\hline
8+10 & 8 & 26 & 0 & 0 & 1.1027 & 0 & 0 & 0 & 0 & 0 & 0 & 0  \\
  & 8+1 & 27 & 5/2 & 0.3440 & 0.7345 & 0 & 0.0518 & 0 & 0 & 0 & 0 & 0  \\
  & 8+2 & 28 & 0 & 0.4190 & 0.7319 & 0.9628 & 0.2438 & 0 & 0 & 1.2756 & 0 & 0  \\
  & 8+3 & 29 & 5/2 & 0.7149 & 0.6249 & 0.6217 & 0.3209 & 0 & 0.0122 & 0.9665 & 0 & 0  \\
  & 8+4 & 30 & 0 & 0.5338 & 0.5817 & 1.0284 & 0.3205 & 0 & 0.0245 & 1.5659 & 0 & 0.2224  \\
  & 8+5 & 31 & 5/2 & 0.4705 & 0.6386 & 0.3292 & 0.4902 & 0 & 0.0296 & 0.4796 & 0 & 0.0115  \\
  & 8+6 & 32 & 0 & 0.4776 & 0.7703 & 0.5530 & 0.4299 & 0 & 0.0301 & 0.8962 & 0 & 0.0896  \\
  & 8+7 & 33 & 1/2 & 0.3598 & 0.7703 & 0.3237 & 0.4528 & 0 & 0.0293 & 0.5203 & 0 & 0.0149  \\
  & 8+8 & 34 & 0 & 0.3249 & 0.7766 & 0.6328 & 0.3683 & 0 & 0.0125 & 0.9936 & 0 & 0.1014  \\
  & 8+9 & 35 & 3/2 & 0.6815 & 0.5298 & 0.2490 & 0.5609 & 0 & 0.0096 & 0.5418 & 0 & 0  \\
  & 8+10 & 36 & 0 & 0.8395 & 0.4933 & 0.4933 & 0.5043 & 0 & 0 & 1.2094 & 0 & 0  \\
  & 8+11 & 37 & 3/2 & 0.5435 & 0.6506 & 0 & 0.1663 & 0 & 0 & 0 & 0 & 0  \\
  & 8+12 & 38 & 0 & 0 & 1.1041 & 0 & 0 & 0 & 0 & 0 & 0 & 0  \\
\hline
\hline 
\multicolumn{3}{|c||}{ Potassium (K) }  \\ 
\hline 
\hline
8+11 & 8 & 27 & 3/2 & 0 & 0 & 0 & 0 & 0 & 0 & 0 & 0 & 0  \\
  & 8+1 & 28 & 3 & 0.9394 & 0 & 0 & 0 & 0 & 0 & 0 & 0 & 0  \\
  & 8+2 & 29 & 3/2 & 0.2866 & 0 & 1.0046 & 0.2110 & 0 & 0 & 0 & 0 & 0  \\
  & 8+3 & 30 & 2 & 0.6562 & 0 & 0.2752 & 0.6458 & 0 & 0.0161 & 0.0737 & 0 & 0  \\
  & 8+4 & 31 & 3/2 & 0.3932 & 0 & 1.0275 & 0.2854 & 0 & 0.0382 & 0.1251 & 0 & 0.1883  \\
  & 8+5 & 32 & 2 & 0.9363 & 0 & 0.1559 & 0.1656 & 0 & 0.0106 & 0.0877 & 0 & 0.0062  \\
  & 8+6 & 33 & 3/2 & 0.2899 & 0 & 0.4321 & 0.0680 & 0 & 0.0220 & 0.0417 & 0 & 0.0532  \\
  & 8+7 & 34 & 1 & 0.8913 & 0 & 0.1872 & 0.1289 & 0 & 0.0071 & 0.1259 & 0 & 0.0079  \\
  & 8+8 & 35 & 3/2 & 0.2088 & 0 & 0.4540 & 0.0606 & 0 & 0.0090 & 0.0325 & 0 & 0.0484  \\
  & 8+9 & 36 & 2 & 0.9671 & 0 & 0.1194 & 0.1425 & 0 & 0.0026 & 0.0128 & 0 & 0  \\
  & 8+10 & 37 & 3/2 & 0.5435 & 0 & 0.6506 & 0.1663 & 0 & 0 & 0 & 0 & 0  \\
  & 8+11 & 38 & 1 & 1.1358 & 0 & 0 & 0 & 0 & 0 & 0 & 0 & 0  \\
  & 8+12 & 39 & 3/2 & 0 & 0 & 0 & 0 & 0 & 0 & 0 & 0 & 0  \\
\hline
\hline
\end{tabularx}
\label{tab:tanglesClArK}
\end{table}
\begin{table}[!t]
\caption{
Summed values of $n$-tangles, $\overline{\tau}^{(n)}_{\pi\nu}$, $\overline{\tau}^{(n)}_{\pi}$ and $\overline{\tau}^{(n)}_{\nu}$, for $n=4,6,8$, in the $J_z=J$ ground states of the $sd$-shell nucleus
calcium.
The implicit isospin symmetry of the Hamiltonian implemented in {\tt BIGSTICK} 
gives rise to exact relations between the $n$-tangles among different nuclei. 
All entries result from exact calculations.
}
\renewcommand{\arraystretch}{1.4}
\begin{tabularx}{\textwidth}{|Y| Y | Y || Y || Y| Y| Y|| Y| Y| Y|| Y| Y| Y|  }
\hline
 Z & N & A & J &  \multicolumn{3}{c||}{summed $4$-tangle}   &   \multicolumn{3}{c||}{summed $6$-tangle}   &   \multicolumn{3}{c|}{summed $8$-tangle} \\
   &  &  &  &  $\overline{\tau}^{(4)}_{\pi\nu}$   &   $\overline{\tau}^{(4)}_{\pi}$    &   $\overline{\tau}^{(4)}_{\nu}$ &  $\overline{\tau}^{(6)}_{\pi\nu}$   &   $\overline{\tau}^{(6)}_{\pi}$   &   $\overline{\tau}^{(6)}_{\nu}$ &  $\overline{\tau}^{(8)}_{\pi\nu}$    &   $\overline{\tau}^{(8)}_{\pi}$   &   $\overline{\tau}^{(8)}_{\nu}$\\
\hline
\hline 
\multicolumn{3}{|c||}{ Calcium (Ca) }  \\ 
\hline 
\hline
8+12 & 8 & 28 & 0 & 0 & 0 & 0 & 0 & 0 & 0 & 0 & 0 & 0  \\
  & 8+1 & 29 & 5/2 & 0 & 0 & 0 & 0 & 0 & 0 & 0 & 0 & 0  \\
  & 8+2 & 30 & 0 & 0 & 0 & 1.4111 & 0 & 0 & 0 & 0 & 0 & 0  \\
  & 8+3 & 31 & 5/2 & 0 & 0 & 1.0971 & 0 & 0 & 0.0114 & 0 & 0 & 0  \\
  & 8+4 & 32 & 0 & 0 & 0 & 1.7202 & 0 & 0 & 0.0232 & 0 & 0 & 0.3099  \\
  & 8+5 & 33 & 5/2 & 0 & 0 & 0.2706 & 0 & 0 & 0.0132 & 0 & 0 & 0.0081  \\
  & 8+6 & 34 & 0 & 0 & 0 & 0.3501 & 0 & 0 & 0.0077 & 0 & 0 & 0.0314  \\
  & 8+7 & 35 & 1/2 & 0 & 0 & 0.2530 & 0 & 0 & 0.0098 & 0 & 0 & 0.0089  \\
  & 8+8 & 36 & 0 & 0 & 0 & 0.3719 & 0 & 0 & 0.0038 & 0 & 0 & 0.0311  \\
  & 8+9 & 37 & 3/2 & 0 & 0 & 0.1564 & 0 & 0 & 0.0013 & 0 & 0 & 0  \\
  & 8+10 & 38 & 0 & 0 & 0 & 1.1041 & 0 & 0 & 0 & 0 & 0 & 0  \\
  & 8+11 & 39 & 3/2 & 0 & 0 & 0 & 0 & 0 & 0 & 0 & 0 & 0  \\
  & 8+12 & 40 & 0 & 0 & 0 & 0 & 0 & 0 & 0 & 0 & 0 & 0  \\
\hline
\hline
\end{tabularx}
\label{tab:tanglesCa}
\end{table}

\begin{figure}[!th]
 \includegraphics[width=0.49\textwidth]{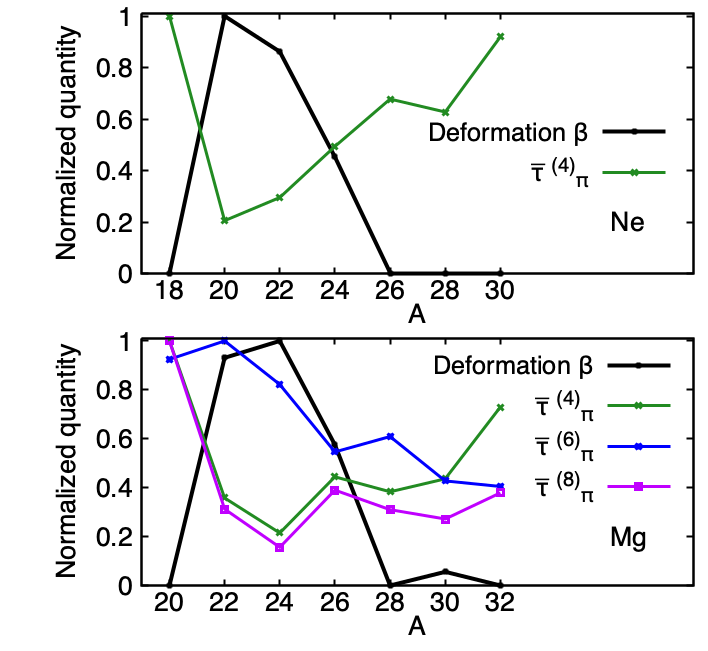}
  \includegraphics[width=0.49\textwidth]{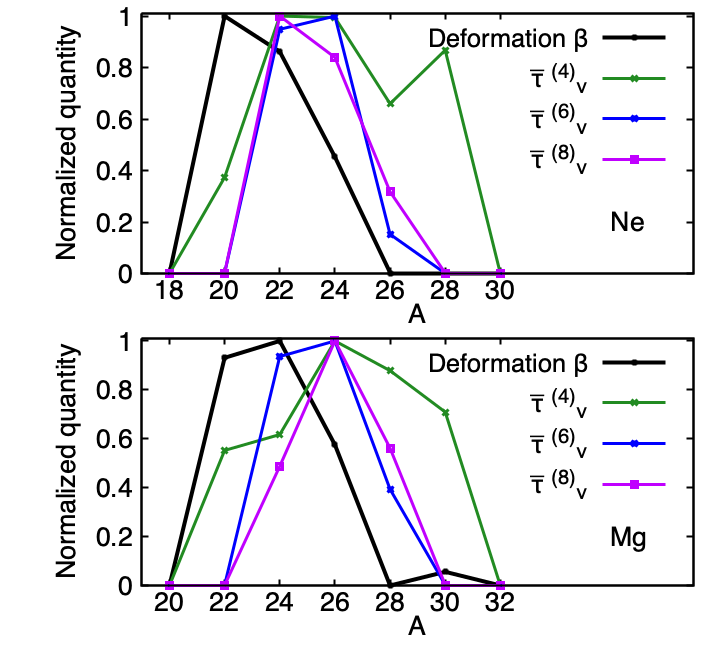}
   \includegraphics[width=0.49\textwidth]{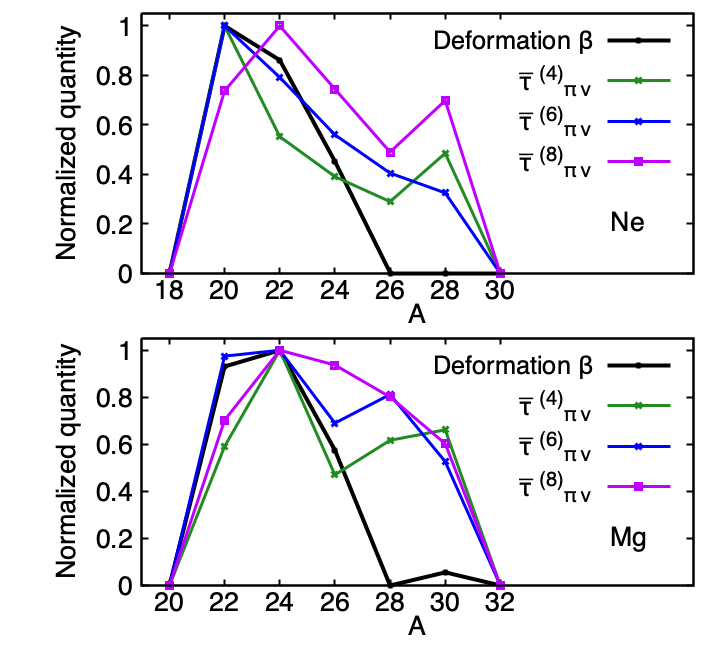}
     \caption{The summed $n=4,6,8$-tangles in the pure proton sector $\bar{\tau}_{\pi}^{(n)}$ (top left), pure neutron sector $\bar{\tau}_{\nu}^{(n)}$ (top right) and mixed proton-neutron sector $\bar{\tau}_{\pi\nu}^{(n)}$ (bottom), compared to the deformation parameter $\beta$ in the
   neon isotopic chain $^{18}$Ne -  $^{30}$Ne and the 
   magnesium isotopic chain $^{20}$Mg -  $^{32}$Mg. 
   The values of $\beta$ were taken from 
   \href{https://www-phynu.cea.fr/science_en_ligne/carte_potentiels_microscopiques/tables/HFB-5DCH-table_eng.htm}{Summary Tables}~\cite{PhysRevC.81.014303} reproduced at
   the website~\cite{phynu-cea}.
   Each quantity has been normalized to its maximum value in the chain.}
     \label{fig:tangles_deformation_Ne_Mg}
 \end{figure}

\clearpage
To complement Fig.~\ref{fig:comparison}, 
the behavior of the pure proton, pure neutron, and mixed proton-neutron components of the summed $n$-tangles in the Ne and Mg chains is shown in Fig.~\ref{fig:tangles_deformation_Ne_Mg}.
As discussed in the main text, the proton-neutron tangles, similarly to the magic, exhibit maxima which coincide with maximum deformation, while also extending towards the region of shape co-existence. The pure proton and pure neutron components of the summed tangles, however, exhibit different behaviours. In the proton sector, the summed $n$-tangles, in particular $\bar{\tau}^{4}_\pi$ and $\bar{\tau}^{8}_\pi$, appear somewhat anti-correlated with the deformation. The neutron components, on the other hand, 
show a similar trend but shifted towards larger values of the neutron number $N$.
%

%\clearpage

%\color{black}
%%%%%%%%%%%%%%%%%%%%%%%

%%%%%%%%%%%%%%%%%%%%%%%
\section{Pauli-String Evaluations in Hierarchical Wavefunctions -- The PSIZe-MCMC Method}
\label{appendix:psize}
\noindent
In this appendix, we describe the Pauli-String $\hat I\hat Z$ exact MCMC  (PSIZe-MCMC) method
for evaluating measures of magic in many-body wavefunctions.

 \begin{figure}[!th]
 \includegraphics[width=0.6\linewidth, keepaspectratio]{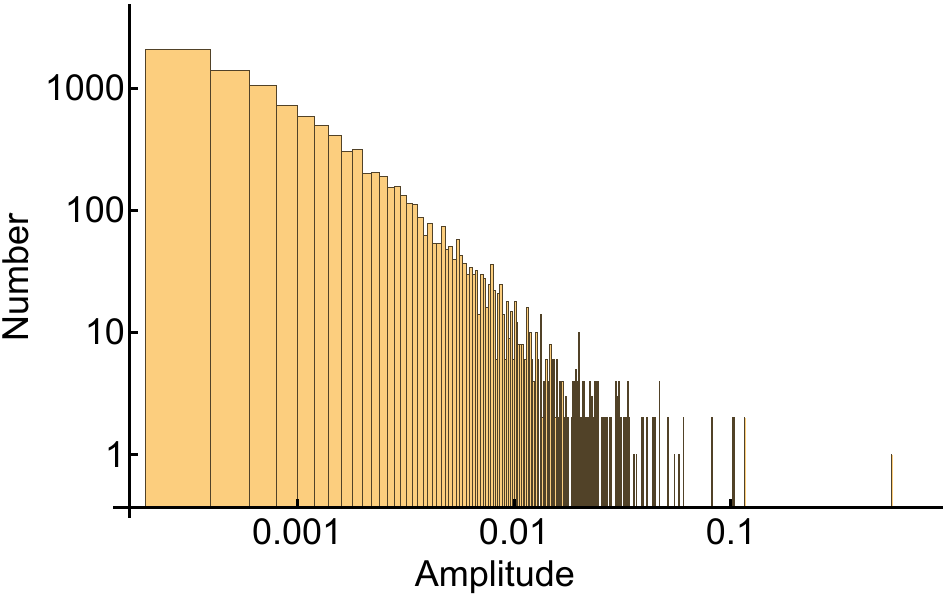}
     \caption{The distribution of positive amplitudes in $^{26}$Al wavefunction. The negative amplitudes exhibit a similar distribution.}
     \label{fig:26Alamps}
 \end{figure}
One of the convergence issues facing the Monte Carlo evaluation of magic in nuclei
is the distribution of amplitudes in the  shell-model wavefunctions.  
In particular, 
in addition to a modest number of natural size amplitudes,
there are a large number of small-amplitude components
in the deformed nuclei,
an example of which is shown in Fig.~\ref{fig:26Alamps}.
An effect of such a wavefunction structure is that Monte Carlo evaluations of observables
have small probability configurations that make large contributions to expectation values,
including the values of Pauli strings (and the $\Xi_P$).
This can give rise to long thermalization times in the Monte Carlo.
Not all nuclei in the periodic table have such poorly conditioned density matrices, 
but for those that do, the direct implementation of MCMC 
to evaluate the magic in the nuclear wavefunction is inefficient.
This is compounded by the nature of the Pauli strings being evaluated in the wavefunction.
In particular, the $d=2^{n_Q}$ Pauli strings with only $\hat I, \hat Z$ (diagonal) operators 
are all non-vanishing, and are typically large, for example, $\hat I^{\otimes n_Q}$ gives rise to $c_{\hat I^{\otimes n}}=1$. 
In contrast, the other $d^2-d$ (off-diagonal) Pauli strings are suppressed.

Matrix elements of the (exponentially-smaller) set  of $\hat I, \hat Z$ operators
acting on 24 qubits that are required to be computed,
can all be evaluated exactly using classical computing.
However, in working with larger systems, or using configurations beyond the naive shell model,
an amplitude cut-off, $\lambda$,  could be imposed to retain 
the components of a wavefunction larger than  $\lambda$~\footnote{
It is straightforward to show that the measures of magic converge with systematic cut-off errors
scaling as ${\cal O}(\lambda^2)$, which can be reduced below a given target precision with associated increasing classical computing resources.
However, it is more practical to evaluate the measures of magic over a range of 
$\lambda$, and perform an extrapolation to $\lambda=0$ using polynomial fits to the sets of 
results obtained for the $\{ \ \lambda_i\ \}$.}.

Contributions from the remaining $d^2-d$ operators that, by construction, 
contain $\hat X$ and $\hat Y$ operators, are evaluated using MCMC.
Their evaluation typically requires smaller computing resources than required for a complete MCMC
because the combinations of amplitudes contributing to 
the expectation values of these Pauli strings have a smaller range, 
and introduce smaller fluctuations in the MCMC evaluation.

To  summarize, the PSIZe-MCMC method involves:
\begin{enumerate}
    \item Evaluating all of the $c_P$, $\Xi_P$,  $\Xi_P^2$ and measures of magic
    explicitly for Pauli strings of $\hat I, \hat Z$
    operators only,
\begin{eqnarray}
{\cal P}^{(d)} & = &  {\rm Tr}_d\  \Xi_P 
\ ,\ 
{\rm Tr}_d\  \Xi_P^2 
\ ,\ 
{\cal M}^{(d)}_{\rm lin} 
\ ,\ 
{\cal M}^{(d)}_1 
\ ,\ 
{\cal M}^{(d)}_2 
\nonumber\ .
\end{eqnarray}
    \item Evaluating the remaining $d^2-d$ expectation values of Pauli strings and measures of magic 
    using the MCMC methods discussed above,

\begin{eqnarray}
\left[ {\rm Tr}\  \Xi_P ^2  \right]^{(PSIZe)}
& = &   
{\rm Tr}_d\  \Xi_P^2
\ +\ \left(1-{\cal P}^{(d)} \right) {\rm Tr}\  \left(\Xi_P^{(d^2-d)}\right)^2
\ ,\ 
\left[ {\cal M}_1 \right]^{(PSIZe)} \ =\    
{\cal M}_1^{(d)} 
\ +\ \left(1-{\cal P}^{(d)} \right) {\cal M}_1^{(d^2-d)}
\ .
\label{eq:M1_PSIZe-MCMC}
\end{eqnarray}

\end{enumerate}
This method can be extended to include explicit evaluations of all Pauli strings that are known to contribute to expectation values in a given wavefunction.
The uncertainty in the measures of magic in the 
nuclei that we have examined 
are substantially reduced in certain cases, e.g. $^{26}$Al.
In the case of $^{26}$Al, where an exact evaluation of the magic measures could not be performed, 
we found a reduction in statistical errors of $\approx 8$ for
${\cal M}_{\rm lin}$ and ${\cal M}_2$, and a factor of $\approx 2$ for 
${\cal M}_1$, using PSIZe-MCMC compared with a full MCMC for comparable classical computing resources.
The PSIZe-MCMC is also largely advantageous for computing magic in nuclei with reduced collectivity and one component dominating their wavefunction. In that case, the $\hat{I}\hat{Z}$ contribution ${\mathcal P}^{(d)} = {\rm Tr}_d\  \Xi_P $ can be above $30 \% $, which largely reduces the error on the magic measure, according to e.g. Eq.~\eqref{eq:M1_PSIZe-MCMC}.

%%%%%%%%%%%%%%%%%%%%%%%%%%%%%%%%%%%%%%%%%%%%%%%%
\subsection{Toy Examples: The PSIZe-MCMC Method for a Two-Qubit Wavefunction}
\label{appendix:psize2Q}
\noindent
To provide a simple example of the PSIZe-MCMC method, we now consider a hierarchical two-qubit wavefunction of the form
\begin{eqnarray}
    |\psi\rangle & = & 
    {1\over\sqrt{442}}
    \left[\ 
    20 |00\rangle + 5|01\rangle + 4 |10\rangle + 1 |11\rangle
    \ \right]
    \ ,
\end{eqnarray}
which gives rise to a set of $c_P$ with values,
separated into those from the $d=4$ 
$\hat I , \hat Z$ operators and those from the remaining 
$d^2-d=12$ operators,
\begin{eqnarray}
c_P^{(d)}
& = & \{ 1 , {15\over 17} , {12\over 13}  , {180\over 221} \ \}
\ =\ 
\{ 1 , 0.8824 , 0.9231  , 0.8145 \ \}
\nonumber\\
\overline{c}_P^{(d^2-d)} 
& = & \{  {40\over 221} , {75\over 221}  , {5\over 13}  , {96\over 221}  , {8\over 17} \ \}
\ =\ \{  0.1810 , 0.3394  , 0.3846  , 0.4344  , 0.4706 \ \}
\nonumber\ ,
\end{eqnarray}
where only the non-zero matrix elements are shown.
These give
\begin{eqnarray}
{\rm Tr}\  \Xi_P & = &  1.0
\ ,\ 
{\rm Tr}\  \Xi_P^2 \ =\  0.1808
\ ,\ 
{\cal M}_{\rm lin} \ =\  0.2767
\ ,\ 
{\cal M}_1 \ =\ 0.6837
\ ,\ 
{\cal M}_2 \ =\ 0.4674
\nonumber\ ,
\end{eqnarray}
while the contributions from 
$c_P^{(d)}$ alone are 
\begin{eqnarray}
{\cal P}^{(d)} & = &  {\rm Tr}_d\  \Xi_P \ =\   0.8235
\ ,\ 
{\rm Tr}_d\  \Xi_P^2 \ =\  0.1733
\ ,\ 
{\cal M}^{(d)}_{\rm lin} \ =\  0.3069
\ ,\ 
{\cal M}^{(d)}_1 \ =\ 0.2177
\ ,\ 
{\cal M}^{(d)}_2 \ =\ 0.5290
\nonumber\ .
\end{eqnarray}
A full MCMC evaluation of these quantities using 
$10^3$ samples, skipping  $10^3$ steps between samples,
gives
\begin{eqnarray}
{\cal M}_{\rm lin}^{(MC)} & = & 0.2789(66)
\ ,\ 
{\cal M}_1^{(MC)} \ =\ 0.690(30)
\ ,\ 
\\
{\cal M}_2^{(MC)} & = &  0.472(13)
\ ,\ 
{\rm Tr}\  \left(\Xi_P^{(MC)}\right)^2 \ =\  0.1796(18)
\nonumber\ ,
\end{eqnarray}
while the PSIZe-MCMC evaluation, with the same parameters as the full MCMC, gives
\begin{eqnarray}
\left[{\cal M}_{\rm lin}\right]^{(PSIZe)} & = & 0.27659(34)
\ ,\ 
\left[ {\cal M}_1 \right]^{(PSIZe)} \ =\    
{\cal M}_1^{(d)} 
\ +\ \left(1-{\cal P}^{(d)}\right) {\cal M}_1^{(d^2-d)}
\ =\  0.6825(43)
\ ,\ 
\nonumber\\
\left[  {\cal M}_2 \right]^{(PSIZe)} & = &  0.46712(68)
\ ,\ 
\left[ {\rm Tr}\  \Xi_P ^2  \right]^{(PSIZe)}
\ =\  
{\rm Tr}_d\  \Xi_P^2
\ +\ \left(1-{\cal P}^{(d)}\right) {\rm Tr}\  \left(\Xi_P^{(d^2-d)}\right)^2
\ =\   0.18085(8)
\ .
\end{eqnarray}
The uncertainties in the measures of magic are substantially smaller using the PSIZe-MCMC
than using the full MCMC evaluation.
One should keep in mind that this is a contrived  example.

These results should be contrasted with the situation where the wavefunction is without a significant hierarchy, including the case of a large degree of entanglement, for which PSIZe-MCMC is not expected to furnish a noticeable advantage.
For a wavefunction of the form, 
for example,
\begin{eqnarray}
    |\psi\rangle & = & 
    0.4327 |00\rangle + 0.4760|01\rangle + 0.5193 |10\rangle + 0.5626 |11\rangle
    \ ,
\end{eqnarray}
the sums and measures of magic are
\begin{eqnarray}
{\rm Tr}\  \Xi_P & = &  1.0
\ ,\ 
{\rm Tr}\  \Xi_P^2 \ =\  0.2409
\ ,\ 
{\cal M}_{\rm lin} \ =\  0.0361
\ ,\ 
{\cal M}_1 \ =\ 0.1283
\ ,\ 
{\cal M}_2 \ =\ 0.0531
\nonumber\ ,
\end{eqnarray}
while the contributions from the 
$\hat I , \hat Z$ operators  are 
\begin{eqnarray}
{\cal P}^{(d)} & = &  {\rm Tr}_d\  \Xi_P \ =\   0.2593
\ ,\ 
{\rm Tr}_d\  \Xi_P^2 \ =\  0.0626
\ ,\ 
{\cal M}^{(d)}_{\rm lin} \ =\  0.7498
\ ,\ 
{\cal M}^{(d)}_1 \ =\ 0.0510
\ ,\ 
{\cal M}^{(d)}_2 \ =\ 1.9987
\nonumber\ .
\end{eqnarray}
A full MCMC evaluation of these quantities with the same parameters as above,
gives
\begin{eqnarray}
{\cal M}_{\rm lin}^{(MC)} & = & 0.0.0367(30)
\ ,\ 
{\cal M}_1^{(MC)} \ =\ 0.130(18)
\ ,\ 
\\
{\cal M}_2^{(MC)} & = &  0.0539(45)
\ ,\ 
{\rm Tr}\  \left(\Xi_P^{(MC)}\right)^2 \ =\  0.24083(75)
\nonumber\ ,
\end{eqnarray}
while the PSIZe-MCMC evaluation gives
\begin{eqnarray}
\left[{\cal M}_{\rm lin}\right]^{(PSIZe)} & = & 0.0365(32)
\ ,\ 
\left[ {\cal M}_1 \right]^{(PSIZe)} \ =\     0.130(17)
\ ,\nonumber\\
\left[  {\cal M}_2 \right]^{(PSIZe)} & = &  0.0536(48)
\ ,\ 
\left[ {\rm Tr}\  \Xi_P ^2  \right]^{(PSIZe)}
\ =\   0.24088(80)
\ .
\end{eqnarray}
In this example, the precision of the estimators for the measures of magic are 
comparable between the two methods.

\section{Markov-Chain Monte Carlo Sampling procedure}
\label{appendix:magic}
\noindent
Statistical estimates of the values of 
the measures of magic, ${\cal M}_{\alpha}^{(d^2 -d)}$, 
are determined using the (well-known) Metropolis-Hastings (MH) algorithm~\cite{metropolis1949journ,b1878965-730b-33c7-b1ad-dfd3acb6f61b}. 
The MH algorithm is used to efficiently sample the expectation values of  the $d^2 -d$ Pauli strings  with $\hat X$ and/or $\hat Y$ in proportion to the probability distribution, $\Xi_{P}$. 
This allows ${\cal M}_{\alpha}^{(d^2 -d)}$, for different values of $\alpha$, to be determined from the same samples. 

By rewriting the sum in Eq.~(\ref{eq:Renyi_entropy_def1}),
an unbiased estimator for ${\cal M}_{\alpha}^{(d^2 -d)}$ can be found  
(below we omit the subscript $(d^2-d)$):
\begin{eqnarray}
    \sum_{\hat{P} \in \widetilde{\mathcal{G}}_{n_Q}}  \Xi_P^{\alpha} & = & 
    \sum_{\hat{P} \in \widetilde{\mathcal{G}}_{n_Q}}  \frac{\expval{P}{\Psi}^{2(\alpha-1)}}{d^{(\alpha-1)}}\frac{\expval{P}{\Psi}^2}{d} \  \approx \ 
    \left\langle\frac{\expval{P}{\Psi}^{2(\alpha-1)}}{d^{(\alpha-1)}}\right\rangle_{\Xi_P} 
    \ =\  \mathbb{E}(\alpha)
\nonumber\\
    {\cal M}_{\alpha} & \approx & \frac{1}{1-\alpha}\log_2 \mathbb{E}(\alpha)-\log_2 d  
    \ .
\end{eqnarray}
For ${\cal M}_2$, this corresponds to
\begin{eqnarray}
    {\cal M}_2 &  \approx & -\log_2 \left( d\  \mathbb{E}(2) \right)
    \ =\    -\log_2 \left\langle\expval{P}{\Psi}^2\right\rangle_{\Xi_P}
    \ ,
\end{eqnarray}
and for ${\cal M}_1$, the sum in Eq.~(\ref{eq:Renyi_entropy_defM1}) can be rewritten as 
\begin{eqnarray}
    \sum_{\hat{P} \in \widetilde{\mathcal{G}}_{n_Q}}  \Xi_P\log_2 \Xi_P = \left\langle \log_2\Xi_P \right\rangle_{\Xi_P}
    \ \ \ {\rm and}\ \ \ 
    {\cal M}_1 \ \approx \ -\left\langle\log_2 \expval{P}{\Psi}^2 \right\rangle_{\Xi_P} 
    \ .
\end{eqnarray}
Finally, for ${\cal M}_{\rm lin}$, 
given in Eq.~(\ref{eq:Renyi_entropy_def3}), the summation becomes
\begin{eqnarray}
    {\cal M}_{lin} & \approx 1\ -\ d\  \mathbb{E}(2) 
    \ =\  1- \left\langle \expval{P}{\Psi}^2 \right\rangle_{\Xi_P} 
    \ \ .
\end{eqnarray}

Due to the high dimensionality of the sampling space, an implicit component of our MH algorithm are the 
{\it physical conditions}, which we define to be strings that can lead to states with correct global properties. For instance, a selected string, when acting on the wavefunction is required to preserve baryon number (the number of $\hat X$ and $\hat Y$ operators must be even in this mapping), isospin projection (the number of $\hat X$ and $\hat Y$ operators in the neutron and proton sectors must each be even in this mapping), to yield a real wavefunction.  These conditions are detailed below.

Defining $n_X^{(prot)}$ to be the number of $\hat X$-gates acting on proton qubits in a Pauli string,
and $n_Y^{(prot)}$ to be the corresponding number of $\hat Y$-gates, and similarly 
$n_X^{(neut)}$ and $n_Y^{(neut)}$ for the neutron qubits,
the Pauli string is selected if it satisfies:
\begin{itemize}
    \item $n_X=n_X^{(prot)}+n_X^{(neut)}$ even;
    \item $n_Y=n_Y^{(prot)}+n_Y^{(neut)}$ even;
    \item $n_X^{(prot)}+n_Y^{(prot)} $ even;
    \item $n_X^{(neut)}+n_Y^{(neut)} $ even;
    \item $n_X^{(neut)}+n_Y^{(neut)} $ $\leq 2 \min(N_{\rm val}^{(neut)},n_Q^{(neut)}-N_{\rm val}^{(neut)})$;
    \item $n_X^{(prot)}+n_Y^{(prot)} $ $\leq 2 \min(N_{\rm val}^{(prot)},n_Q^{(prot)}-N_{\rm val}^{(prot)})$;
%    \item $n_X+n_Y$  $\leq 2\min(N_p,N_q-N_p)$,
\end{itemize}
where 
$n_Q^{(neut)}$ and $n_Q^{(prot)}$ are the number of qubits in the register of neutrons and protons respectively,
$N_{\rm val}^{(neut)}$ and $N_{\rm val}^{(prot)}$ are the number of active neutrons and protons,
and 
$n_Q^{(neut)}-N_{\rm val}^{(neut)}$ and 
$n_Q^{(prot)}-N_{\rm val}^{(prot)}$ are the number of holes in the neutron and proton active spaces.

%%%%%%%%%%%%%%%%%%%%%%%%%%
\subsection{Error analysis of MC}
\label{app:MCMCerrors}
\noindent
Our results are typically determined from 128 independent MCMC chains of length $5\times 10^3$ steps, 
which have been {\it thinned} by a further $5\times 10^3$ intervening steps.
The thinning reduces  autocorrelation lengths of the samples and improves calculated statistics.
Therefore, our chain lengths (number of steps) 
of $5\times10^{3}$ are in effect of lengths  $2.5\times10^{7}$.
In order to further mitigate the effects of autocorrelations, 
these chains are branched from the average of 128 thermalized \emph{burn-in} chains of $5\times 10^3$ steps with the same thinning.

Standard statistical diagnostic techniques are applied to quantify the effectiveness of the MH sampler.
Most notably we compute the following statistics: 
the Monte-Carlo Standard Error (MCSE) to quantify an error bound on $\mathcal{M}_{\alpha}$ estimates; 
the potential scale reduction factor, $\widehat{R}$, to quantify the degree of thermalization of the Markov chains by comparing the intra- and inter-chain variances ~\cite{vehtari2021rank}; 
the effective sample size (ESS) of the bulk and tail of the posterior distribution to quantify the effects of autocorrelation amongst our samples; 
and the acceptance rate of the MH-sampler, $r$. 
We utilize the \texttt{ArviZ}~\cite{ARVIZ} package to extract these statistical quantities~\cite{Kumar2019}.
Following Ref.~\cite{vehtari2021rank}, the `ranked' equivalent of the $\widehat{R}$ statistic used to quantify the mixing of the MCMC samples. This is chosen to overcome the shortcomings of the original $\widehat{R}$ statistic, which can fail to detect convergence issues in the event of a posterior distribution having, amongst other scenarios, heavy tails~\cite{vehtari2021rank}. Furthermore, the ranked-$\widehat{R}$ holds a tighter threshold of 1.01, as opposed to 1.1 for traditional $\widehat{R}$ values. 
By computing the ESS for both the bulk \emph{and} the tail of the posterior distribution, 
the effects of autocorrelation in the Markov process on the estimates of $\mathcal{M}_{\alpha}$
across the entirety of the posterior distribution can be quantified\footnote{For reference, the `tail' of the posterior is defined at the 5\% and 95\% quantile values with the $ESS$-$tail$ defined as the minimum of the $ESS$ at both quantile values, i.e.~$ESS$-$tail$ $\coloneqq \min$($ESS$-5\%, $ESS$-95\%).}.
Finally, the MCSE is used to quantify the error of the Monte Carlo estimator.

The mean values of the estimators of the measures of magic result from averaging the last values of the accumulated means of each of the 128 chains.
The uncertainties in these mean values are determined by bootstrap resampling the accumulated means, and forming the standard deviation of the bootstrap resamples. 
These results are given in the subsequent tables.

%%%%%%%%%%%%%%%%%%%%%%%%%%%%%%%%%%%%%%%%%%%%
\subsection{Representative MC chains}
\label{app:chains}
\begin{figure}[!ht]
\includegraphics[width=.49\textwidth]{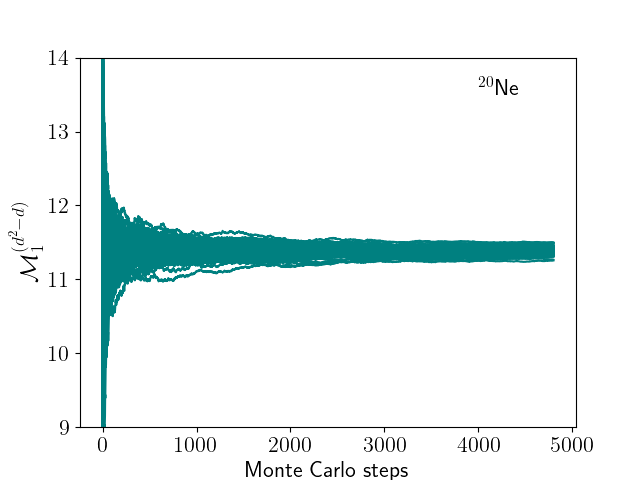}
\includegraphics[width=.49\textwidth]{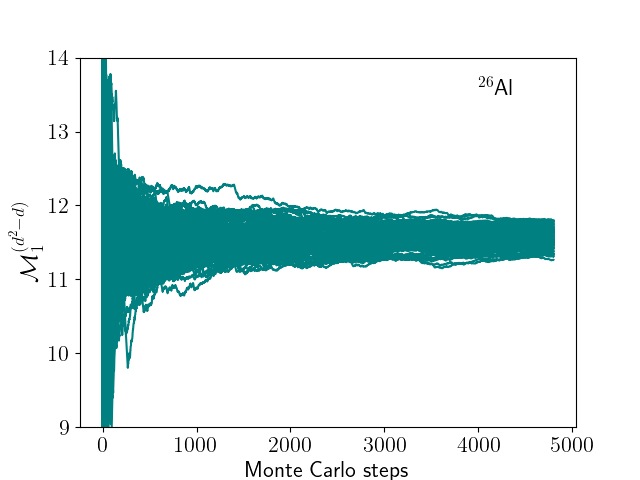}
\caption{
Representative PSIZe-MCMC chains used to compute the magic measure $\mathcal{M}_1$ in $^{20}$Ne (left panels) and $^{26}$Al (right panels). The average of these chains provide the partial value $\mathcal{M}_1^{(d^2 - d)}$ obtained from sampling strings with $X$ and/or $Y$ Pauli matrices.}
\label{fig:chains}
\end{figure}

\noindent
The structure of the nuclear wavefunction determines the behavior of the MCMC sampling, 
and the extracted uncertainties in the estimates of the measures of magic,
for given classical computing resources.  Representative results obtained from  the MCMC part of 
our PSIZe-MCMC evaluations are shown in Fig.~\ref{fig:chains}. 

\begin{figure}[!ht]
\includegraphics[width=.49\textwidth]{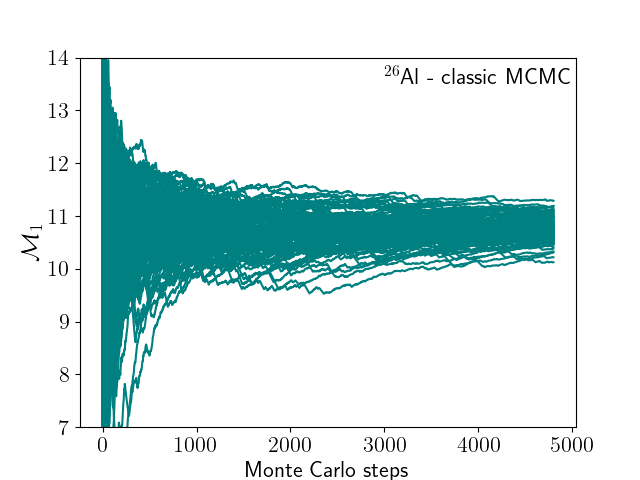}
\includegraphics[width=.49\textwidth]{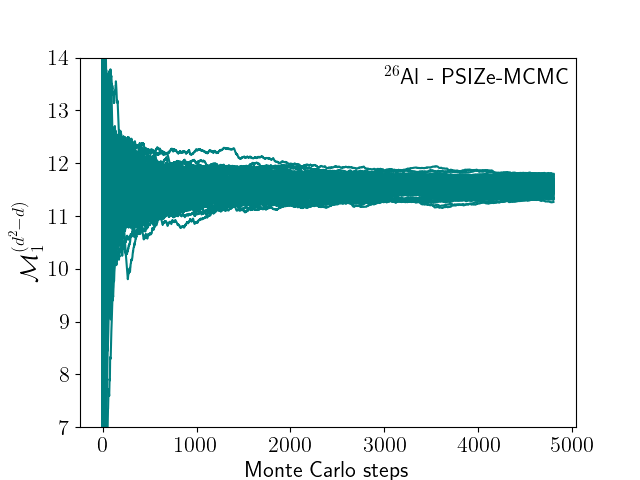}
\caption{Comparison of the chains obtained with the "classic" MCMC algorithm, i.e. sampling over all Pauli strings (left panel) and those obtained with the PSIZe-MCMC algorithm (right panel), for the case of $^{26}$Al.}
\label{fig:26Al_comp_chains}
\end{figure}

In order to illustrate the utility of the PSIZe-MCMC method, Fig.~\ref{fig:26Al_comp_chains} shows a same-scale comparison of the chains obtained with the "classic" MCMC algorithm (sampling over all Pauli strings) and those obtained with the PSIZe-MCMC algorithm, for the case of $^{26}$Al, using the same number of MC steps. 
In this nucleus, the contribution of $\hat I \hat Z$ strings to the total probability distribution is of the order of $10\%$ (${\cal P}^{(d)} = {\rm Tr}_d\  \Xi_P \ \approx 0.095$), representative of the degree of collectivity. 
It is clear that the chains 
thermalize faster using PSIZe-MCMC.
The time to run the algorithm is also
approximately 
$2.6$ times faster (for the same number of MC steps) and the error on the magic reduced by up to an order of magnitude, as shown in table~\ref{tab:26Al_comp_MCMC}. 

\begin{table}[!h]
    \centering
        \caption{Comparison of the magic values and errors obtained using a classic MCMC algorithm and the PSIZe-MCMC algorithm for the case of $^{26}$Al.}
    \begin{tabular}{|c|c|c|c|c|}
    \hline
      method & ${\cal M}_{\rm lin}$ &${\cal M}_1$ & ${\cal M}_2$   & $\widehat{R}$ \\
      \hline
      classic MCMC  & 0.9860(3) & 10.754(18) & 6.157(30) & 1.01 \\  
      PSIZe-MCMC & 0.9861(0) & 10.7673(93) & 6.1711(32) & 1.00 \\
      \hline
    \end{tabular}
    \label{tab:26Al_comp_MCMC}
\end{table}

%%%%%%%%%%%%%%%%%%%%%%%%%%%%%%%%%%%%%%%%%%%%%%%%
\section{Tables of Results}
\label{appendix:resulttables}
\noindent
In this appendix, we give the numerical results for quantities shown in the figures in the main text.

%%%%%%%%%%%%%%
\subsection{The magic in p-shell nuclei}
\label{appendix:pshellmagic}
\noindent
The results 
for $p$-shell nuclei
displayed in Fig.~\ref{fig:M12chart} in the main text are given in the tables in this section.
The numerical values for Helium, Lithium, Beryllium and Boron are given in Table~\ref{tab:magicresultsHeLiBeB}.

\begin{table}[!h]
\caption{
Measures of magic in the $J_z=J$ ground states of the $p$-shell nuclei
Helium, Lithium, Beryllium and Boron.
The implicit isospin symmetry of the Hamiltonian implemented in {\tt BIGSTICK} 
gives rise to exact relations between the measures of magic among different nuclei,
as indicated in the entry.
The second to last column indicates the technique used in the evaluation: ``Exact'' indicates an exact computation, while "PSIZe-MCMC" indicates evaluation using the PSIZe-MCMC algorithm, in which case the last column indicates the $\widehat{R}$ value.
}
\renewcommand{\arraystretch}{1.4}
\begin{tabularx}{\textwidth}{|Y| Y | Y || Y || Y| Y| Y||  Y| }
  \hline
 Z & N & A & J & ${\cal M}_{\rm lin}$ &${\cal M}_1$ & ${\cal M}_2$ & Method \\
 \hline\hline
\multicolumn{3}{|c||}{Helium (He)}  \\
\hline
\hline
2 & 2 & 4 & 0  &  0 &  0 & 0  &  Exact \\
  & 2+1 & 5 & $3/2$  & 0  & 0  &  0 &  Exact \\
  & 2+2 & 6 & 0  & 0.4439  & 1.0422  & 0.8465  &  Exact \\
  & 2+3 & 7 & $3/2$  & 0.2937  & 0.7006  & 0.5016  &  Exact \\
  & 2+4 & 8 & 0  & 0.4441  & 1.0430  & 0.8472  &  Exact \\
  & 2+5 & 9 & $1/2$  & 0  & 0  & 0  &  Exact \\
  & 2+6 & 10 &  0 &  0 &  0 &  0 &  Exact \\
 \hline\hline
 \multicolumn{3}{|c||}{Lithium (Li)}  \\
\hline
\hline
2+1 & 2 & 5 & $3/2$  &  0 &  0 & 0  &  Exact \\
  & 2+1 & 6 & 1  & 0.8333  & 3.1400  & 2.5846  &  Exact \\
  & 2+2 & 7 & $3/2$  & 0.8741  & 3.8837  & 2.9893  &  Exact \\
  & 2+3 & 8 & 2  & 0.8154  & 3.3972  & 2.4375  &  Exact \\
  & 2+4 & 9 & $3/2$  & 0.6830  & 2.4551  & 1.6573  &  Exact \\
  & 2+5 & 10 & 1  & 0.4815   &  1.5728  & 0.9476   &  Exact \\
  & 2+6 & 11 & $3/2$  &  0 &  0 &  0 &  Exact \\
 \hline\hline
 \multicolumn{3}{|c||}{Berylium (Be)}  \\
\hline
\hline
2+2 & 2 & 6 & 0  &  0.4439  &  1.0422  &  0.8465  &  Exact \\
  & 2+1 & 7 & $3/2$  &  0.8741  & 3.8837  & 2.9893  &  Exact \\
  & 2+2 & 8 & 0  &  0.9490 & 5.6194  & 4.2940  &  Exact \\
  & 2+3 & 9 &  $3/2$ & 0.9161  & 4.8221  & 3.5748  &  Exact \\
  & 2+4 & 10 & 0  & 0.8153  & 3.7548  & 2.4365  &  Exact \\
  & 2+5 & 11 & $1/2$  &  0.5558  &  2.0018  &  1.1706  &  Exact \\
  & 2+6 & 12 &  0 &  0.3306  &  0.7923  &  0.5790  &  Exact \\
 \hline\hline
 \multicolumn{3}{|c||}{Boron (B)}  \\
\hline
\hline
2+3 & 2 & 7 & $3/2$  & 0.2937  & 0.7006  & 0.5016 &  Exact \\
  & 2+1 & 8 & 2  & 0.8154  & 3.3972  & 2.4375 &  Exact \\
  & 2+2 & 9 &  $3/2$ & 0.9161  & 4.8221  & 3.5748 &  Exact \\
  & 2+3 & 10 & 3  &  0.8047 & 3.2040  & 2.3562  &  Exact \\
  & 2+4 & 11 &  $3/2$ & 0.8580  &  4.1249 &  2.8164 &  Exact \\
  & 2+5 & 12 &  1 &  0.7708  &  3.2887  &  2.1253  &  Exact \\
  & 2+6 & 13 & $3/2$  &  0.2791  &  0.7023  &  0.4721  &  Exact \\
 \hline\hline
\end{tabularx}
 \label{tab:magicresultsHeLiBeB}
\end{table}
The numerical values for Carbon, Nitrogen and Oxygen are given in Table~\ref{tab:magicresultsCNO}.
\begin{table}[!t]
\caption{
Measures of magic in the $J_z=J$ ground states of the $p$-shell nuclei
Carbon, Nitrogen and Oxygen.
The implicit isospin symmetry of the Hamiltonian implemented in {\tt BIGSTICK} 
gives rise to exact relations between the measures of magic among different nuclei,
as indicated in the entry.
The second to last column indicates the technique used in the evaluation: ``Exact'' indicates an exact computation, while "PSIZe-MCMC" indicates evaluation using the PSIZe-MCMC algorithm, in which case the last column indicates the $\widehat{R}$ value.
}
\renewcommand{\arraystretch}{1.4}
\begin{tabularx}{\textwidth}{|Y| Y | Y || Y || Y| Y| Y||  Y| }
  \hline
 Z & N & A & J & ${\cal M}_{\rm lin}$ &${\cal M}_1$ & ${\cal M}_2$ & Method \\
 \hline\hline
\multicolumn{3}{|c||}{Carbon (C)}  \\
\hline
\hline
2+4 & 2 & 8 &  0  & 0.4441  & 1.0430  & 0.8472  &  Exact \\
  & 2+1 & 9 & $3/2$  & 0.6830  & 2.4551  & 1.6573  &  Exact \\
  & 2+2 & 10 & 0  & 0.8153  & 3.7548  & 2.4365  &  Exact \\
  & 2+3 & 11 & $3/2$ & 0.8580  &  4.1249 &  2.8164 &  Exact \\
  & 2+4 & 12 &  0 & 0.8982  &  4.7009 &  3.2957 &  Exact \\
  & 2+5 & 13 &  $1/2$ & 0.6079  & 2.0995  & 1.3507  &  Exact \\
  & 2+6 & 14 & 0  &  0.3523 & 0.8364  &  0.6267 &  Exact \\
 \hline\hline
 \multicolumn{3}{|c||}{Nitrogen (N)}  \\
\hline
\hline
2+5 & 2 & 9 & $1/2$  &  0 &  0 & 0  &  Exact \\
  & 2+1 & 10 &  1  & 0.4815   &  1.5728  & 0.9476 &  Exact \\
  & 2+2 & 11 & $1/2$  &  0.5558  &  2.0018  &  1.1706 &  Exact \\
  & 2+3 & 12 & 1 &  0.7708  &  3.2887  &  2.1253 &  Exact \\
  & 2+4 & 13 & $1/2$ & 0.6079  & 2.0995  & 1.3507 &  Exact \\
  & 2+5 & 14 & 0  &  0.2298  &  0.6863  &  0.3767  &  Exact \\
  & 2+6 & 15 & $1/2$ &  0 &  0 &  0 &  Exact \\
 \hline\hline
 \multicolumn{3}{|c||}{Oxygen (O)}  \\
\hline
\hline
2+6 & 2 & 10 &  0 &  0  &  0  &  0  &  Exact \\
  & 2+1 & 11 &  $1/2$ & 0  & 0  & 0  &  Exact \\
  & 2+2 & 12 & 0 &  0.3306  &  0.7922  &  0.5790  &  Exact \\
  & 2+3 & 13 & $3/2$  &  0.2791  &  0.7023  &  0.4721 &  Exact \\
  & 2+4 & 14 & 0  &  0.3523 & 0.8364  &  0.6267 &  Exact \\
  & 2+5 & 15 &  $1/2$ &  0  &  0  &  0  &  Exact \\
  & 2+6 & 16 & 0  &  0  &  0  &  0  &  Exact \\
 \hline\hline
\end{tabularx}
 \label{tab:magicresultsCNO}
\end{table}
%

%%%%%%%%%%%%%%
\subsection{The magic in sd-shell nuclei}
\label{appendix:sdshellmagic}
\noindent

The results 
for $sd$-shell nuclei
displayed in Fig.~\ref{fig:M12chart} in the main text are given in the tables in this section.
The numerical values for 
Oxygen, Fluorine and Neon are given in Table~\ref{tab:magicresultsOFNe}.
\begin{table}[!t]
\caption{
Measures of magic in the $J_z=J$ ground states of the $sd$-shell nuclei
Oxygen, Fluorine and Neon.
The implicit isospin symmetry of the Hamiltonian implemented in {\tt BIGSTICK} 
gives rise to exact relations between the measures of magic among different nuclei,
as indicated in the entry.
The second to last column indicates the technique used in the evaluation: ``Exact'' indicates an exact computation, while "MC" indicates evaluation using the PSIZe-MCMC algorithm, in which case the last column indicates the $\widehat{R}$ value.
}
\renewcommand{\arraystretch}{1.2}
\begin{tabularx}{\textwidth}{|Y| Y | Y || Y|| Y| Y| Y||  Y| Y| }
  \hline
 Z & N & A & J & ${\cal M}_{\rm lin}$ &${\cal M}_1$ & ${\cal M}_2$ & Method & $\widehat{R}$ \\
 \hline\hline
\multicolumn{3}{|c||}{Oxygen (O)}  \\
\hline
\hline
8 & 8 & 16 & 0 & 0 & 0 & 0 &  Exact &\\
  & 8+1 & 17 & ${5/2}$ & 0 & 0 & 0 & Exact &\\
  & 8+2 & 18 & 0 & 0.7077 & 2.2185 & 1.7744 & Exact &\\
  & 8+3 & 19 & ${5/2}$ & 0.4887 & 1.7752 & 0.9677 & Exact &\\
  & 8+4 & 20 & 0 & 0.7394 & 2.9584 & 1.9404 & Exact &\\
  & 8+5 & 21 & ${5/2}$ &0.5676 & 2.2996 & 1.2096 & Exact &\\
  & 8+6 & 22 & 0 & 0.6276 & 2.5758 & 1.4250 & Exact &\\
  & 8+7 & 23 & ${1/2}$ & 0.3851 & 1.6762 & 0.7016 & Exact &\\
  & 8+8 & 24 & 0 & 0.3046 & 1.1691 & 0.5240 & Exact &\\
  & 8+9 & 25 & ${3/2}$ & 0.1818 & 0.7135 & 0.2895 & Exact &\\
  & 8+10 & 26 & 0  & 0.1929 & 0.7031 & 0.3092 & Exact &\\
  & 8+11 & 27 & ${3/2}$ & 0  &  0 & 0 & Exact &\\
  & 8+12 & 28 & 0 &0 & 0 & 0 & Exact &\\
 \hline\hline
\multicolumn{3}{|c||}{Fluorine (F)}  \\
\hline
\hline
8+1 & 8 & 17 & ${5/2}$ & 0  &  0 & 0  &  Exact & ($^{17}$O) \\
 %& 8+1 & 18 & 1 &  0.93214(3) & 4.923(1) & 3.881(1)   & MC & 1.00       \\
 & 8+1 & 18 & 1 &  0.9452(2) & 4.970(8) & 4.191(5)   & MC & 1.00       \\
 & 8+2 & 19 & ${1/2}$ &  0.9786(0) & 7.4530(28) & 5.5452(14) & MC & 1.00   \\
 & 8+3 & 20 & 2 &  0.9849(0) & 8.2920(35) & 6.0524(22) & MC& 1.00 \\
 & 8+4 & 21 & ${5/2}$ &  0.9569(1) & 7.1205(36) & 4.5364(20) & MC& 1.00 \\
 & 8+5 & 22 &  4  &  0.8294(3) & 5.0629(46) & 2.5516(22) & MC& 1.00\\
 & 8+6 & 23 & ${5/2}$ &  0.8594(0) & 5.1226(28) & 2.8302(4) & MC& 1.00     \\
 & 8+7 & 24 & 3 & 0.6927(0) & 3.6049(18) & 1.7024(1) & MC& 1.00   \\
 & 8+8 & 25 & ${5/2}$ &  0.5802(0) & 2.7345(13) & 1.2521(0) & MC& 1.00   \\
 & 8+9 & 26 & 1 &  0.7150(3) & 3.4319(41) & 1.8110(13) & MC& 1.00 \\
 & 8+10 & 27 & ${5/2}$ & 0.5645(2) & 2.2058(21) & 1.1991(6) & MC & 1.00 \\
 %& 8+10 & 27 & ${5/2}$ & 0.589(2)& 2.45(2) & 1.282(9) & MC & 1.00    \\
 & 8+11 & 28 & 3  & 0.0718(3) & 0.268(4) & 0.1074(5) & MC & 1.00   \\
  %& 8+11 & 28 & 3  & 0.0727(3) & 0.245(2) & 0.1088(4) & MC & 1.00    \\
 & 8+12 & 29 & ${5/2}$ & 0 & 0 & 0 & Exact &\\
  \hline\hline
\multicolumn{3}{|c||}{Neon (Ne) }  \\
\hline
\hline
8+2 & 8 & 18 &  0 & 0.7077 & 2.2185 & 1.7744 & Exact & ($^{18}$O) \\
    & 8+1 & 19 & ${1/2}$ & 0.9786(0) & 7.4530(28) & 5.5452(14) & MC &   ($^{19}$F)  \\
    & 8+2 & 20 & 0  &  0.9970(0) & 11.3135(38) & 8.3792(34) & MC& 1.00  \\
    & 8+3 & 21 &  ${3/2}$ &  0.9979(0) & 11.9268(39) & 8.8808(49) & MC& 1.00     \\
    & 8+4 & 22 & 0  &  0.9956(0) & 11.7996(62) & 7.8280(86) & MC& 1.00 \\
    & 8+5 & 23 & ${5/2}$  &  0.9843(0) & 9.9095(74) & 5.9926(74) & MC& 1.00      \\
    & 8+6 & 24 & 0  & 0.9725(1) & 9.114(11) & 5.1821(73) & MC& 1.00     \\
    & 8+7 & 25 & ${1/2}$  &  0.9380(2) & 7.649(14) & 4.0110(57)& MC & 1.01     \\
    & 8+8 & 26 & 0  & 0.9179(3) & 6.609(12) & 3.6059(47) & MC& 1.01    \\
    & 8+9 & 27 & ${3/2}$  &  0.8873(3) & 5.8035(93) & 3.1494(37) & MC& 1.00       \\
    & 8+10 & 28 & 0 &  0.8496(4) & 4.9394(76) & 2.7328(35) & MC& 1.00    \\
    & 8+11 & 29 & ${3/2}$ &  0.7431(2) & 3.2535(31) & 1.9608(10) & MC& 1.00    \\
    & 8+12 & 30 & 0 & 0.5648 &  1.7100  &  1.2002   &  Exact   &\\
\hline
\hline
\end{tabularx}
 \label{tab:magicresultsOFNe}
\end{table}
The numerical values for 
Sodium, Magnesium and Aluminium are given in Table~\ref{tab:magicresultsNaMgAl}.
\begin{table}[!t]
\caption{
Measures of magic in the $J_z=J$ ground states of the $sd$-shell nuclei
Sodium, Magnesium and Aluminium.
The implicit isospin symmetry of the Hamiltonian implemented in {\tt BIGSTICK} 
gives rise to exact relations between the measures of magic among different nuclei,
as indicated in the entry.
The second to last column indicates the technique used in the evaluation: ``Exact'' indicates an exact computation, while "MC" indicates evaluation using the PSIZe-MCMC algorithm, in which case the last column indicates the $\widehat{R}$ value.
}
\renewcommand{\arraystretch}{1.4}
\begin{tabularx}{\textwidth}{|Y| Y | Y || Y|| Y| Y| Y||  Y| Y| }
  \hline
 Z & N & A & J & ${\cal M}_{\rm lin}$ &${\cal M}_1$ & ${\cal M}_2$ & Method & $\widehat{R}$\\
 \hline\hline
\multicolumn{3}{|c||}{Sodium (Na)}  \\
\hline
\hline
8+3 & 8 & 19 & ${5/2}$ & 0.4887 & 1.7752 & 0.9677 & Exact &\\
    & 8+1 & 20 & 2 & 0.9849(0) & 8.2920(35) & 6.0524(22) & MC & ($^{20}$F) \\  
    & 8+2 & 21 & ${3/2}$ &  0.9979(0) & 11.9268(39) & 8.8808(49) & MC& ($^{21}$Ne)     \\
    & 8+3 & 22 & 3  &  0.9963(0) & 11.6542(58) & 8.0867(99) & MC& 1.00 \\ 
    & 8+4 & 23 & ${3}/{2}$  &  0.9990(0) & 13.7406(48) & 10.0296(83) & MC& 1.00   \\ 
    & 8+5 & 24 & 4  &  0.9908(1) & 10.7501(2) & 6.7618(90) & MC& 1.00   \\ 
    & 8+6 & 25 & ${5}/{2}$  &  0.9718(4) & 10.006(19) & 5.147(19) & MC& 1.01  \\ 
    & 8+7 & 26 & 1  &  0.9372(7) & 8.469(25) & 3.995(15) & MC& 1.01 \\ 
    & 8+8 & 27 & ${3}/{2}$  &  0.9492(5) & 8.370(20) & 4.300(15) & MC& 1.01   \\ 
    & 8+9 & 28 &  2 &  0.9750(1) & 8.4581(89) & 5.3232(54) & MC& 1.00   \\ 
    & 8+10 & 29 & ${5}/{2}$  &  0.8408(7)  &     5.620(13)       & 2.6512(64) & MC& 1.00   \\ 
    & 8+11 & 30 & 2  &  0.8838(1) & 4.8886(35) & 3.1054(10)& MC& 1.00 \\
    & 8+12 & 31 & ${5}/{2}$ & 0.2395  & 1.0050  &  0.3949 & Exact &\\ 
    
 \hline\hline
\multicolumn{3}{|c||}{Magnesium (Mg)}  \\
\hline
\hline
8+4 & 8 & 20 & 0 & 0.7394 & 2.9584 & 1.9404 & Exact &\\
    & 8+1 & 21 & ${5/2}$ & 0.9569(1) & 7.1205(36) & 4.5364(20) & MC& ($^{21}$F) \\
    & 8+2 & 22 & 0  &  0.9956(0) & 11.7996(62) & 7.8280(86) & MC& ($^{22}$Ne)  \\
    & 8+3 & 23 & ${3}/{2}$  &  0.9990(0) & 13.7406(48) & 10.0296(83) & MC& ($^{23}$Na)   \\ 
    & 8+4 & 24 & 0  &  0.9994(0) & 15.0727(59) & 10.598(14)&  MC& 1.00    \\ 
    & 8+5 & 25 & ${5}/{2}$  &  0.9973(0) & 13.3024(77) & 8.5124(9)&  MC& 1.00    \\
    & 8+6 & 26 & 0  & 0.9926(1) & 12.411(16) & 7.085(16) & MC & 1.00 \\
    %& 8+6 & 26 & 0  & 0.9930(3) & 12.54(6) & 7.15(9) & MC & 1.01  \\ 
    & 8+7 & 27 & ${1}/{2}$  &  0.9856(1) & 11.152(18) & 6.121(13)&  MC& 1.01   \\ 
    & 8+8 & 28 & 0 &  0.9884(1) & 10.963(17) & 6.435(17)&  MC& 1.01   \\
    & 8+9 & 29 & ${1}/{2}$  &  0.9738(2) & 9.178(14) & 5.256(11)&  MC& 1.01  \\ 
    & 8+10 & 30 & 0 & 0.9386(4) & 7.309(14) & 4.0251(85) &  MC& 1.01  \\ 
    & 8+11 & 31 & ${3}/{2}$  & 0.8672(2) & 4.9779(48) & 2.9129(18) & MC& 1.00  \\ 
    & 8+12 & 32 & 0 & 0.6459  & 2.3703  &  1.4979 & Exact &\\ 
\hline
\hline
\multicolumn{3}{|c||}{Aluminium (Al)}  \\
\hline
\hline
8+5 & 8 & 21 & ${5}/{2}$ &0.5676 & 2.2996 & 1.2096 &  Exact &\\
    & 8+1 & 22 & 4  &  0.8294(3) & 5.0629(46) & 2.5516(22) & MC& ($^{22}$F)\\
    & 8+2 & 23 & ${5}/{2}$  &  0.9843(0) & 9.9095(74) & 5.9926(74) & MC& ($^{23}$Ne)      \\
    & 8+3 & 24 & 4  &  0.9908(1) & 10.7501(2) & 6.7618(90) & MC& ($^{24}$Na)    \\ 
    & 8+4 & 25 & ${5}/{2}$  &  0.9973(0) & 13.3024(77) & 8.5124(9)&  MC& ($^{25}$Mg)   \\
    & 8+5 & 26 & 5  & 0.9861(0) & 10.7673(93) & 6.1711(32) &  MC & 1.00  \\ 
    & 8+6 & 27 & ${5}/{2}$  & 0.9913(0) & 12.322(10) & 6.8376(32) & MC & 1.00 \\
    %& 8+6 & 27 & ${5}/{2}$  &  0.9912(1) & 12.35(3) & 6.82(7)  &  MC & 1.00  \\ 
    & 8+7 & 28 & 2    & 0.9743(0) & 10.1967(10) & 5.2814(17) & MC &1.00 \\
    %& 8+7 & 28 & 2    &  0.9750(1) & 10.54(4) & 5.32(5)   &  MC & 1.01  \\ 
    & 8+8 & 29 & ${5}/{2}$ & 0.9748(2) & 9.861(18) & 5.311(14) & MC & 1.01   \\ 
    & 8+9 & 30 & 3  & 0.9420(6) & 8.259(24) & 4.107(14)&  MC & 1.01  \\ 
    & 8+10 & 31 & ${5}/{2}$  & 0.9160(3) & 6.879(12) & 3.5741(56) &  MC& 1.01  \\ 
    & 8+11 & 32 & 2  &  0.8072(3) & 4.7921(74) & 2.3749(25) &  MC& 1.00   \\ 
    & 8+12 & 33 & ${5}/{2}$ &  0.3027 &  1.2690 & 0.5202 & Exact &\\ 
\hline
\hline
\end{tabularx}
 \label{tab:magicresultsNaMgAl}
\end{table}
The numerical values for 
Silicon, Phosphorus and Sulfur are given in Table~\ref{tab:magicresultsSiPS}.
\begin{table}[!t]
\caption{
Measures of magic in the $J_z=J$ ground states of the $sd$-shell nuclei Silicon, Phosphorus and Sulfur.
The implicit isospin symmetry of the Hamiltonian implemented in {\tt BIGSTICK} 
gives rise to exact relations between the measures of magic among different nuclei,
as indicated in the entry.
The second to last column indicates the technique used in the evaluation: ``Exact'' indicates an exact computation, while "MC" indicates evaluation using the PSIZe-MCMC algorithm, in which case the last column indicates the $\widehat{R}$ value.
}
\renewcommand{\arraystretch}{1.4}
\begin{tabularx}{\textwidth}{|Y| Y | Y || Y|| Y| Y| Y||  Y| Y| }
  \hline
 Z & N & A & J & ${\cal M}_{\rm lin}$ &${\cal M}_1$ & ${\cal M}_2$ & Method & $\widehat{R}$\\
 \hline\hline
\multicolumn{3}{|c||}{Silicon (Si)}  \\
\hline
\hline
8+6 & 8 & 22 & 0 & 0.6276 & 2.5758 & 1.4250 &  Exact &\\
    & 8+1 & 23 & ${5}/{2}$  &   0.8594(0) & 5.1226(28) & 2.8302(4) & MC& ($^{23}$F)     \\
    & 8+2 & 24 & 0  &  0.9725(1) & 9.114(11) & 5.1821(73) & MC& ($^{24}$Ne)      \\
    & 8+3 & 25 & ${5}/{2}$  &  0.9718(4) & 10.006(19) & 5.147(19) & MC& ($^{25}$Na)   \\
    & 8+4 & 26 &  0 &  0.9930(3) & 12.54(6) & 7.15(9) & MC & ($^{26}$Mg)   \\ 
    & 8+5 & 27 & ${5}/{2}$  &  0.9912(1) & 12.35(3) & 6.82(7)  &  MC & ($^{27}$Al)  \\ 
    & 8+6 & 28 & 0  & 0.9964(0) & 13.9613(82) & 8.1134(43) &  MC & 1.00 \\
    %& 8+6 & 28 & 0  & 0.9963(1) & 13.69(5) & 8.1(1) & MC & 1.01    \\ 
    & 8+7 & 29 & ${1}/{2}$  & 0.9842(0) & 11.599(10) & 5.9880(24) & MC & 1.00 \\
    %& 8+7 & 29 & ${1}/{2}$  & 0.9841(1)& 11.52(4)& 5.98(5)& MC & 1.01    \\ 
    & 8+8 & 30 & 0 & 0.9778(2) & 10.433(21) & 5.494(12) & MC & 1.01 \\
    %& 8+8 & 30 & 0 & 0.9763(7) & 10.21(8) & 5.40(8) & MC & 1.01    \\
    & 8+9 & 31 & ${3}/{2}$  & 0.9523(2) & 8.495(16) & 4.3891(56) & MC& 1.01 \\
    & 8+10 & 32 & 0  &  0.9319(4) & 7.558(17) & 3.8770(89) &  MC& 1.01   \\ 
    & 8+11 & 33 &  ${3}/{2}$ &  0.6134(0) & 3.1782(19) & 1.3712(1)&  MC& 1.00    \\ 
    & 8+12 & 34 & 0 &  0.2995 & 1.2799  & 0.5136 & Exact &\\
 \hline\hline
\multicolumn{3}{|c||}{Phosphorus (P)}  \\
\hline
\hline
8+7 & 8 & 23 & ${1/2}$ & 0.3851 & 1.6762 & 0.7016 &  Exact &\\
    & 8+1 & 24 &  3 &  0.6927(0) & 3.6049(18) & 1.7024(1) & MC& ($^{24}$F)   \\
    & 8+2 & 25 & ${1}/{2}$  &  0.9380(2) & 7.649(14) & 4.0110(57)& MC & ($^{25}$Ne)     \\
    & 8+3 & 26 & 1  &  0.9372(7) & 8.469(25) & 3.995(15) & MC& ($^{26}$Na) \\ 
    & 8+4 & 27 & ${1}/{2}$  &  0.9856(1) & 11.152(18) & 6.121(13)&  MC& ($^{27}$Mg)  \\
    & 8+5 & 28 & 2  &  0.9750(1) & 10.54(4) & 5.32(5)   &  MC & ($^{28}$Al)  \\  
    & 8+6 & 29 & ${1}/{2}$  &  0.9841(1)& 11.52(4)& 5.98(5)& MC & ($^{29}$Si)    \\ 
    & 8+7 & 30 & 1  & 0.9672(1) & 10.103(15) & 4.9299(43) & MC & 1.01 \\
    %& 8+7 & 30 & 1  & 0.9658(5) & 9.83(7) & 4.87(8) & MC & 1.01 \\
    & 8+8 & 31 & ${1}/{2}$  & 0.9719(1) & 10.076(17) & 5.1531(70)& MC& 1.01 \\
    & 8+9 & 32 & 1  &  0.9607(3) & 9.223(20) & 4.670(11)& MC& 1.01 \\
    & 8+10 & 33 & ${1}/{2}$  & 0.8994(7) & 7.157(21) & 3.313(10) & MC& 1.01  \\ 
    & 8+11 & 34 & 1  &  0.7157(4) & 3.8849(72) & 1.8147(23)& MC& 1.00   \\ 
    & 8+12 & 35 & ${1}/{2}$ &  0.2792 & 1.2195  & 0.4724 & Exact &\\
\hline
\hline
\multicolumn{3}{|c||}{Sulfur (S)}  \\
\hline
\hline
8+8 & 8 & 24 & 0 & 0.3046 & 1.1691 & 0.5240 &  Exact &\\
    & 8+1 & 25 & ${5}/{2}$  &  0.5802(0) & 2.7345(13) & 1.2521(0) & MC& ($^{25}$F)   \\
    & 8+2 & 26 & 0  &  0.9179(3) & 6.609(12) & 3.6059(47) & MC& ($^{26}$Ne)     \\
    & 8+3 & 27 & ${3}/{2}$  &  0.9492(5) & 8.370(20) & 4.300(15) & MC& ($^{27}$Na)    \\  
    & 8+4 & 28 & 0  &  0.9884(1) & 10.963(17) & 6.435(17)&  MC& ($^{28}$Mg)    \\
    & 8+5 & 29 & ${5}/{2}$  &  0.9748(2) & 9.861(18) & 5.311(14) & MC & ($^{29}$Al)   \\ 
    & 8+6 & 30 & 0  &  0.9763(7) & 10.21(8) & 5.40(8) & MC & ($^{30}$Si)     \\
    & 8+7 & 31 & ${1}/{2}$ &  0.9719(1) & 10.076(17) & 5.1531(70)& MC& ($^{31}$P) \\
    & 8+8 & 32 & 0  & 0.9820(1) & 10.572(15) & 5.7932(68) & MC& 1.00 \\
    & 8+9 & 33 & ${3}/{2}$  & 0.9328(1) & 7.885(11) & 3.8955(18)& MC& 1.01  \\ 
    & 8+10 & 34 & 0  & 0.9047(7) & 7.148(18) & 3.391(10)& MC& 1.01 \\
    & 8+11 & 35 &  ${3}/{2}$ & 0.5569(0) & 2.6453(18) & 1.1744(1) & MC& 1.00 \\
    & 8+12 & 36 & 0 & 0.3089  & 1.1254  & 0.5330 & Exact &\\
\hline
\hline
\end{tabularx}
 \label{tab:magicresultsSiPS}
\end{table}
The numerical values for Chlorine, Argon and Potassium are given in Table~\ref{tab:magicresultsClArK}.
\begin{table}[!t]
\caption{
Measures of magic in the $J_z=J$ ground states of the $sd$-shell nuclei
Chlorine, Argon and Potassium.
The implicit isospin symmetry of the Hamiltonian implemented in {\tt BIGSTICK} 
gives rise to exact relations between the measures of magic among different nuclei,
as indicated in the entry.
The second to last column indicates the technique used in the evaluation: ``Exact'' indicates an exact computation, while "MC" indicates evaluation using the PSIZe-MCMC algorithm, in which case the last column indicates the $\widehat{R}$ value.
}
\renewcommand{\arraystretch}{1.4}
\begin{tabularx}{\textwidth}{|Y| Y | Y || Y|| Y| Y| Y||  Y| Y| }
  \hline
 Z & N & A & J & ${\cal M}_{\rm lin}$ &${\cal M}_1$ & ${\cal M}_2$ & Method & $\widehat{R}$\\
 \hline\hline
\multicolumn{3}{|c||}{Chlorine (Cl)}  \\
\hline
\hline
8+9 & 8 & 25 & ${3/2}$ & 0.1818 & 0.7135 & 0.2895 &  Exact &\\
    & 8+1 & 26 & 1  &  0.7150(3) & 3.4319(41) & 1.8110(13) & MC& ($^{26}$F) \\
    & 8+2 & 27 & ${3}/{2}$  &  0.8873(3) & 5.8035(93) & 3.1494(37) & MC& ($^{27}$Ne)      \\
    & 8+3 & 28 & 2  &  0.9750(1) & 8.4581(89) & 5.3232(54) & MC& ($^{28}$Na)   \\ 
    & 8+4 & 29 & ${1}/{2}$  &  0.9738(2) & 9.178(14) & 5.256(11)&  MC& ($^{29}$Mg)  \\ 
    & 8+5 & 30 & 3  &  0.9420(6) & 8.259(24) & 4.107(14)&  MC & ($^{30}$Al)  \\ 
    & 8+6 & 31 & ${3}/{2}$  &  0.9523(2) & 8.495(16) & 4.3891(56) & MC& ($^{31}$Si)\\
    & 8+7 & 32 & 1  & 0.9607(3) & 9.223(20) & 4.670(11)& MC& ($^{32}$P) \\
    & 8+8 & 33 & ${3}/{2}$  &  0.9328(1) & 7.885(11) & 3.8955(18)& MC& ($^{33}$S)  \\ 
    & 8+9 & 34 & 1  & 0.8792(0) & 6.2101(60) & 3.0492(7) & MC& 1.00   \\ 
    & 8+10 & 35 & ${3}/{2}$  &  0.9370(2) & 7.5731(88) & 3.9883(47) & MC& 1.01    \\ 
    & 8+11 & 36 & 2  & 0.4716(5) & 2.4962(52) & 0.9204(13) & MC& 1.00    \\ 
    & 8+12 & 37 & ${3}/{2}$ & 0.1792  & 0.6842  & 0.2845  & Exact &\\
 \hline\hline
\multicolumn{3}{|c||}{Argon (Ar)}  \\
\hline
\hline
8+10 & 8 & 26 & 0  & 0.1929 & 0.7031 & 0.3092 &  Exact &\\
    & 8+1 & 27 & ${5}/{2}$  & 0.589(2)& 2.45(2) & 1.282(9) & MC & ($^{27}$F)    \\
    & 8+2 & 28 & 0  &  0.8496(4) & 4.9394(76) & 2.7328(35) & MC& ($^{28}$Ne)    \\
    & 8+3 & 29 &  ${5}/{2}$ &  0.8408(7)  &     5.620(13)       & 2.6512(64) & MC& ($^{29}$Na)   \\ 
    & 8+4 & 30 & 0  &  0.9386(4) & 7.309(14) & 4.0251(85) &  MC& ($^{30}$Mg)   \\ 
    & 8+5 & 31 & ${5}/{2}$  &  0.9160(3) & 6.879(12) & 3.5741(56) &  MC& ($^{31}$Al)  \\ 
    & 8+6 & 32 & 0  &  0.9319(4) & 7.558(17) & 3.8770(89) &  MC& ($^{32}$Si)   \\ 
    & 8+7 & 33 & ${1}/{2}$ & 0.8994(7) & 7.157(21) & 3.313(10) & MC& ($^{33}$P)  \\
    & 8+8 & 34 & 0  &  0.9047(7) & 7.148(18) & 3.391(10)& MC& ($^{34}$S)  \\
    & 8+9 & 35 & ${3}/{2}$  &  0.9370(2) & 7.5731(88) & 3.9883(47) & MC& ($^{35}$Cl)     \\ 
    & 8+10 & 36 & 0  & 0.9454(1) & 7.9601(59) & 4.1961(37)& MC & 1.00 \\
    & 8+11 & 37 &  ${3}/{2}$ & 0.7137(2) & 3.3795(30) & 1.8043(9) & MC& 1.00   \\ 
    & 8+12 & 38 & 0 &  0.1946 & 0.6895  &  0.3123 & Exact &\\
\hline
\hline
\multicolumn{3}{|c||}{Potassium (K)}  \\
\hline
\hline
8+11 & 8 & 27 & ${3}/{2}$  & 0  & 0  & 0  &  Exact &\\
    & 8+1 & 28 & 3  &  0.0727(3) & 0.245(2) & 0.1088(4) & MC & ($^{28}$F)    \\
    & 8+2 & 29 & ${3}/{2}$  &   0.7431(2) & 3.2535(31) & 1.9608(10) & MC& ($^{29}$Ne)    \\
    & 8+3 & 30 & 2  &  0.8838(1) & 4.8886(35) & 3.1054(10)& MC& ($^{30}$Na)  \\
    & 8+4 & 31 & ${3}/{2}$  &  0.8672(2) & 4.9779(48) & 2.9129(18) & MC& ($^{31}$Mg)   \\
    & 8+5 & 32 & 2  & 0.8072(3) & 4.7921(74) & 2.3749(25) &  MC& ($^{32}$Al)   \\ 
    & 8+6 & 33 & ${3}/{2}$  &  0.6134(0) & 3.1782(19) & 1.3712(1)&  MC& ($^{33}$Si)    \\
    & 8+7 & 34 & 1  &  0.7157(4) & 3.8849(72) & 1.8147(23)& MC& ($^{34}$P)    \\
    & 8+8 & 35 & ${3}/{2}$  &   0.5569(0) & 2.6453(18) & 1.1744(1) & MC& ($^{35}$S)  \\
    & 8+9 & 36 & 2  &  0.4716(5) & 2.4962(52) & 0.9204(13) & MC& ($^{36}$Cl)    \\ 
    & 8+10 & 37 & ${3}/{2}$  & 0.7137(2) & 3.3795(30) & 1.8043(9) & MC&  ($^{37}$Ar)   \\ 
    & 8+11 & 38 & 1  &  0.0545(0) & 0.2251(3) & 0.0809(0) & MC & 1.00   \\ 
    & 8+12 & 39 & ${3}/{2}$ & 0  & 0  &  0 & Exact &\\
\hline
\hline
\end{tabularx}
 \label{tab:magicresultsClArK}
\end{table}
Finally, the numerical values for 
Calcium are given in Table~\ref{tab:magicresultsCa}.
\begin{table}[!t]
\caption{
Measures of magic in the $J_z=J$ ground states of the $sd$-shell nuclei
Calcium.
The implicit isospin symmetry of the Hamiltonian implemented in {\tt BIGSTICK} 
gives rise to exact relations between the measures of magic among different nuclei,
as indicated in the entry.
The second to last column indicates the technique used in the evaluation: ``Exact'' indicates an exact computation, while "MC" indicates evaluation using the PSIZe-MCMC algorithm, in which case the last column indicates the $\widehat{R}$ value.
}
\renewcommand{\arraystretch}{1.4}
\begin{tabularx}{\textwidth}{|Y| Y | Y || Y|| Y| Y| Y||  Y| Y| }
  \hline
 Z & N & A & J & ${\cal M}_{\rm lin}$ &${\cal M}_1$ & ${\cal M}_2$ & Method & $\widehat{R}$\\
 \hline\hline
\multicolumn{3}{|c||}{Calcium (Ca)}  \\
\hline
\hline
8+12 & 8 & 28 & 0  &  0 & 0  & 0  &  Exact &\\
    & 8+1 & 29 & ${5}/{2}$  &  0 & 0  & 0  & Exact &\\ 
    & 8+2 & 30 & 0 & 0.5648 &  1.7100  &  1.2002  & Exact &\\ 
    & 8+3 & 31 & ${5}/{2}$ & 0.2395  & 1.0050  &  0.3949 & Exact &\\ 
    & 8+4 & 32 & 0  & 0.6459  & 2.3703  & 1.4979  & Exact &\\ 
    & 8+5 & 33 &  ${5}/{2}$ &  0.3027 &  1.2690 & 0.5202  & Exact &\\ 
    & 8+6 & 34 &  0 &  0.2995 & 1.2799  & 0.5136  & Exact &\\ 
    & 8+7 & 35 &  ${1}/{2}$ &  0.2792 & 1.2195  & 0.4724  & Exact &\\ 
    & 8+8 & 36 &  0 & 0.3089  & 1.1254  & 0.5330  & Exact &\\ 
    & 8+9 & 37 &  ${3}/{2}$ & 0.1792  & 0.6842  & 0.2845  & Exact &\\ 
    & 8+10 & 38 &  0 &  0.1946 & 0.6895  &  0.3123 & Exact &\\ 
    & 8+11 & 39 & ${3}/{2}$  & 0  & 0  & 0  & Exact &\\ 
    & 8+12 & 40 & 0 & 0  & 0  &  0 & Exact &\\
 \hline\hline
\end{tabularx}
\label{tab:magicresultsCa}
\end{table}

\end{document}